\documentclass[aps,pra,reprint,showkeys,longbibliography]{revtex4-2} 
\usepackage{amsmath}
\usepackage{amssymb}
\usepackage{graphicx}
\usepackage{bm}
\usepackage{xcolor}
\usepackage[toc,page]{appendix}
\usepackage{braket}

\makeatletter                           
\@ifclasswith{revtex4-2}{draft}{%
    \usepackage[notref,notcite]{showkeys}
    
}{}
\makeatother

\newcommand{\hateq}{\;\hat{=}\;}
\newcommand{\nn}{\nonumber}
\newcommand{\gl}{\big{(}}
\newcommand{\gr}{\big{)}}

\newcommand{\tr}{\mathrm{tr}}
\newcommand{\subt}[1]{_{\text{#1}}}

\newcommand{\qq}[1]{``#1''}

\newcommand{\eps}{\varepsilon}
\newcommand{\Shat}{\widehat{S}}
\newcommand{\ie}{i.e.\ }
\newcommand{\eg}{e.g.\ }
\newcommand{\rhs}{r.h.s.\ }

\begin{document}

\title{Complex wave functions, CPT and quantum field theory for\protect\\ classical generalized Ising models}

\author{Christof Wetterich}
\affiliation{Institut f\"ur Theoretische Physik\\
    Universit\"at Heidelberg\\
    Philosophenweg 16, D-69120 Heidelberg}

\begin{abstract}

    The quantum or quantum field theory concept of a complex wave function is useful for understanding the information transport in classical statistical generalized Ising models. 
    We relate complex conjugation to the discrete transformations charge conjugation ($C$), parity ($P$) and time reversal ($T$).
    A subclass of generalized Ising models are probabilistic cellular automata (PCA) with deterministic updating and probabilistic initial conditions.
    Simple two-dimensional PCA correspond to discretized quantum field theories for Majorana--Weyl, Weyl or Dirac fermions.
    Momentum and energy are conserved statistical observables.
    For PCA describing free massless fermions we investigate the vacuum and field operators for particle excitations.
    For the correlation function one finds the Lorentz-invariant Feynman propagator of quantum field theory.
    Furthermore, these automata admit probabilistic boundary conditions that correspond to thermal equilibrium with the quantum Fermi--Dirac distribution.
    PCA with updating sequences of propagation and interaction steps can realize a rich variety of discrete quantum field theories for fermions with interactions.
    For information theory the quantum formalism for PCA sheds new light on deterministic computing or signal processing with probabilistic input.

\end{abstract}

\maketitle


\section{Introduction}
\label{sec:introduction}

For a long time the quantum laws for probabilities have been considered as fundamentally different from classical statistics.
Bell's inequalities \cite{BELL, CHSH} for classical correlation functions and the experimental establishment of their violation in quantum systems \cite{ASDA} have been interpreted as a no go theorem for the embedding of quantum mechanics in classical statistics.
In the present paper we demonstrate that quantum concepts as complex wave functions, operators for observables and a linear time evolution law are very useful for an understanding of the transport of probabilistic information in classical statistical systems.
We focus on generalized Ising models, as often used in information theory \cite{SHA}.
In particular we demonstrate that probabilistic cellular automata with deterministic invertible updating are equivalent to discrete quantum field theories for fermions.
For these models quantum systems follow indeed from a classical probabilistic setting.

\indent Ising type models \cite{LENZ, ISING, BINDER} have been used for a description of a large variety of physical systems.
They are at the basis of information theory.
The Ising spins with values $s_i = 1,-1$ can be associated to bits or fermionic occupation numbers $n_i = 1,0$ by the simple relation $n_i=(s_i+1)/2$.
The bits or Ising spins are placed on a lattice and we assume that the lattice can be ordered as a sequence of hypersurfaces.
The hypersurfaces can be labeled by a discrete time coordinate, defining probabilistic time as an ordering structure for observables \cite{CWPT}.
We investigate the large class of models for which interactions occur only for Ising spins on neighboring hypersurfaces.

\indent Our focus is on the boundary problem: Assume that boundary conditions are specified by probabilistic information about the configurations of the Ising spins on the hypersurfaces at initial and final time $t_{in}$ and $t_f$.
What are the expectation values and correlations for spins in the bulk at arbitrary time $t$, $t_{in} < t < t_f$?
This is a problem of information transport from the boundary to the interior of the bulk.
The probabilistic information at $t_{in}$ and $t_f$ is encoded in boundary terms of the generalized Ising models.
If the boundary term at $t_f$ does not fully specify the probabilistic distribution for the configurations of spins at $t_f$, one can also ask how the probability distribution of the \qq{receiver} at $t_f$ is influenced by the one of the \qq{sender} at $t_{in}$.

\indent These issues of information transport can be tackled in a sequential way by an evolution law how the probabilistic information is transported from one hypersurface to the next, or from $t$ to $t+\eps$.
A linear evolution law can be formulated in terms of wave functions or probability amplitudes \cite{CWIT}.
The corresponding generalized quantum formalism for classical statistics \cite{CWQF} computes expectation values of observables in terms of operators acting on wave functions in close analogy to quantum mechanics.
The evolution law involves the step evolution operator which corresponds to a particular normalization of the transfer matrix \cite{BAX, SUZ, FUCH}.
In a sense, the quantum formalism for classical statistics corresponds to a Schrödinger picture, while the usual transfer matrix formalism may be associated to a Heisenberg picture.
In general, the operators for observables do not commute with the step evolution operator. Contrary to widespread prejudice non-commuting matrices play an important role in classical statistics.

\indent The quantum formalism for the information transport in generalized Ising models starts with real wave functions \cite{CWQPCS}, and a real step evolution operator.
Quantum mechanics involves a complex wave function.
This complex structure allows for the powerful tool of complex basis transformations.
It permits eigenstates of momentum or energy.
The phases of the complex wave function are crucial for phenomena as interference. 
A natural question asks if these features can be realized in classical statistical systems. 
Indeed, a complex structure can also be implemented for the wave functions of classical statistical systems \cite{CWPW2020}, provided that the step evolution operator is compatible with two suitable discrete transformations.

\indent The focus of the present paper is a detailed discussion of a complex structure associated to the particle-hole transformation.
This transformation flips the sign of all Ising spins, $s_i \to -s_i$.
Equivalently, it interchanges particles with $n_i=1$ and holes with $n_i=0$, or \qq{occupied} with \qq{empty} bits.
In the complex picture this transformation is mapped to complex conjugation.
It is closely related to charge conjugation in quantum field theories.
This applies to particle physics or to statistical systems with excitations of a vacuum, or more general equilibrium state,
consisting of particles and holes.
For the present paper we consider two-dimensional lattices.

\indent We supplement the discussion of charge conjugation $C$ with an investigation of the discrete transformations parity $P$ and time reversal $T$. 
Again, these symmetries are realized directly on the level of the Ising spins by suitable transformations $s_i(t,x) \to s_i (t,-x)$ or $s_i(t,x) \to s_i (-t,x)$, with $x$ denoting the location in space.
This induces the action of $P$ and $T$ on wave functions.
If the discrete symmetries $C$, $P$, $T$, or combinations thereof as $PT$ or $CPT$, are realized in the bulk (except the boundary conditions), the evolution equation for the probabilistic information respects these symmetries. 
This restricts the form of the step evolution operator.
We discuss the interesting possibility that beside the particle-hole conjugation the complex structure also involves the $CPT$-transformation.
For such a choice complex conjugation amounts to the $PT$-transformation.
The most commonly used version for time reversal in particle physics uses $\tilde{T} = CT$.
Then complex conjugation corresponds to the $CP \tilde{T}$-transformation.

\indent Having established the formalism for complex wave functions and the discrete symmetries $C$, $P$ and $T$, we turn in the second part to more concrete applications.
For this purpose we discuss particularly simple generalized Ising models, namely probabilistic cellular automata (PCA). Cellular automata \cite{JVN, ULA, ZUS} have found applications in wide areas of science \cite{HED, GAR, RICH, AMPA,LIRO, HPP, TOOM, DKT, WOLF, VICH, PREDU, CREU, TOMA, FLN}. They have been proposed for a deterministic understanding of quantum mechanics \cite{GTH, ELZE, HOOFT2, HOOFT3, HOOFT4, TEL2}.
We consider here probabilistic cellular automata \cite{VER, DOM, DK, GJH, RU, GD, LMS, PA, FU, MM, RLK, PE} of a particular type: The updating rule is deterministic and invertible, while the probabilistic aspects enter only by probabilistic initial conditions.
The deterministic updating guarantees a unitary evolution law.
Since the step evolution operator is orthogonal we can express it in terms of a hermitian Hamiltonian.
We discuss here classical PCA and not quantum cellular automata in the sense of ref.~\cite{Arrighi2019}.
Nevertheless, this type of PCA describes discrete quantum systems \cite{CWIT,CWPW2020}.
With the complex structure introduced in the first part, all features of the quantum formalism with complex wave functions are realized.
The probabilistic setting is crucial for smooth wave functions, as eigenstates of momentum or energy.

\indent There exists a close general relation between fermions and generalized Ising models \cite{CWFCS, CWFGI}.
It is based on the observation that in a real formulation the Hilbert space for fermions in the occupation number basis is the same as for Ising models, with basis states given by the configurations of occupation numbers.
Correspondingly, the discrete quantum systems obtained from the formulation of PCA as generalized Ising models are particular quantum field theories for fermions in a discrete setting \cite{CWPCA, CWFCB, CWNEW, CWPCAQP, CWFPPCA}.
Every PCA with a deterministic and invertible updating rule constitutes a discrete quantum field theory for fermions.
This holds in a generalized sense which does not require a complex structure or Lorentz symmetry.
(See also ref. \cite{PLECH, BER1, BER2, SAM, ITS, PLE1} for different relations between fermions and Ising spins for particular cases.) 

\indent For updating rules that involve some form of interaction the precise form of the quantum Hamiltonian may not easily be found.
One is guaranteed, however, that a hermitian Hamiltonian $H$ and the associated unitary evolution exist. 
In case of time-translation invariance this implies the existence of a conserved energy observable.
For $H$ not explicitly known as for an example of automata with stochastically distributed scattering points, the quantum formalism is crucial for unveiling the properties of $H$ and the associated evolution \cite{Kreuzkamp2024}.

\indent For conserved $H$ one can investigate probabilistic initial conditions which correspond to eigenstates of the Hamiltonian.
The time-evolution of the probabilistic information is then periodic, thus realizing a crucial feature of quantum systems.
This holds in general, even for very complex PCA.
It may be difficult to find the energy eigenstates in practice, but their existence dictates general features of the probabilistic evolution, see ref.~\cite{Kreuzkamp2024} for an example.

\indent For the step evolution operator for generalized Ising models which describes free massless fermions in one time and one space dimension one finds a simple expression of the Hamiltonian in terms of fermionic annihilation and creation operators in momentum space.
This simplicity requires the possibility of complex basis transformations, as realized for PCA with a suitable complex structure.
It allows the construction of a \qq{particle physics vacuum} for which all excitations have positive energy.
The Hamiltonian $H$ typically has positive and negative eigenvalues.
The particle physics vacuum corresponds to a complete filling of all states with negative energy, implying a non-vanishing \qq{ground state energy} $E_0$. 
This vacuum is characterized by a minimum of $H$.
Both particle- and antiparticle-(hole) excitations have energy larger than $E_0$.
We evaluate the Feynman propagator for this vacuum, which takes in the continuum limit the standard form for free massless Weyl- or Dirac-fermions.

\indent For simple transport automata corresponding to a quantum field theory for free massless fermions the evolution of the wave function for the one-particle excitations obeys the relativistic Dirac equation and therefore realizes Lorentz symmetry.
We also discuss boundary conditions that correspond to thermal equilibrium states.
For these states the momentum distribution amounts to the Fermi--Dirac distribution.
We may interpret this as a sign of the quantum nature of the classical statistical system realizing the PCA.
As the temperature is taken to zero the thermal equilibrium state approaches the particle physics vacuum.

\indent In the opposite direction we can view the transport automata as a lattice formulation or free chiral fermions in two dimensions.
The discretization is done directly for a \qq{euclidean functional integral}, without invoking any analytic continuation for a description of the evolution in \qq{real} time.
For this discretization we have found no sign of \qq{fermion doublers}.

\indent Interactions among the fermions can be introduced by automata that feature a sequence of propagation and interaction steps, reminiscent of the setting in Feynman's path integral.
The interaction part is typically local in space and realizes simple permutations of spin configurations in a local cell.
The interaction part of the step evolution operator does not commute with the propagation part.
Furthermore, the interactions are not small.
These properties turn the establishment of a continuum limit into a difficult task. 
We limit the present paper to a discussion of two examples.
One features scattering with particle production and a modified propagation.
The other realizes Lorentz symmetry in the naive continuum limit.

\indent On the formal side we introduce field operators for fermions in close analogy to the more standard formulation of quantum field theory.
These field operators interpolate continuously between the annihilation and creation operators on the lattice sites.
They introduce a continuous space-time manifold even for the discrete setting without taking the continuum limit.
The continuous field operators permit the realization of continuous symmetries as chiral symmetry or Lorentz symmetry.
We also show how the usual complex functional integral for quantum field theories is related to the real (\qq{euclidean}) functional integral for the generalized Ising models describing PCA.

\indent In sect.~\ref{sec:generalized_ising_models} we introduce wave functions for generalized Ising models and operators for observables.
This constitutes the quantum formalism for classical statistics.
Sect.~\ref{sec:complex_structure} turns to the complex structure which is related to the particle-hole conjugation. We discuss the meaning of the phases of the complex wave function and the linear time evolution law in the complex picture.
Sect.~\ref{sec:fermion_picture} is devoted to the fermionic description of generalized Ising models.
We express the step evolution operator (transfer matrix) and the associated Hamiltonian in terms of fermionic annihilation and creation operators.
In momentum space the Hamiltonian for transport automata finds a simple expression.
We discuss the real wave function for Majorana--Weyl fermions and the complex wave function for Weyl fermions.
In two dimensions they describe right-movers or left-movers, which are combined to Dirac fermions in sect.~\ref{sec:dirac_automaton}. 
Sect.~\ref{sec:parity_and_time_reversal} addresses parity and time reversal, while charges and charge conjugation are discussed in sect.~\ref{sec:charges_and_charge_conjugation}.

\indent In sect.~\ref{sec:vacuum} we turn to the concept of vacuum.
For PCA we focus on the half-filled \qq{particle physics vacuum} for which all excitations have positive energy. 
Sect.~\ref{sec:one-particle-states} investigates the one-particle excitations for this vacuum.
In sec.~\ref{sec:field-operators} we formulate field operators which depend continuously on space and time.
Applying products of those to the vacuum produces multi-particle states.
By use of the field operators we compute the Feynman propagator for the particle physics vacuum.
The continuous field operators permit the formulation of continuous symmetries as Lorentz symmetry or chiral symmetry even for the discrete setting of the generalized Ising models without taking a continuum limit.
Sect.~\ref{sec:Density_matrix_thermal_equilibrium} introduces the density matrix and describes the thermal equilibrium state for PCA.
In sect.~\ref{sec:interactions} we discuss PCA which are equivalent to discrete quantum field theories for fermions with interactions.
They are described by sequences of updating for propagation and interaction, in analogy to Feynman's path integral.
Sect.~\ref{sec:complex_functional_integral} establishes the connection between the euclidean functional integral and the more familiar complex functional integral for quantum field theories.
In sect.~\ref{sec:conclusions} we present our conclusions.

\section{Generalized Ising models}
\label{sec:generalized_ising_models}

We consider a two-dimensional square lattice with lattice points labeled by $(t,x)$ and lattice distance $\eps$. 
On each lattice point we place Ising spins $s_\gamma(t,x)=\pm1$, with $\gamma$ denoting different species or colors. 
The action $S$ is a function of the spin configurations $\omega=\{s_\gamma(t,x)\}$. 
We choose boundary conditions which are periodic in $x$, while for $t$ we specify boundary terms $\mathcal{B}\subt{in}[s_\gamma(t\subt{in},x)]$ and $\mathcal{B}\subt{f}[s_\gamma(t\subt{f},x)]$ which only depend on the spins at the initial time $t\subt{in}$ or final time $t\subt{f}$. 
The overall probability distribution for $\{p_\omega\}$ the spin configurations is given by
the set of overall probabilities $p_\omega$,
\begin{equation}
    \label{eq:G1}
    p_\omega = \mathcal{B}\subt{f}(\omega)\exp\big\{-S(\omega)\big\}
    \mathcal{B}\subt{in}(\omega)\,.
\end{equation}
Here we choose $S$, $\mathcal{B}\subt{in}$ and $\mathcal{B}\subt{f}$ such that
$p_\omega\geq0$ and $\sum_\omega p_\omega = 1$. This can be achieved by additive shifts in $S$ and multiplicative factors in $\mathcal{B} = \mathcal{B}\subt{f} \mathcal{B}\subt{in}$. 
Positivity of $p_\omega$ requires $\mathcal{B} > 0$, while $\mathcal{B}_f$ or $\mathcal{B}\subt{in}$ can take both positive or negative values.
We emphasize that $\mathcal{B}\subt{in}$ alone does not specify the expectation values of the Ising spins at $t\subt{in}$.
Those also involve $S$ and $\mathcal{B}_f$.
The boundary term $\mathcal{B}\subt{in}$ is not a time-local probability distribution for the spins $s_\gamma(t\subt{in})$.
We will see that it rather corresponds to a type of wave function at $t\subt{in}$.
We may equivalently encode the overall probability distribution in the function $\bar{p}(\omega)$, which maps each spin configuration $\omega$ to a probability $p_\omega$.

\indent The expectation value of an observable $A$ which takes the value $A_\omega$ for the spin configuration $\omega$ obeys the standard classical statistical rule
\begin{equation}
    \label{eq:G1A}
    \langle A \rangle = \sum_\omega A_\omega p_\omega\,.
\end{equation}
Besides this defining relation for expectation values we will use no other additional axioms.
The probabilities for the quantum system corresponding to PCA will follow from eq.~\eqref{eq:G1A}.
We take a finite number of lattice points $N_t$ in the $t$-direction, and a finite odd number $N_x$ in the $x$-direction. 
The \qq{functional integral} over spin configurations amounts to the finite sum $\sum_\omega$. 
A continuum limit of diverging $N_t$, $N_x$, or $\eps\to0$ for fixed $L = N_x\eps$, $\Delta t = N_t\eps$, can be taken at the end.

\subsection*{Local chains}
We consider the large class of generalized Ising models which are \qq{local
    chains}, where
\begin{equation}
    \label{eq:G2}
    S(\omega) = \sum_{t=t\subt{in}}^{t\subt{f}-\eps} \mathcal{L}(t)\,,
\end{equation}
with $\mathcal{L}(t)$ only involving spins at two neighboring time layers
\begin{equation}
    \label{eq:G3}
    \mathcal{L}(t) = \mathcal{L}\big[ s_\gamma(t+\eps,x), s_\gamma(t,x)\big]\,.
\end{equation}
Two examples with rather different properties are the nearest neighbor Ising
model (with constant $\Delta$),
\begin{align}
    \label{eq:G4}
    \mathcal{L}\subt{Is}(t) = & \, -\beta\sum_x\big\{ s(t+\eps,x)s(t,x) + \frac12
    s(t,x+\eps)s(t,x) \nn                                                                   \\
                              & + \frac12s(t+\eps,x+\eps)s(t+\eps,x)  + \Delta - 2\big\}\,,
\end{align}
or the Majorana--Weyl automaton for which
\begin{equation}
    \label{eq:G5}
    \mathcal{L}\subt{MW} = \lim_{\beta\to\infty} \left(-\beta\sum_x\big\{
    s(t+\eps,x+\eps)s(t,x) - 1\big\} \right)\,.
\end{equation}

\indent The second model has diagonal interactions only in one direction. 
Consider the spin configuration at $t$, denoted by $\tau = [s(t,x)]$, and the one at $t+\eps$, denoted by $\rho = [s(t+\eps,x)]$.
The expression $\mathcal{L}\subt{MW}$ depends both on $\tau$ and on $\rho$.
If $\rho$ equals precisely the configuration $\tau$ displaced by one position to the right, all factors $s(t+\eps,x+\eps)s(t,x)$ equal one and $\mathcal{L}\subt{MW}(t) = 0$. 
On the other hand, for all other spin configurations $\rho$ at $t+\eps$ not obeying this simple displacement rule at least one of the terms in the sum \eqref{eq:G5} equals $(-1)$.
One then finds a diverging $\mathcal{L}\subt{MW}(t)$ due to the limit $\beta\to\infty$, and $\exp(-\mathcal{L}\subt{MW})=0$.
As a result, the factor $\exp(-S)$ in eq.~\eqref{eq:G1} vanishes and such configurations have zero probability. 
An overall spin configuration $\omega = \{s(t,x)\}$ with non-zero probability can be labeled by an initial configuration $[s(t\subt{in},x)]$.
The corresponding overall spin configuration $\omega$ is the one for which $[s(t\subt{in},x)]$ is displaced by one position to the right after each time step $\eps$. 
For all other $\omega$ one has $p_\omega = 0$.
This model describes a probabilistic cellular automaton (PCA). The updating rule is deterministic. 
Each spin configuration $[s(t,x)]$ is displaced by one position to the right at each time step. 
The probabilistic aspect enters only by the boundary term $\mathcal{B}$. 
It associates to each initial configuration $[s(t\subt{in},x)]$ a probability.

\indent For a PCA the initial probabilities extend to the overall probability distribution ${p_\omega}$, since they are transported to subsequent time steps by the deterministic updating. 
A given initial spin configuration $[s(t\subt{in},x)]$ defines a \qq{trajectory} in the space of overall spin configurations $\omega = \{s(t,x)\}$. 
(In our notation we distinguish time-local spin configurations $\tau = [s(t,x)]$ for spins at a given time $t$ and overall configurations $\omega = \{s(t,x)\}$ for spins at all times.) 
The trajectory is the sequence of time-local spin configurations $\{[s(t\subt{in},x)], [s(t\subt{in}+\eps,x)], [s(t\subt{in}+2\eps,x)],\dots\}$ that obtains from the initial configuration $[s(t\subt{in},x)]$ by applying the updating rule of the automaton.
Each initial spin configuration $[s(t\subt{in},x)]$ defines a trajectory. 
The probability for the trajectory equals the probability of the initial spin configuration from which it originates. 
The probabilities $p_\omega$ vanish for all overall spin configurations which do not belong to a trajectory. 
As a consequence, the configuration sum $\sum_\omega$ reduces to a sum over trajectories. 
As compared to the Ising model this is an enormous simplification. Instead of summing all overall spin configurations one only needs to sum trajectories or equivalently, time-local spin configurations at $t\subt{in}$.

\indent We will employ a formalism that describes the time evolution of the probabilistic information for both models \eqref{eq:G4} and \eqref{eq:G5}.
It is based on the concept of wave functions for classical statistics.
The properties of the evolution will be rather different for the two models, however.
For the next neighbor Ising model \eqref{eq:G4} the wave function approaches an equilibrium wave function as one proceeds from the boundary to the bulk.
The rate of this approach to equilibrium involves the correlation length (or \qq{correlation time}).
Deep inside the bulk, at distances from the boundaries much bigger than the correlation length, the boundary information is lost.
Expectation values of Ising spins deep inside the bulk are given by an equilibrium distribution independently of the boundary term $\mathcal{B}$.
In contrast, for the PCA of model \eqref{eq:G5} the boundary information is never lost.
The expectation values for spins at all $t$ depend on the boundary terms $\mathcal{B}$.
The probability for a time-local spin configuration $[s(t,x)]$ equals the one for the initial spin configuration at $t\subt{in}$ from which it originates by the updating rule.
The conservation of the initial probabilistic information is characteristic for the unitary time evolution in quantum mechanics.

\subsection*{Wave functions for generalized Ising models}
We are interested in the time-local probabilistic information at a given time $t$. 
This obtains by summing in a convenient way over configurations at $t'\neq t$. 
We will see that the time-local probabilistic information is best described by a pair of classical wave functions. 
Their evolution with time obeys a simple linear evolution equation in close analogy to the Schrödinger equation in quantum mechanics.

\indent For local chains we can split at any given $t$ the action into
\begin{align}
    \label{eq:G6}
    S =     & \, S_{<} + S_{>}\,,\nn                        \\
    S_{<} = & \, \sum_{t'<t}\mathcal{L}(t')\,,\quad S_{>} =
    \sum_{t'\geq t}\mathcal{L}(t')\,.
\end{align}
One may define a pair of \qq{wave functions for the universe} by
\begin{equation}
    \label{eq:G7}
    \tilde{\phi}(\omega) = \exp\gl-S_{<}\gr \mathcal{B}\subt{in}\,,\quad
    \bar{\phi}(\omega) = \mathcal{B}\subt{f} \exp\gl-S_{>}\gr\,,
\end{equation}
with
\begin{equation}
    \label{eq:G8}
    \bar{p}(\omega) = \bar{\phi}(\omega) \tilde{\phi}(\omega)\,.
\end{equation}

\indent Time-local wave functions obtain by integrating out variables $s_\gamma(t'<t,x)$
for
\begin{equation}
    \label{eq:G9}
    \tilde f(t;[s_\gamma(t,x)]) = \sum_{\{s_\gamma(t'<t,x)\}} \tilde{\phi}(\omega)\,,
\end{equation}
and $s_\gamma(t'>t,x)$ for
\begin{equation}
    \label{eq:G10}
    \bar{f}(t;[s_\gamma(t,x)]) = \sum_{\{s_\gamma(t'>t,x)\}} \bar{\phi}(\omega)\,.
\end{equation}
Both wave functions $\tilde{f}(t)$ and $\bar{f}(t)$ depend on the time-local configuration $\tau = [s_\gamma(t,x)]$. 
The time-local probabilities $p(t;[s_\gamma(t,x)])$ obtain from the overall probability distribution $\{p_\omega\}$ by integrating over all spins with $t'\neq t$.
They obey
\begin{equation}
    \label{eq:G11}
    \begin{aligned}
        p(t;[s_\gamma(t,x)]) &= \sum_{\{s_\gamma(t' \ne t, x)\}} \bar{p}(\omega) \\
                             &= \bar{f}(t;[s_\gamma(t,x)])\tilde{f}(t;[s_\gamma(t,x)])\,.
    \end{aligned}
\end{equation}
The pair of wave functions $\bar{f}$, $\tilde{f}$ contains more time-local probabilistic information than the time-local probability distribution. 
This will be important for formulating a simple evolution law.

\indent We may expand arbitrary wave functions in terms of some fixed basis functions
$h_\tau[s_\gamma(t,x)]$,
\begin{align}
    \label{eq:G12}
    \tilde{f}(t;[s_\gamma(t,x)]) = & \, \sum_\tau \tilde q_\tau(t)h_\tau[s_\gamma(t,x)]\,,\nn \\
    \bar{f}(t;[s_\gamma(t,x)]) =   & \, \sum_\tau \bar q_\tau(t)h_\tau[s_\gamma(t,x)]\,.
\end{align}
A basis well adapted for our purpose is the \qq{occupation number basis}~\cite{CWIT, CWQF}. 
The basis function $h_\tau$ equals one for the particular spin configuration $\tau$ and vanishes for all other spin configurations.
The sets $\{\tilde q_\tau(t)\}$, $\{\bar q_\tau(t)\}$ are then the wave functions in this basis.
The time-local probabilities obey
\begin{equation}
    \label{eq:G12A}
    p_\tau(t) = \bar q_\tau(t)\tilde q_\tau(t)\,.
\end{equation}

\indent Following the time evolution from $\tilde q(t)$ to $\tilde q(t+\eps)$ one only adds an additional factor $\exp(-\mathcal{L}(t))$ in the definition~\eqref{eq:G7} and an additional integration over $[s_\gamma(t,x)]$ in eq.~\eqref{eq:G9}. 
This results in a linear evolution law~\cite{CWIT,CWQF,CWPW2020},
\begin{align}
    \label{eq:G13}
    \tilde{q}_\tau(t+\eps) & = \sum_\rho \Shat(t)_{\tau\rho}\tilde{q}_\rho(t)\nn             \\
    \bar{q}_\tau(t+\eps)   & = \sum_\rho \gl\Shat^T(t)\gr^{-1}_{\tau\rho} \bar{q}_\rho(t)\,.
\end{align}
The step evolution operator $\Shat$ \cite{CWIT} corresponds to the transfer matrix~\cite{BAX, SUZ, FUCH} in a
normalization where the largest absolute value of its eigenvalues equals one.
This normalization is achieved by an additive shift in $\mathcal{L}(t)$, as by
$\Delta$ in eq.~\eqref{eq:G4}. In the occupation number basis there exists a
simple algorithm for computing $\Shat(t)$ from $\mathcal{L}(t)$~\cite{CWIT, CWQF, CWPW2020}.

\subsection*{Probabilistic cellular automata}
For invertible automata the step evolution operator takes the simple form of a \qq{unique jump operator}. It is a matrix which has in each row and column only a single non-vanishing element $\pm1$,
\begin{equation}
    \label{eq:G14}
    \Shat_{\tau\rho} = \pm\delta_{\tau,\tau(\rho)}\,.
\end{equation}
Here $\tau(\rho)$ denotes the spin configuration that obtains from $\rho$ by the deterministic updating rule of the automaton. 
For automata the step evolution operator is orthogonal, such that $\tilde q(t)$ and $\bar q(t)$ obey the same evolution law. 
For time-local probabilities eq.~\eqref{eq:G12A} implies
\begin{equation}
    \label{eq:G15}
    p_\tau(t+\eps) = \delta_{\tau,\tau(\rho)} p_\rho(t) = p_{\rho(\tau)}(t)\,,
\end{equation}
where $\rho(\tau)$ denotes the configuration from which $\tau$ originates by the updating. 
For the example of the Majorana--Weyl automaton~\eqref{eq:G5} the configuration $\tau(\rho)$ obtains from $\rho$ by displacing all spins by one position to the right. 
For all other overall spin configurations $[s_\gamma(t,x)]$ not on the trajectory of $\rho$ given by by the updating the probability vanishes due to the $\delta$-function in eq.~\eqref{eq:G15}.

\indent For automata a further important simplification occurs.
We can choose boundary conditions such that $\bar q(t\subt{in}) = \tilde q(t\subt{in})$.
Since $\Shat^T\Shat = 1$ both $\bar q$ and $\tilde q$ obey the same time evolution.
Therefore this boundary condition implies for all $t$
\begin{equation}
    \label{eq:G16}
    \bar q(t) = \tilde q(t) = q(t)\,.
\end{equation}
We are left with a single real wave function $q(t)$. The components of the wave
function are the square root of the probabilities
\begin{equation}
    \label{eq:G17}
    p_\tau(t) = q_\tau^2(t)\,.
\end{equation}
The normalization of the probabilities implies that $q$ is a unit vector
\begin{equation}
    \label{eq:G18}
    \sum_\tau q_\tau^2 = 1\,.
\end{equation}
The final boundary condition $\mathcal{B}\subt{f}$ translates in the occupation number basis to $\bar q(t\subt{f})$ and similarly $\tilde q(t\subt{in})$ encodes $\mathcal{B}\subt{in}$.
The choice of $\bar q(t\subt{f})$ which leads to the identification~\eqref{eq:G16} is actually no restriction.
It is sufficient to choose $\mathcal{B}\subt{in}$ such that $\tilde q_\tau^2(t\subt{in}) = p_\tau(t\subt{in})$.
Eq.~\eqref{eq:G12A} implies $\bar q_\tau(t\subt{in}) = \tilde q_\tau (t\subt{in})$ for all $\tilde q_\tau(t\subt{in})\neq 0$.
Trajectories generated for initial configurations $\tau$ with $\tilde q_\tau(t\subt{in}) = 0$ have zero probabilities and do not contribute to the overall probability distribution $p_\omega$.
This is independent of the choice of $\bar q_\tau(t\subt{f})$ for the configurations $\tau$ reached by these trajectories at $t\subt{f}$.
The time evolution becomes an initial value problem.
For automata no detailed knowledge about the future is needed~\cite{CWPW2020}.

\indent Instead of the Ising spins $s_\gamma(t,x)=\pm1$ we may use fermionic occupation numbers or bits with values $n_\gamma(t,x) = 1,0$,
\begin{equation}
    \label{eq:G19}
    n_\gamma(t,x) = \frac12\gl s_\gamma(t,x) + 1 \gr\,.
\end{equation}
The Majorana--Weyl automaton describes in this language an arbitrary number of fermions which move one position to the right at each time step.
These right-movers correspond to a discretized two-dimensional quantum field theory for Majorana--Weyl or Weyl fermions in the occupation number basis.
Majorana--Weyl fermions can indeed be described by real wave functions.
For smooth initial wave functions the continuum limit $\eps\to0$ at fixed $N_t\eps$, $N_x\eps$ can be
taken~\cite{CWPW2020}.
We will see below that for suitable vacuum states this automaton can also describe the
complex wave function for Weyl fermions.

\subsection*{Operators for observables}
Classical time-local observables associate to each configuration of spins $\tau = [s_\gamma(t)]$ at a given time $t$ a value $A_\tau$.
One can define an associated diagonal operator $\hat{A}_{\tau\rho} = A_\tau \delta_{\tau\rho}$.
The expectation value can be computed from the wave functions by the (generalized) quantum rule
\begin{equation}
    \label{eq:OP1}
    \langle A \rangle
    = \sum_{\tau} A_{\tau} p_\tau(t)
    = \sum_{\tau,\rho} \bar{q}_\rho(t) \hat{A}_{\rho\tau} \tilde{q}_\tau(t)
    = \left\langle \bar{q}(t) \hat{A} \tilde{q}(t) \right\rangle\,.
\end{equation}
For cellular automata \eqref{eq:G16} this is the usual quantum rule in a real formulation.
The quantum rule for expectation values can be extended to more general non-diagonal operators that we will encounter below \cite{CWPW2020}.

\indent For local observables we can express the expectation value $\langle A(t) \rangle$ in terms of a chain of step evolution operators \cite{CWIT, CWQF, CWPW2020}
\begin{equation}
    \label{eq:OP2}
    \langle A(t) \rangle = \tr\left\{\Shat(t_f-\eps)\dots\Shat(t) \hat{A}\, \Shat(t-\eps) \dots \Shat(t_{\text{in}}) \hat{B}\right\}\,,
\end{equation}
where we did not write the sums over indices in the matrix expression.
The matrix $\hat{B}$ involves the boundary terms.
If $\hat{A}$ does not commute with the step evolution $\Shat$ the position of the operator insertion matters.

\section{Complex structure}
\label{sec:complex_structure}

Quantum mechanics and quantum field theories are generally based on complex wave functions. The Schrödinger equation is a complex equation. Complex wave
functions are crucial for basis transformations as the Fourier transform or for
an understanding of interference effects. It is, of course, always possible to
cast quantum mechanics into a real formulation by taking the real and imaginary
parts of the complex wave functions as components of a real wave function with
twice the number of components. We are interested in the inverse of this
procedure where a suitable complex structure maps real wave functions to complex
wave functions with half the number of components. This requires the presence of
two anticommuting discrete linear transformations $K$ and $I$ with
\begin{equation}
    \label{eq:C1}
    K^2 = 1\,,\quad I^2 = -1\,,\quad K I = - I K\,.
\end{equation}
In the complex language $K$ will translate to complex conjugation, and $I$ to
the multiplication by $i$. There are typically many possible choices for
$(K,I)$. Useful complex structures should be compatible with the evolution.
We want to represent the wave function as a complex vector such that the evolution law becomes
\begin{equation}
    \label{eq:C2}
    \varphi(t+\eps) = V(t)\varphi(t)\,,
\end{equation}
with $V(t)$ a complex step evolution operator. For orthogonal $\Shat(t)$ the
matrix $V(t)$ is unitary
\begin{equation}
    \label{eq:C3}
    \varphi(t+\eps) = U(t)\varphi(t)\,,\quad U^\dagger(t)U(t) = 1\,.
\end{equation}
This is the case for automata.

\indent We focus here on a particular complex structure based on the particle-hole transformation.
In the appendix~\ref{app:A} we investigate general complex structures.
The mapping of operators in the real picture to complex operators in the complex picture is discussed in appendix~\ref{app:B}.

\subsection*{Particle-hole transformation and complex wave function}

We propose here a general complex structure which is based on the particle-hole
transformation. This transformation flips the sign of all Ising spins. It maps
fermions at each given site $(t,x)$, \eg $n_\gamma(t,x) = 1$, to holes at the
same site, $n_\gamma(t,x) = 0$, and vice versa. For generalized Ising models
this complex structure is compatible with the evolution provided that the action
is particle-hole symmetric. This is the case for our
examples~\eqref{eq:G4},~\eqref{eq:G5}.

We start our discussion for a single site $x$.
We assume the presence of two real wave functions $\tilde q$ and $\bar q'$ for given $t$.
The wave functions have two components
\begin{equation}
    \label{eq:C4}
    \tilde q = \begin{pmatrix} \tilde q_1 \\ \tilde q_0 \end{pmatrix}\,, \quad
    \bar q' = \begin{pmatrix} \bar q_1' \\ \bar q_0' \end{pmatrix}\,,
\end{equation}
where the upper components correspond to $n=1$, and the lower components to $n=0$,
for a bit or corresponding Ising spin. One may distinguish $\tilde{q}$ and $\bar{q}'$ by the value of a second bit, while in case of a single bit $\bar{q}'$ and $\tilde{q}$ may be identified.
We define the complex wave function by
\begin{equation}
    \label{eq:C5}
    \varphi = \frac12\big\{ \tilde q + \hat q + i\tau_3(\tilde q - \hat q) \big\}\,,
\end{equation}
where $\hat q$ is the particle-hole conjugate of $\bar q'$,
\begin{equation}
    \label{eq:C6}
    \hat q = \tau_1 \bar q' = \begin{pmatrix} \bar q_0' \\ \bar q_1' \end{pmatrix}\,.
\end{equation}
In the real formulation the complex conjugation $K$ maps
\begin{equation}
    \label{eq:C7}
    K(\tilde q) = \hat q = \tau_1 \bar q'\,,\quad K(\hat q) = \tilde q\,,\quad K(\bar
    q') = \tau_1\tilde q\,.
\end{equation}
It corresponds to a map from $\tilde q$ to $\bar q'$ combined with a
particle-hole transformation. The multiplication by $i$ is realized in the real
picture by the map $I$,
\begin{equation}
    \label{eq:C8}
    I(\tilde q) = i\tau_2 \bar q'\,,\quad I(\bar q') = i\tau_2 \tilde q\,.
\end{equation}

Both maps $K$ and $I$ are compatible with the identification $\bar q' = \tilde q
    = q$. General $\tilde q$ and $\bar q'$ are mapped to a two-component complex wave
function, corresponding to the four real numbers in $\tilde q$ and $\bar q'$. For
the identification $\tilde q = \bar q' = q$ this two-component description is
redundant, since only one complex component of $\varphi$ is independent,
\begin{equation}
    \label{eq:C9}
    \varphi_1 = \varphi_0 = \frac12\big\{ q_1 + q_0 + i(q_1 - q_0) \big\}\,.
\end{equation}
This single complex component corresponds to the two real components of $q$.

This complex structure generalizes to an arbitrary number of sites $x$. The
vectors $\tilde q$ and $\bar q'$ have now $2^{N_x}$ components.
For the representation of the distribution $\{\tilde{q}_\tau\}$ as a vector, we
employ a direct product basis, $x=j\eps$, $N_x$ odd,
\begin{align}
    \label{eq:C10}
    E_\tau & = \sigma_\tau\prod_x e(x) = \sigma_\tau\prod_j e(j)\nn \\
    =      & \, \sigma_\tau e\left( -\frac{N_x-1}{2}\right) \otimes
    e\left(-\frac{N_x-3}{2}\right)\otimes \dots e\left(\frac{N_x-1}{2}\right)
\end{align}
where $e(j)$ takes the values
$\left(\begin{smallmatrix}1\\0\end{smallmatrix}\right)$ or
$\left(\begin{smallmatrix}0\\1\end{smallmatrix}\right)$ for occupied or empty
bits at the site $x$.  The choice of signs $\sigma_\tau = \pm1$ will be discussed later.
The wave functions in this basis are given by the vectors
\begin{equation}
    \label{eq:C11}
    \tilde q = \tilde q_\tau E_\tau\,,\quad \bar q' = \bar q'_\tau E_\tau\,.
\end{equation}
From now on we employ the convention of summation over double indices.

We define the particle-hole conjugate of
$\tilde{q}$ by
\begin{equation}
    \label{eq:YY1}
    \tilde{q}^c = B \tilde{q}\,,\quad B = D T_1\,.
\end{equation}
Here
\begin{equation}
    \label{eq:YYA}
    T_1 = \gl\tau_1\otimes \tau_1\otimes \dots \otimes \tau_1\gr
\end{equation}
flips between occupied and empty bits at every position $x$. The matrix $D$ is a diagonal matrix with eigenvalues $\pm 1$, $D^2 = 1$. It governs the relative signs between $\sigma_\tau$ and $\sigma_{\tau^c}$. We will take
\begin{equation}
    \label{eq:YY2}
    D=(1\otimes \tau_3 \otimes 1 \otimes \tau_3 \otimes \dots \otimes \tau_3 \otimes 1)\,,
\end{equation}
and $N_x = 1\;\mathrm{mod}\; 4$, such that
\begin{equation}
    [D,T_1]=0 \,,\quad B^2 = (DT_1)^2 = 1\,.
\end{equation}
Similarly, we employ the particle-hole conjugate of $\bar{q}'$
\begin{equation}
    \label{eq:C12}
    \hat q = B\bar q' = \bar q'_\tau B E_\tau\,.
\end{equation}

\indent The complex wave function is defined by
\begin{equation}
    \label{eq:C13}
    \varphi = \frac12\big\{ \tilde q + \hat q + i T_3 ( \tilde q -  \hat q) \big\}\,,
\end{equation}
where
\begin{align}
    \label{eq:C14}
    T_3\tilde q = \tilde q_\tau T_3 E_\tau\,,\quad T_3\hat q = \hat q_\tau T_3
    E_\tau\,,\nn \\
    T_3 = \gl \tau_3\otimes\tau_3\otimes \dots \otimes\tau_3\gr\,,\quad\{T_3,B\}=0\,.
\end{align}
The anticommutation of $B$ and $T_3$ follows for odd $N_x$ since $\{T_1, T_3\} = 0$ in this case.
Complex conjugation of $\varphi$ corresponds to the map $q\leftrightarrow \hat q$.
The components of $\varphi$ involve the particle-hole conjugate of $\bar{q}'$,
\begin{equation}
    \label{eq:YY4}
    \varphi = \varphi_{\tau} E_\tau\,,\quad
    \hat{q}_{\tau} E_\tau = \bar{q}^{\prime c}_{\tau} E_\tau = \bar{q}^{\prime}_{\tau} B E_\tau\,.
\end{equation}
The normalization is given by
\begin{equation}
    \label{eq:40A}
    \varphi^\dagger \varphi = \varphi_\tau^* \varphi_\tau = \frac12 (\tilde{q}_\tau \tilde{q}_\tau + \bar{q}'_\tau \bar{q}'_\tau) =1
\end{equation}
One can generalize the complex structure~\eqref{eq:C13} by replacing $\hat q$ by some other real wave
function with the same number of components as $\tilde q$.

\subsection*{Constraints for complex wave functions}

\indent We discuss two different settings.
For the first the wave function $\bar{q}'$ is independent of $\tilde{q}$.
For example, we may distinguish $\tilde{q}$ and $\bar{q}'$ by an additional bit.
In sect.~\ref{sec:parity_and_time_reversal} we discuss the possibility to associate $\bar{q}'$ with the $T$-or $PT$-transform of $\tilde{q}$ and, in particular, in sect.~\ref{sec:charges_and_charge_conjugation} with the $CPT$-transform, which seems to be a very natural choice.
For the second setting $\bar{q}'$ is identical to $\tilde{q}$ or can be expressed in terms of $\tilde{q}$, $\bar{q}' = \tilde{G} \tilde{q}$.
In this case there is a constraint on the most general form of $\varphi$ beyond the normalization.
The complex $2^{N_x}$-component vector $\varphi$ can be expressed in terms of $2^{N_x}-1$ real numbers.

\indent For the case $\bar q' = \tilde q = q$ one has
\begin{equation}
    \label{eq:C15}
    \varphi = \frac12\big\{ q + B q + iT_3(q - B q) \big\}\,.
\end{equation}
In the real formulation the maps $K$ and $I$ read in this case
\begin{equation}
    \label{eq:C16}
    K(q) = B q\,,\quad I(q) = T_3 B q\,.
\end{equation}
Complex conjugation corresponds then to particle-hole conjugation. The complex wave function is normalized according to
\begin{equation}
    \label{eq:C17}
    \varphi^\dagger\varphi = q^Tq = q_\tau q_\tau = 1\,.
\end{equation}
Due to the identification $\bar q' = \tilde q$ only $2^{N_x-1}$ components of
$\varphi$ are independent.
Indeed, the complex wave function \eqref{eq:C15} obeys the identity
\begin{equation}
    \label{eq:C18}
    B \varphi = \varphi\,.
\end{equation}

For general $\tilde q$ and $\bar q'$ we may use the projectors $(1\pm B)/2$,
\begin{equation}
    \label{eq:C18A}
    \varphi^{\pm} = \frac{1\pm B}{2} \varphi\,.
\end{equation}
For $\bar q' = \tilde q$ the part $\varphi^-$ vanishes and we can consider $\varphi^+$ as
a complex wave function with $2^{N_x-1}$ independent components (up to
normalization). It can be written in the form
\begin{equation}
    \label{eq:C18B}
    \varphi^+(t) = \varphi^+_\tau(t) \frac{1+B}{2} E_\tau\,.
\end{equation}
For general $\bar{q}'$ the complex conjugation of $\varphi$ involves a map from $\tilde{q}$ to $\bar{q}'$ in addition to the particle-hole conjugation. The transformation $\varphi\to B\varphi$ corresponds in the real picture to
\begin{equation}
    \label{eq:49A}
    \varphi \to B \varphi
    \;\;\hateq\;\; \tilde{q} \to \bar{q}'\,.
\end{equation}

\subsection*{(Quantum) phases}

The phases of the components $\varphi_\tau$ determine the relation to the real
picture.
For real $\varphi_\tau$ the components $\tilde{q}_\tau$ and $\bar{q}_{\tau}^{\prime c}$ are equal, while for purely imaginary
$\varphi_\tau$ their signs are opposite. We may write
\begin{equation}
    \label{eq:C18C}
    \varphi_\tau = \frac12\big\{\tilde{q}_\tau + \bar{q}_{\tau}^{\prime c} + i\eta(\tau)(\tilde{q}_\tau -
    \bar{q}_{\tau}^{\prime c})\big\}\,,
\end{equation}
with signs $\eta(\tau)$,
\begin{equation}
    \label{eq:C18D}
    \eta(\tau) = (-1)^{m_d}\,,
\end{equation}
where $m_d$ is the number of spins down (or empty bits) for the configuration
$\tau$. One infers
\begin{align}
    \label{eq:C18E}
    \tilde{q}_\tau =            & \frac12\big[\varphi_\tau + \varphi_\tau^* - i\eta(\tau)(\varphi_\tau -
    \varphi_\tau^*)\big]\,, \nn                                                                  \\
    \bar{q}_{\tau}^{\prime c} = & \frac12\big[\varphi_\tau + \varphi_\tau^* + i\eta(\tau)(\varphi_\tau -
    \varphi_\tau^*)\big]\,.
\end{align}

\indent A phase rotation of $\varphi_\tau$ corresponds to a rotation between the component
of the real wave function $\tilde{q}$ for the configuration $\tau$ and the component of the real wave function $\bar{q}'$ for the particle-hole conjugate configuration $\tau^c$,
\begin{align}
    \label{eq:49A0}
    \varphi_\tau                  & \to \exp(i\alpha(\tau)) \;\;\hateq \nn                                                             \\
    \tilde{q}_\tau            & \to \cos(\alpha(\tau))\tilde{q}_\tau + \sin(\alpha(\tau))\eta(\tau)\bar{q}_{\tau}^{\prime c}\,,\nn \\
    \bar{q}_{\tau}^{\prime c} & \to \cos(\alpha(\tau))\bar{q}_{\tau}^{\prime c} - \sin(\alpha(\tau))\eta(\tau)\tilde{q}_\tau\,.
\end{align}
Similar to quantum mechanics, the phase of the complex wave function can play an important role.
For the particular case $\tilde{q}_\tau = \bar{q}_\tau'= q_\tau$ the rotation involves $q_\tau$ and $q_\tau^c$.

\indent In particular, for the probabilities of PCA one has
\begin{align}
    \label{eq:68A}
    p_\tau &= \tilde{q}_\tau^2\,, \quad
    \bar{p}_{\tau^c} = (\bar{q}_{\tau^c}')^2\,, \nn\\
    p_\tau + \bar{p}_{\tau^c} &= 2\varphi_\tau^{*} \varphi_\tau\,,\nn\\
    p_\tau - \bar{p}_{\tau^c} &= -i\eta(\tau) \left(\varphi_\tau^2 - \varphi_\tau^{*2}\right)\,.
\end{align}
Under a phase rotation of $\varphi_\tau$ the sum $p_\tau + \bar{p}_{\tau^c}$ remains invariant.
The difference varies according to 
\begin{equation}
    \label{eq:68B}
    \begin{aligned}
        p_\tau - \bar{p}_{\tau^c} \to &\cos(2\alpha(\tau)) (p_\tau - \bar{p}_{\tau^c}) \\ 
        &+2\eta(\tau) \sin(2\alpha(\tau)) \tilde{q}_\tau \bar{q}_\tau^{\prime c}
    \end{aligned}
\end{equation}
with $\tilde{q}_\tau \bar{q}_\tau^{\prime c} = \pm \sqrt{p_\tau \bar{p}_{\tau^c}}$.
For more general Ising type models the probabilities also involve the conjugate complex wave function $\bar{\varphi}$.

\subsection*{Time evolution of complex wave function}
For the Majorana--Weyl automaton~\eqref{eq:G5} the compatibility of the complex structure~\eqref{eq:C15},~\eqref{eq:C16} with the time evolution follows from the simple updating rule that all spin configurations are displaced by one position to the right.
This makes no difference between particles and holes.
We may take any choice of $\bar q'$ such that $\hat{q}$ has the same time evolution as $\tilde q$.
Then, the evolution operator $U$ for $\varphi$ equals the real step evolution
operator in the real picture,
\begin{equation}
    \label{eq:C20}
    U(t) = \Shat(t)\,.
\end{equation}
More in detail, the matrix $\hat{S}(t)$ in the basis~\eqref{eq:C10} is given by
\begin{equation}
    \label{eq:YY5}
    \hat{S}(t) = \hat{S}_{\tau \rho}(t) E_\tau \otimes E_\rho\,.
\end{equation}
If $\bar{q}'(t)$ has the same time evolution as $\tilde{q}(t)$, the compatibility of the complex structure~\eqref{eq:C13} with the time evolution requires choices of signs $\sigma_\tau$ in eq.~\eqref{eq:C10} such that
\begin{equation}
    \label{eq:YY6}
    [\hat{S}(t),T_3] = 0\,,\quad
    [\hat{S}(t),B] = 0\,.
\end{equation}

\indent For the Ising model the wave functions $\bar q$ and $\tilde q$ of sect.~\ref{sec:generalized_ising_models} do not follow the same evolution law.
It is therefore not possible to identify $\bar{q}'$ with $\bar{q}$.
We can define a complex wave function which is compatible with the time evolution by using $\hat q(t) = B \tilde{q}(t)$ or $\hat{q}(t) = B \bar{q}'(t)$.
A similar complex wave function $\bar\varphi$ replaces $\tilde q(t)$ by the conjugate wave function $\bar q(t)$ of sect.~\ref{sec:generalized_ising_models} and $\hat{q}(t)$ by $B\hat{q}(t)$ or $B\tilde{q}'(t)$.
We briefly discuss in sect.~\ref{sec:parity_and_time_reversal} the possibility to associate $\bar{q}'(t)$ with $\bar{q}(-t)$ and $\tilde{q}'(t)$ with $\tilde{q}(-t)$.
One has then two complex wave functions $\varphi$ and $\bar\varphi$.
For the complex wave function $\varphi$ the operator $V(t)$ can be identified with $\Shat(t)$, while for $\bar\varphi$ one has $\bar{V}(t) = (\Shat^T(t))^{-1}$.

For generalized Ising models the real and the complex picture are equivalent.
The complex formulation offers the advantage that complex basis transformations as the Fourier transform can be implemented in a straightforward way.
For the derivation of a complex Schrödinger-type evolution equation we write
\begin{equation}
    \label{eq:C21}
    V(t) = e^{-i\eps G(t)}\,,\quad G(t) = H(t) + iJ(t)\,,
\end{equation}
with Hermitian matrices $H^\dagger = H$, $J^\dagger = J$. With a discrete time
derivative $\partial_t\varphi$ eq.~\eqref{eq:C2} yields
\begin{align}
    \label{eq:C22}
    i\partial_t\varphi(t) = & \, \frac{i}{\eps}[\varphi(t+\eps) - \varphi(t)]\nn                           \\
    =                   & \,\frac{i}{\eps}\big[\exp\{-i\eps [H(t) + iJ(t)]\} - 1\big]\varphi(t)\,.
\end{align}
For smooth enough wave functions one can take the limit of continuous time
$\eps\to0$,
\begin{equation}
    \label{eq:C23}
    i\partial_t\varphi(t) = [H(t) + iJ(t)]\varphi(t)\,.
\end{equation}
For unitary $V$, as for automata, the piece $J(t)$ vanishes and one finds the
usual continuous Schrödinger equation,
\begin{equation}
    \label{eq:C24}
    i\partial_t\varphi(t) = H(t)\varphi(t)\,.
\end{equation}
This continuum limit is possible for the Majorana--Weyl automaton since a smooth wave function remains smooth if displaced one position to the right. 
For the Ising model~\eqref{eq:G4} the non-vanishing piece $J(t)$ leads to an approach to an equilibrium wave function~\cite{CWPT, CWPW2020}.

\indent For a unitary step evolution operator the Hamiltonian is defined by the relation
\begin{equation}
    \label{eq:63AA}
    U(t) = \exp\{-i \eps H(t)\}\,,
\end{equation}
where we may take $H(t)$ piecewise constant between $t$ and $t+\eps$.
We may also choose a continuous $H(t)$ as
\begin{equation}
    \label{eq:63BB}
    U(t) = \exp \left\{-i \int_t^{t+\eps} \mathrm{d} t'\, H(t') \right\}\,.
\end{equation}
For a given $U(t)$ the solution $H(t)$ of eq.~\eqref{eq:63AA} or \eqref{eq:63BB} is not unique.
This only concerns the continuous evolution between two discrete time steps.
For characteristic times much larger than the discrete time steps the precise form of the interpolation plays no role.
This is the case for the continuum limit.
For $U(t)$ independent of $t$ one employs naturally $H$ independent of $t$.
Finding in this case a simple form of $H$ has the advantage of establishing a simple conserved quantity.

\section{Fermion picture}
\label{sec:fermion_picture}

The wave functions of generalized Ising models can be interpreted as wave
functions of multi-fermion systems in the occupation number basis. It may
therefore not be a surprise that we can express operators in terms
of the fermionic creation and annihilation operators.

\subsection*{Fermionic annihilation and creation operators}

We start by constructing these operators in the real picture. Basic building
blocks are the real $2\times 2$-matrices
\begin{align}
    \label{eq:F1}
    a =   & \, \begin{pmatrix} 0 & 0 \\ 1 & 0 \end{pmatrix}\,,\quad
    a^T = \begin{pmatrix} 0 & 1 \\ 0 & 0 \end{pmatrix}\,,\nn        \\
    a^2 = & \, (a^T)^2 = 0\,,\quad \{a,a^T\} = 1\,.
\end{align}
They obey the identities
\begin{align}
    \label{eq:F2}
    \tau_3a =         & \, -a\tau_3 = -a\,,\quad \tau_3a^T = -a^T\tau_3 = a^T\,,\nn \\
    \tau_1 a \tau_1 = & \, a^T\,,
\end{align}
and
\begin{equation}
    \label{eq:F3}
    \hat n = a^Ta = \begin{pmatrix} 1 & 0 \\ 0 & 0 \end{pmatrix}\,,\quad 1-\hat a =
    aa^T = \begin{pmatrix} 0 & 0 \\ 0 & 1 \end{pmatrix}\,.
\end{equation}

For an odd number $N_x$ labeled by $j$ in the interval $[-(N_x-1)/2,(N_x-1)/2]$
we define the annihilation and creation numbers $a(j)$ and $a^\dagger(j)$ for a
given site $j$. They obey
\begin{equation}
    \label{eq:F4}
    \begin{aligned}
        \{a(j),a(j')\} = \{a^\dagger(j),a^\dagger(j')\} = 0\,, \\
        \{a(j),a^\dagger(j')\}                          = \delta_{jj'}\,.
    \end{aligned}
\end{equation}
This is realized by
\begin{align}
    \label{eq:F5}
    a(j) =         & \, \tau_3\otimes\tau_3\otimes \tau_3 \otimes a \otimes 1\otimes 1\otimes
    \dots 1\,,\nn                                                                             \\
    a^\dagger(j) = & \, \tau_3\otimes\tau_3\otimes \tau_3 \otimes a^T \otimes 1\otimes
    1\otimes \dots 1\,,
\end{align}
where the factors $a$, $a^T$ stand at the position $j$. The $\tau_3$-factors at
the positions $j' = -(N_x-1)/2,\dots j-1$ guarantee the anticommutation
relations~\eqref{eq:F4}.
We observe the relations
\begin{equation}
    \label{eq:YY7}
    T_3 a(j) T_3 = -a(j)\,,\quad T_3 a^\dagger(j) T_3 = -a^\dagger(j)\,,
\end{equation}
and
\begin{equation}
    \label{eq:F10}
    T_1 a(j) T_1 = (-1)^{p(j)} a^\dagger(j)\,,\quad p(j) = j-\frac{N_x-1}{2}\,,
\end{equation}
where $p(j)$ counts the number of $\tau_3$ factors in eq.~\eqref{eq:F5}.
With the choice \eqref{eq:YY2} for $D$ the relation
\begin{equation}
    \label{eq:YY8}
    D a(j) D = (-1)^{p(j)} a(j)
\end{equation}
implies
\begin{equation}
    \label{eq:YY9}
    B a(j) B = a^\dagger(j)\,,\quad Ba^\dagger(j) B = a(j)\,.
\end{equation}

\indent The occupation number operators $\hat n(j)$ at the site $j$ are diagonal
\begin{equation}
    \label{eq:F6}
    \hat n(j) = a^\dagger(j)a(j) = 1\otimes 1\otimes \dots \hat n \otimes 1\otimes 1
    \otimes \dots 1\,.
\end{equation}
Their expectation value can be computed by the generalized quantum rule
\begin{align}
    \label{eq:F7}
    \langle \hat n(j)\rangle (t) = & \, \bar q^T(t)\hat n(j) \tilde q(t) = \bar q_\tau
    (t) \gl\hat n(j)\gr_{\tau\rho} \tilde q_\rho(t)\nn                                 \\
    =                              & \,\sum_\tau \hat n(j)_{\tau\tau} p_\tau(t)\,.
\end{align}
The eigenvalues of the operator $\hat n(j)$ equal one if for the spin
configuration $\tau$ the site $j$ is occupied, and zero if it is empty. This
generalized quantum rule expresses the expectation values as a bilinear in the
wave functions. It follows directly from the expectation value as evaluated by
the standard classical statistical rule~\eqref{eq:G1A} for the overall
probability distribution~\cite{CWIT,CWQF,CWPT}. This extends to arbitrary functions of the
occupation numbers at a given time $t$ which are associated to operators which
are functions of $\hat n(j)$. In the language of Ising spins this covers
observables which are arbitrary functions of the Ising spins at a given time
$t$.

Starting from the totally empty state,
\begin{equation}
    \label{eq:H1}
    \lvert0\rangle_E = \begin{pmatrix}0\\1\end{pmatrix}\otimes
    \begin{pmatrix}0\\1\end{pmatrix}\otimes \dots
    \begin{pmatrix}0\\1\end{pmatrix}\,,
\end{equation}
as a \qq{reference state}, we can obtain an arbitrary basis state~\eqref{eq:C10}
by applying suitable products of creation operators
\begin{equation}
    \label{eq:H2}
    \lvert j_1,j_2,\dots j_K\rangle = \pm a^\dagger(j_1)a^\dagger(j_2)\dots
    a^\dagger(j_K)\lvert 0\rangle_E\,,
\end{equation}
where the bits at positions $(j_1,j_2,\dots j_K)$ are occupied and all other bits are empty.
Arbitrary real wave functions obtain as suitable linear combinations of basis states.
Similarly, by using products of $a^\dagger(j)$ and $a(j)$ we can construct any
wave function from an arbitrary reference state.

\indent The wave functions \eqref{eq:H2} are antisymmetric in the exchange of positions.
This follows directly from the anti-commutation relations of the creation operators.
This characteristic feature for fermions is related to the fact that the occupation numbers can only take the values $n=1,0$.
Usually, this property of the occupation numbers (Pauli principle) is derived from the antisymmetry of the multi-fermion wave function.
We follow here the opposite direction by starting from the occupation number basis.
The relation between fermionic occupation numbers and Ising spins is very direct.
The antisymmetry of wave functions becomes also very apparent if one employs the map between generalized Ising models and Grassmann functional integrals \cite{CWFCS, CWFGI, CWFCB, CWNEW, CWFPPCA}.

\subsection*{Fermionic operators for Fourier modes}

We define annihilation and creation operators $a(k)$ and $a^\dagger(k)$ in
Fourier space by a discrete Fourier transform
\begin{equation}
    \label{eq:H3}
    a(k) = \sum_j D(k,j)a(j)\,,\quad a^\dagger(k) = \sum_j
    a^\dagger(j)D^{-1}(j,k)\,,
\end{equation}
with unitary matrices
\begin{equation}
    \label{eq:H4}
    D(k,j) = \frac{1}{\sqrt{N_x}}\exp\left\{ -\frac{2\pi i}{N_x}kj\right\}\,.
\end{equation}
Here $k$ takes integer values in the same range as $j$, $k = -(N_x-1)/2,\dots
    (N_x-1)/2$. The momentum $q = 2\pi k/(N_x\eps)$ is periodic, with $k+N_x$
identified with $k$, as usual for the discrete Fourier transform.
The unitarity of $D$ follows from the relation
\begin{equation}
    \label{eq:IP1}
    \frac{1}{N_x} \sum_j \exp\left(\frac{2\pi i k j}{N_x}\right) = \hat{\delta}(k)\,,
\end{equation}
with $\hat{\delta}(k) = 1$ for $k=0 \mod N_x$, and $\hat{\delta}(k) = 0$ otherwise.

\indent From the anticommutation relations for $a(j)$, $a^\dagger(j)$ one infers the
anticommutation relations for $a(k)$, $a^\dagger(k)$,
\begin{align}
    \label{eq:H5}
    \{a(k'),a(k)\} =         & \, \{a^\dagger(k'),a^\dagger(k)\} = 0\,,\nn \\
    \{a^\dagger(k'),a(k)\} = & \, \delta_{kk'}\,.
\end{align}
The occupation number operator for the mode $k$ obeys
\begin{equation}
    \label{eq:H6}
    \hat n(k) = a^\dagger(k)a(k)\,,
\end{equation}
with eigenvalues $(1,0)$.
One has the relations
\begin{equation}
    \label{eq:YY10}
    \{T_3,a(k)\} = 0\,,\quad \{T_3,a^\dagger(k)\} = 0\,.
\end{equation}
and
\begin{alignat}{3}
    \label{eq:YY10a}
    B a^*\!(k) B = & \, a^\dagger(k)\,,\quad  &  & B a^T\!(k) B     &  & = a(k)\,,\nn \\
    B a(k) B =     & \, a^\dagger(-k)\,,\quad &  & B a^\dagger(k) B &  & = a(-k)\,.
\end{alignat}

\subsection*{Hamiltonian for Majorana--Weyl automaton}

We will establish below that the step evolution operator for the Majorana--Weyl
automaton can be written in terms of a simple Hamiltonian for right movers \cite{CWPW2020}, 
\begin{equation}
    \label{eq:H7}
    \Shat = \exp\left\{-i\eps H^{(R)}\right\}\,,\quad H^{(R)} = \sum_k\frac{2\pi k}{N_x\eps}
    \hat n(k)\,.
\end{equation}
The Hamiltonian $H^{(R)}$ describes independent fermionic oscillators.
Expressed by $a(j)$, $a^\dagger(j)$ it takes a non-local form ($L=N_x\eps$)
\begin{align}
    \label{eq:H8}
    H^{(R)} & = \sum_j\sum_{m>0} \frac{i\pi(-1)^m}{L\sin\left(\frac{\pi
    m\eps}{L}\right)} a^\dagger(j)[a(j+m)-a(j-m)]\nn                                           \\
    \approx & \sum_x\sum_y{}^{\prime}\frac{i\pi\eps}{L\sin\left( \frac{\pi(y-x)}{L}
    \right)} a^\dagger(x)\bigg\{\frac{a(y+\eps)-a(y)}{\eps} \nn                                \\
            & \!\!+\frac{a(2x-y) - a((2x-y) -\eps)}{\eps} \nn                                  \\
            & \!\!-\frac{\pi}{L} \mathrm{cotg}\left(\frac{\pi(y-x)}{L}\right) \big[a(y+\eps) -
        a(2x - (y+\eps)\big] \bigg\}\,.
\end{align}
Here the second expression holds for $\eps/L\ll1$, and $\sum_y^{\prime}$ sums over positive and odd $(y-x)/\eps = m$.
The non-local form of the Hamiltonian is of no worry since after a time step $\eps$ the expression $\exp(-i\eps H^{(R)})$ equals the step evolution operator $\Shat$ which simply displaces the whole spin configuration one position to the right.
In a discrete setting $\Shat$ is a local operator.
The vanishing commutators
\begin{equation}
    \label{eq:YY11}
    [T_3,H^{(R)}]=0\,,\quad [B,H^{(R)}]=0\,,
\end{equation}
guarantee the relations \eqref{eq:YY6}.

The eigenvalues of $H^{(R)}$ are given by the sum over the momenta $q$ for which the modes are occupied.
They can be positive or negative.
For the Majorana--Weyl automaton the negative eigenvalues do not pose a problem.
The simple step evolution operator $\Shat$ guarantees a well behaved discrete evolution.
We are not interested in the evolution for $t$ in-between the discrete time steps.
The Hamiltonian which leads to $\Shat$ by $\Shat = \exp(-i\eps H)$ is actually not uniquely determined.
Different $H$ can yield the same $\Shat$.
A possible continuum behavior for $\eps \to 0$ is rather associated to a very large number $N_t$ of discrete time steps at fixed $\Delta t = N_t\eps$.

\indent With a constant shift by $E_0$ all eigenvalues of $H^{(R)} - E_0$ are positive.
For this purpose we use the anticommutation relations~\eqref{eq:H5} and write $H^{(R)}$
in the form
\begin{equation}
    \label{eq:63A}
    H^{(R)} \!= \frac{2\pi}{L}\!\left(\,\sum_{k>0} k a^\dagger(k)a(k) + \sum_{k<0}
    (-k)\gl a(k)a^\dagger(k) - 1 \gr\!\right).
\end{equation}
We observe that $a(k)a^\dagger(k)$ has eigenvalues one and zero.
Thus $H^{(R)}$ becomes positive if we subtract the constant piece
\begin{equation}
    \label{eq:63B}
    E_0 = \frac{2\pi}{L}\sum_{k<0}k = -\frac{(N_x^2-1)\pi}{4N_x\eps}\,.
\end{equation}

\indent Introducing for $k<0$ creation and annihilation operators for antiparticles
\begin{equation}
    \label{eq:63C}
    b^\dagger(k) = a(-k)\,,\quad b(k) = a^\dagger(-k)
\end{equation}
this yields
\begin{equation}
    \label{eq:63D}
    H^{(R)} - E_0 = \frac{2\pi}{L}\sum_{k>0}\big\{ k\gl a^\dagger(k)a(k) +
    b^\dagger(k)b(k)\gr \big\}\,.
\end{equation}
Thus $H^{(R)}$ is bounded from below with minimum value $E_0$.
A shift of $H^{(R)}$ by $2\pi/\eps$ does not change $\Shat$. With $N_x = 1\mod 8 + 2\tilde{n}_x$, $\tilde{n}_x = (-2,-1,0,1)$ we can write
\begin{equation}
    \label{eq:63E}
    E_0 = -\frac{2\pi}{\eps}\left(\frac{2\tilde{n}_x + 1}{8} - \frac{1}{8N_x}
    \right)\,.
\end{equation}

For a proof of the relation~\eqref{eq:H7} we need to show that $\Shat$ defined
in this way transports arbitrary spin configurations one position to the right.
Using eqs.~\eqref{eq:H5},~\eqref{eq:H6} in a product form of eq.~\eqref{eq:H7}, $\Shat = \prod_k S_k$, one finds
\begin{align}
    \label{eq:H9}
    \Shat a(k) =         & \, \exp\left(\frac{2\pi ik}{N_x}\right) a(k)\Shat\,,\nn \\
    \Shat a^\dagger(k) = & \, \exp\left(-\frac{2\pi ik}{N_x}\right)
    a^\dagger(k)\Shat\,.
\end{align}
Inverting the Fourier transform~\eqref{eq:H3} yields
\begin{equation}
    \label{eq:H10}
    \Shat a(j) = a(j+1)\Shat\,,\quad 
    \Shat a^\dagger(j) = a^\dagger(j+1)\Shat\,.
\end{equation}
For a suitable choice of the signs $\sigma_\tau$ for the basis vectors $E_\tau$ the relation~\eqref{eq:H10} describes indeed that $\Shat$ transports arbitrary spin configurations one position to the right.

\indent In order to show this we take some reference state $\lvert R\rangle$ with $\Shat \lvert R\rangle = \lvert R\rangle$.
Consider now all states which obtain from $\lvert R\rangle$ by applying products of $a(j)$ and $a^\dagger(j)$.
Applying $\Shat$ according to eq.~\eqref{eq:H10} to such a state produces a similar state with the replacement $a(j) \to a(j+1)$, $a^\dagger(j) \to a^\dagger(j+1)$.
This also happens if every spin configuration is displaced by one position to the right.
The issue is therefore only a question of signs.
The basis vectors $E_\tau$ have to be chosen to be compatible with the sign conventions in the definition of $a(j)$ and eq.~\eqref{eq:H10}.
In other words, the relation \eqref{eq:H10} partially fixes the signs for the basis vectors $E_\tau$. 

\indent We construct the basis vectors $E_\tau$ by applying products of factors $a^\dagger(j)$ to the reference state $\lvert R\rangle = \lvert 0\rangle_E$. A given prescription for the ordering of the factors fixes the signs $\sigma_\tau$ in eq.~\eqref{eq:C10}.
Eq.~\eqref{eq:H10} partially fixes this ordering.
For the example of a single occupied bit the basis vectors are $a^\dagger(j)\Ket{0}_E$ without relative signs between different $j$.
For two occupied bits there should be no relative sign between $a^\dagger(j_1) a^\dagger(j_2) \ket{0}_E$ and $a^\dagger(j_1') a^\dagger(j_2')\Ket{0}_E$ if $j_1-j_1' = j_2-j_2'$. 
This includes the periodic behavior.
For example, if $a^\dagger(j) a^\dagger(\frac{N_x-1}{2})\ket{0}_E$ is a basis vector, $a^\dagger(j+1) a^\dagger(-\frac{N_x-1}{2})\ket{0}_E$  is also a basis vector.
The signs of the particle-hole conjugate basis vectors $E_{\tau^c}$ are determined by the signs of $E_\tau$ by the relation
\begin{equation}
    E_{\tau^c}=B E_{\tau}\,.
\end{equation}
The relation $B\Shat B = \Shat$ in eq.~\eqref{eq:YY6} guarantees that the relation $E_{\tau'} = \Shat E_\tau$ translates to $E_{\tau^{\prime c}} = \Shat E_{\tau^c}$.
With this choice of signs eq.~\eqref{eq:H10} implies then that eq.~\eqref{eq:H7} indeed produces the step evolution operator
for the transport to the right of the basis vectors.
This extends to arbitrary
linear combinations of basis vectors.
In our case this covers arbitrary states.

\subsection*{Majorana--Weyl fermions}
The notion of a single-particle or of multi-particle states depends on the vacuum.
Let us first focus on the totally empty vacuum $\lvert 0\rangle_E$.
We can define a total particle number $N_E$ by the number of occupied bits.
In the setting with real wave functions the associated operator is given by
\begin{equation}
    \label{eq:H11}
    \hat N_E = \sum_j\hat n(j) = \sum_k \hat n(k)\,.
\end{equation}
Obviously, the total particle number is preserved by the transport to the right,
corresponding to
\begin{equation}
    \label{eq:H12}
    [\hat N_E,\Shat] = 0\,.
\end{equation}
A general one particle state obtains as
\begin{equation}
    \label{eq:H13}
    q^{(1)}(t) = \sum_j q^{(1)}(t,j)a^\dagger(j) \lvert0\rangle_E\,,
\end{equation}
with real wave function $q^{(1)}(t,j) = q^{(1)}(t,x)$. It obeys the evolution
equation
\begin{equation}
    \label{eq:H14}
    q^{(1)}(t+\eps,x) = q^{(1)}(t,x-\eps)\,.
\end{equation}
Similarly, a two-particle wave function,
\begin{equation}
    \label{eq:H15}
    q^{(2)}(t) = \sum_{j,j'} q^{(2)}(t,j,j')a^\dagger(j)a^\dagger(j')
    \lvert0\rangle_E\,,
\end{equation}
obeys
\begin{align}
    \label{eq:H16}
    q^{(2)}(t,j',j) =     & \, -q^{(2)}(t,j,j')\,,\nn      \\
    q^{(2)}(t+\eps,x,y) = & \, q^{(2)}(t,x-\eps,y-\eps)\,.
\end{align}

\indent For smooth wave functions we can take the continuum limit $\eps\to0$ at fixed
$L=N_x\eps$ and $\Delta t = N_t\eps$. Eq.~\eqref{eq:H14} can be written in the
form
\begin{equation}
    \label{eq:H17}
    \frac{q^{(1)}(t+\eps,x) - q^{(1)}(t,x)}{\eps} =
    \frac{q^{(1)}(t,x-\eps) - q^{(1)}(t,x)}{\eps}\,.
\end{equation}
If $q$ is differentiable in the continuum limit $\eps\to 0$, this reads
\begin{equation}
    \label{eq:H18}
    \partial_t q^{(1)}(t,x) = -\partial_x q^{(1)}(t,x)\,.
\end{equation}
Eq.~\eqref{eq:H18} is precisely the evolution equation for a single right-moving Majorana--Weyl fermion in continuum quantum field theory.
Indeed, Majorana--Weyl fermions can be described by real wave functions. 
Smooth wave functions remain smooth by the transport to the right.
For a valid continuum limit it is therefore sufficient to start at $t\subt{in}$ with an initial smooth wave function.

There is no obvious sign of \qq{lattice doublers}.
This observation may be of interest in the context of the Ninomiya--Nielsen theorem \cite{NINI1,NINI2,FRI}.
The one particle states are very simple. 
They are a superposition of configurations where a single occupied bit moves to the right. 
On the more formal level the absence of doublers is related to the Hamiltonian $\sim\sum_k k \hat n(k)$ rather than $\sim\sum_k \sin\left(\frac{2\pi k}{N_x}\right) \hat n(k)$ or similar. 
The Hamiltonian~\eqref{eq:H7} corresponds to the SLAC lattice derivative~\cite{Drell1976}.

\subsection*{Weyl fermions}

For different vacua the particle excitations have different properties.
For the example of half-filled vacua the single particle excitations can be an additional occupied bit or an additional empty bit (hole) moving to the right.
They can be combined into a Weyl fermion, described by a complex wave function using the complex structure of the preceding section. Complex wave functions offer the advantage that eigenstates of the momentum operator or the Hamiltonian can be defined.

\indent We will discuss the vacuum and associated particle states more in detail in later sections.
Here we emphasize that the notion of particles is not unique.
We may define particle states by applying creation or annihilation operators to a vacuum.
There are different choices of such operators in the complex picture.
The relation to the bit configurations in the real picture further depends on the choice of the complex structure.
Nevertheless, the simple dynamics of propagating free massless Weyl fermions is the same for the different possible choices.

\subsection*{Annihilation and creation operators in the complex picture}
\indent For a description of Weyl fermions we want to define creation and annihilation operators in the complex picture.
We consider first the case of unconstrained $\varphi$, \ie the setting where $\bar{q}'$ is independent of $\tilde{q}$. Arbitrary operators in the complex picture can then be translated to suitable operators in the real picture.
In particular, we may investigate the action of $a(j)$ and $a^\dagger(j)$, as defined by eq.~\eqref{eq:F5}, on the complex wave function $\varphi$, given by eq.~\eqref{eq:C13}.
One finds in the complex picture
\begin{equation}
    \label{eq:AN1}
    a(j) \varphi = \frac{1}{2} \{(1-i T_3) a(j) \tilde q + (1+i T_3) B a^\dagger(j) \bar q'\}\,.
\end{equation}
For $a^\dagger(j) \varphi$ one exchanges $a(j)$ and $a^\dagger(j)$ on the \rhs of eq.~\eqref{eq:AN1}.
One observes that $a(j)\varphi$ acts as an annihilation operator for $\tilde{q}$ and a creation operator for $\bar{q}'$.

\indent One may translate eq.~\eqref{eq:AN1} to
the real picture. Employing the correspondence
\begin{equation}
    \label{eq:AN2}
    \varphi\to i \varphi \;\;\hateq \;\; \tilde q \to T_3 B \bar q'\,,\quad \bar q'\to T_3 B \tilde q\,,
\end{equation}
one infers
\begin{equation}
    \label{eq:AN3}
    \varphi \to a(j) \varphi \;\;\hateq \;\; \tilde q \to a(j) B \bar q'\,,\quad \bar q'\to B a(j) \tilde q\,.
\end{equation}
The relation
\begin{equation}
    \begin{aligned}
        \label{eq:AN4}
        \hat{N} & (j) \varphi = a^\dagger(j) a(j) \varphi                                                                    \\
                & = \frac{1}{2}\{ (1+i T_3) a^\dagger(j) a(j) \tilde q + (1-i T_3) B a(j) a^\dagger(j) \bar q' \}\,,
    \end{aligned}
\end{equation}
is represented in the real picture by the map
\begin{equation}
    \label{eq:AN5}
    \varphi \to a^\dagger(j) a(j) \varphi \;\,\hateq \,\; \tilde q \to a^\dagger(j)a(j)\tilde q\,,\quad \bar q'\to a(j)a^\dagger(j)\bar q'\,.
\end{equation}
This shows that the annihilation and creation operators $a(j)$ and $a^\dagger(j)$ in the complex picture are not compatible with the identification $\bar q' = \tilde q$.

\indent One observes that the action of the operators $a(j)$ or $a^\dagger(j)$ in the complex picture differs from applying $a(j)$ or $a^\dagger(j)$ on $\tilde q$ and $\bar q'$.
In particular, an eigenstate $a^\dagger(j)a(j)\varphi=\varphi$ requires
\begin{equation}
    \label{eq:AN6}
    \hat{n}(j) \tilde q = \tilde q \,,\quad
    (1-\hat{n}(j)) \bar q' = \bar q'\,.
\end{equation}
It is realized for configurations where for $\tilde{q}_\tau$ the spin $s(j)$ is up, while it is down for $\bar{q}'_\tau$. 

\indent The operator $\hat{n}_C(j)  = a^\dagger(j) a(j)$ acting on complex $\varphi$ is represented in the real picture by
\begin{equation}
    \label{eq:127A}
    q = \begin{pmatrix} \tilde q \\ \bar{q}' \end{pmatrix}\,, \quad
    \hat{n}_C(j) = \begin{pmatrix} \hat{n}(j) & 0 \\ 0 & 1-\hat{n}(j) \end{pmatrix}\,.
\end{equation}
It commutes with the matrix $I = \tau_3 \otimes \tau_3 B$, as required for consistency with the complex picture, cf.\ app.~\ref{app:B}.
In contrast, the operator $\hat{n}_R(j) = a^\dagger(j) a(j)$ acting on the real wave functions $\tilde{q}$ and $\bar{q}'$ reads in the above basis \eqref{eq:127A}
\begin{equation}
    \label{eq:127B}
    \hat{n}_R(j) = 1 \otimes \hat{n}(j)\,.
\end{equation}
It does not commute with the matrix $I$
\begin{equation}
    \label{eq:127C}
    I \hat{n}_R(j) I = 1 \otimes B T_3 \hat{n}(j) B T_3 = -1 \otimes (1-\hat{n}(j))\,.
\end{equation}
The operator $\hat{n}_R(j)$ is therefore not compatible with the complex structure.
A representation of $\hat{n}_R(j)$ as a linear map in the complex picture mixes $\varphi$ and $\varphi^*$.

\indent Multiplication of $\varphi$ by $a(j)$ results in the real picture in a multiplication by an operator $\tilde a(j)$ obeying
\begin{equation}
    \label{eq:AN7}
    \begin{pmatrix} \tilde q \\ \hat q \end{pmatrix}
    \to \tilde a(j) \begin{pmatrix} \tilde q \\ \hat q \end{pmatrix}
    = \begin{pmatrix} 0 & a(j) \\ a(j) & 0 \end{pmatrix} \begin{pmatrix} \tilde q \\ \hat q \end{pmatrix}\,,
\end{equation}
Similarly, the transformation $\varphi\to i a(j)\varphi$ leads in the real picture to
\begin{equation}
    \label{eq:AN8}
    \begin{pmatrix} \tilde q \\ \hat q \end{pmatrix} \to \tilde a'(j) \begin{pmatrix} \tilde q \\ \hat q \end{pmatrix} = \begin{pmatrix} T_3 a(j) & 0 \\ 0 & - T_3 a(j) \end{pmatrix} \begin{pmatrix} \tilde q \\ \hat q \end{pmatrix}\,.
\end{equation}
In the complex picture we can multiply $a(j)$ by an arbitrary phase factor. This results in the real picture in a linear combination of $\tilde{a}(j)$ and $\tilde{a}'(j)$.

\indent In the complex picture the phase of $a(j)$ drops out in $a^\dagger(j) a(j)$. In the real picture this operator is represented in the basis $\begin{pmatrix} \tilde q \\ \hat q \end{pmatrix}$ by
\begin{equation}
    \label{eq:AN9}
    \hat{N}(j) = \tilde a^\dagger(j) \tilde a(j) = \tilde{a}^{\prime\dagger}(j) \tilde{a}'(j) = \begin{pmatrix} \hat{n}(j) & 0 \\ 0 & \hat{n}(j) \end{pmatrix}\,.
\end{equation}
The expectation value reads
\begin{equation}
    \label{eq:103A}
    \langle \hat{N}(j) \rangle = \frac{1}{2}\gl \langle \tilde{q} \hat{n}(j) \tilde{q}\rangle + \langle \bar{q}' (1-\hat{n}(j)) \bar{q}'\rangle \gr\,.
\end{equation}
This implies for a particular wave function with $\bar{q}'=\tilde{q}$ the mean value $\langle \hat{N}(j)\rangle=1/2$.

\indent Arbitrary complex wave functions $\varphi$ can be obtained by applying suitable linear combinations of products of $a(j)$ and $a^\dagger(j)$ to some reference state, for example to a particle-hole symmetric half filled vacuum.
The Fourier transform~\eqref{eq:H3} can be adapted to the complex picture.
This extends to the expression~\eqref{eq:H7} for the step evolution operator, provided that the relation~\eqref{eq:H10} in terms of $a(j)$, $a^\dagger(j)$ describes the displacement by one position to the right.
This is realized again by the choice of suitable signs for the basis vectors $E_\tau$.

\subsection*{Different choices of fermionic operators}
\indent The choice of annihilation operators is not unique.
A different set of annihilation operators in the complex picture is given by
\begin{align}
    \label{eq:AA9A}
    A(j) = \tau_1 \otimes \tau_2 \otimes \tau_1 \otimes \dots \otimes \tau_2 \otimes A_+ \otimes 1 \otimes \dots \otimes 1 \nn \\\text{for} \quad j=0 \mod 2\,, \nn\\
    A(j) = \tau_1 \otimes \tau_2 \otimes \tau_1 \otimes \dots \otimes A_- \otimes 1 \otimes 1 \otimes \dots \otimes 1 \nn      \\\text{for} \quad j=1 \mod 2\,,
\end{align}
with
\begin{equation}
    \label{eq:AA9B}
    A_+ = \frac{1}{2} (\tau_3 + i \tau_2)\,,\quad
    A_- = - \frac{1}{2} (\tau_1 + i \tau_3)\,.
\end{equation}
These operators obey the relations \eqref{eq:F4} and
\begin{equation}
    \label{eq:AA9C}
    B A(j) B = -A(j)\,,\quad T_3 A(j) T_3 = A^\dagger(j)\,.
\end{equation}

\indent The particle number operator
\begin{equation}
    \label{eq:134A}
    \hat{N}(j) = A^\dagger(j) A(j)
\end{equation}
is given by a direct product of unit matrices with the exception of the position $j$ where the factor reads
\begin{equation}
    \label{eq:AA9D}
    A^\dagger_+ A_+ = \frac{1}{2} (1+\tau_1)\,,\quad
    A^\dagger_- A_- = \frac{1}{2} (1+\tau_2)\,,
\end{equation}
for $j=0\mod 2$ and $j=1\mod 2$, respectively.
In the real picture the operator $A^\dagger(j) A(j)$ acts in the same way on $\tilde{q}$ and $\bar{q}'$.
For $j=0 \mod 2$ it is a real operator which acts on $\tilde{q}$ or $\bar{q}'$.
The eigenstates of $\hat{N}(j)$ are superpositions of states with a particle or a hole at position $j$.
The local particle number $\hat{N}(j) = A^\dagger(j) A(j)$ differs from the operator $\hat{N}(j) = a^\dagger(j) a(j)$ in eq.~\eqref{eq:AN5}.

\indent We can define the operators $a(k)$ and $a^\dagger(k)$ by replacing in eq.~\eqref{eq:H3} $a(j) \to A(j)$, $a^\dagger(j) \to A^\dagger(j)$.
The expression \eqref{eq:H7} for $H^{(R)}$ remains the same, and this also holds for eqs.~\eqref{eq:H8}\eqref{eq:H10} with the replacements $a \to A$.
As a result one has two different possible Hamiltonians $H^{(R)}$, one with $a(k)$ expressed in terms of $a(j)$, the other using instead $A(j)$. Both Hamiltonians lead to the same step evolution operator.

\indent The non-uniqueness of the Hamilton operator does not affect the discrete time evolution.
Two different Hamilton operators realizing a given step evolution operator express the same probabilistic cellular automaton in two different fermionic versions.
Both Hamilton operators are conserved quantities for the discrete time evolution, and therefore also for the continuum limit for characteristic time scales much larger than $\eps$.
This extends to other observables, as fermionic particle numbers constructed from annihilation and creation operators.
The particle physics vacuum in position space depends on the choice of $A(j)$, $A^\dagger(j)$.
Fermionic quantum field theories constructed from different choices of $A(j)$, $A^\dagger(j)$ are based on different vacuum wave functions for the automaton or generalized Ising model.

\indent For arbitrary unitary matrices $W$, the fermionic operators
\begin{equation}
    \label{eq:AA9E}
    \tilde{A}(j) = W a(j) W^\dagger\,,\quad W^\dagger W = 1\,,
\end{equation}
obey the anticommutation relations \eqref{eq:H5}.
Defining operators in Fourier space similar to eq.~\eqref{eq:H3} and $H^{(R)}$ by eq.~\eqref{eq:H7}, the relation between $\Shat$ and $H^{(R)}$ holds provided that $\tilde{S}=W\Shat W^\dagger = \Shat$.
This holds since eq.~\eqref{eq:H10} is obeyed for $\tilde{A}(j)$ and $\tilde{S}$.
The set \eqref{eq:AA9A} obtains for
\begin{align}
    \label{eq:WoperatorSets}
    W   & = W_+ \otimes W_- \otimes W_+ \otimes W_- \otimes \dots \otimes W_+\,,\nn                   \\
    W_+ & = \frac{1}{\sqrt{2}} (\tau_1 + \tau_3), \quad W_- = \frac{1}{\sqrt{2}} (\tau_2 + \tau_3)\,.
\end{align}

\indent The operator set $A(j)$ in eq.~\eqref{eq:AA9A} is not compatible with the identification $\bar{q}' = \tilde{q}$ for the complex structure \eqref{eq:C13}.
Only operators that commute with $B$ are compatible with the constraint $B \varphi = \varphi$, while $A(j)$ anticommutes with $B$.
Products of $A(j)$, $A^\dagger(j)$ with an even number of factors, as $A^\dagger(j) A(j)$ or $H^{(R)}$ commute with $B$, while products with an odd number of factors map $\varphi^+ \to \varphi^-$ (cf.\ eq.~\eqref{eq:C18A}).
The observables in a fermionic quantum field theory involve all an even number of annihilation or creation numbers.
For the choice \eqref{eq:AA9A} they are compatible with $B\varphi = \varphi$ or with the identification $\bar{q}' = \tilde{q}$.
This distinguishes the set $A(j)$ from the set $a(j)$ in eq.~\eqref{eq:F5}.
If we do not want to associate with the additional bit distinguishing $\tilde{q}$ and $\bar{q}'$ a physical meaning, we may consider it as an auxiliary mathematical structure for the definition of the fermionic operators.
\qq{Physical wave functions} identify then $\bar{q}'$ with $\tilde{q}$, and this identification is preserved by the evolution. 
We will choose the set \eqref{eq:AA9A} in this case, such that all observables with an even number of fermionic operators are compatible with $\bar{q}' = \tilde{q}$.

\indent One may ask if a set $\tilde{A}(j)$ exists which obeys the anticommutation relations \eqref{eq:F4} and commutes with $B$.
For general $\tilde{A}(j)$ in eq.~\eqref{eq:AA9E} the condition $B \tilde{A}(j) B = \tilde{A}(j)$ is equivalent to $[W^\dagger B W, a(j)] = 0$.
This requires the existence of a matrix $\tilde{B} = W^\dagger B W$, $\tilde{B}^2 = 1$, $\tilde{B}^\dagger = \tilde{B}$, with an equal number of eigenvalues $+1$ and $-1$, which commutes with all $a(j)$.
(For this argument we may replace $a(j)$ by $A(j)$ in eq.~\eqref{eq:AA9A}, with a corresponding change of $W$.)
One does not find a set of $N_x$ annihilation operators that commute with $B$.
If such a set would exist, one could construct a complex Hilbert space with $2^{N_x}$ independent basis vectors by applying products of $A^{\dagger}(j)$ to the totally empty vacuum.
For $B \varphi = \varphi$ only $2^{N_x-1}$ independent basis vectors exist.
This contradiction makes it impossible to construct a set of $N_x$ annihilation operators which commute with $B$.

\indent On the other hand, sets of $N_x-1$ annihilation operators commuting with $B$ are easily constructed.
For example, we may modify for $j<(N_x-1)/2$ the set \eqref{eq:AA9A} by replacing for the last bit the unit matrix by $\tau_3$, implying $B A(j) B = A(j)$ without modifying the anticommutators for these operators.
For $j_{\text{max}}=(N_x-1)/2$ we replace in the last bit $A_+$ by $\tau_1$.
Then $A(j_{\text{max}}) = A^\dagger(j_{\text{max}}) = \pm B$ anticommutes with all $A(j)$, $A^\dagger(j)$ for $j<j_{\text{max}}$, but has itself not the canonical commutation relation, obeying $A^2(j_{\text{max}}) = 1$.
In the limit $N_x \to \infty$ it may not matter if the annihilation operator for a single position is \qq{anomalous}. 
In the appendix \ref{app:C} we discuss a further set of $N_x-1$ annihilation and creation operators which can be related to the annihilation or creation of composite particles.

\indent For keeping the discussion simple we focus in the following on the case where $\tilde{q}$ and $\bar{q}'$ are distinguished by an additional bit.
We can then use arbitrary sets of $N_x$ annihilation operators $A(j)$ obeying the fermionic anticommutation relations.

\subsection*{Energy eigenstates and energy-momentum relation}
In the complex picture the Hamiltonian $H^{(R)}$ is a Hermitian operator acting
on a complex wave function. It is a conserved quantity. We can find eigenstates
$\varphi_n$ to the eigenvalues $E_n$ of $H^{(R)}$. Their evolution is periodic
\begin{equation}
    \label{eq:F14}
    \varphi_n(t) = \exp(-iE_nt)\varphi_n(0)\,.
\end{equation}
For discrete time steps $t=m_t\eps$ this evolution produces precisely the
corresponding evolution of the time-local probability distribution for the
Majorana--Weyl automaton.

The total momentum operator $\hat P$ for the multi-fermion system is given by
\begin{equation}
    \label{eq:F15}
    \hat P = \sum_k\frac{2\pi k}{N_x\eps} a^\dagger(k) a(k) = \sum_p p\hat n(p)\,.
\end{equation}
Eq.~\eqref{eq:H7} entails the simple relation
\begin{equation}
    \label{eq:F16}
    H^{(R)} = \hat P\,.
\end{equation}
We observe that both $\hat P$ and $H^{(R)}$ have positive and negative
eigenvalues.

A similar Majorana--Weyl automaton for left-movers moves at every time step each
configuration $\tau$ by one position to the left. The corresponding Hamiltonian
obtains from $H^{(R)}$ by switching the sign,
\begin{equation}
    \label{eq:78A}
    H^{(L)} = -\hat P = -\sum_k\frac{2\pi k}{N_x\eps} a^\dagger(k)a(k)\,.
\end{equation}
For left-movers the particles have negative $k$, and we define for the positive
$k$-modes creation and annihilation operators for antiparticles similar to
eq.~\eqref{eq:63C}. This yields
\begin{equation}
    \label{eq:83A}
    H^{(L)} - E_0 = \frac{2\pi}{L}\sum_{k<0} (-k)\gl a^\dagger(k)a(k) +
    b^\dagger(k)b(k)\gr\,,
\end{equation}
with $E_0$ given by eq.~\eqref{eq:63B}.

While we have focused in this section on the Majorana--Weyl automaton, the fermionic annihilation and creation operators in the real or complex picture can be used for arbitrary generalized Ising models. 
For example, one could express the step evolution operator of the Ising model in terms of these operators. 
For generalized Ising models there also exists a map to a Grassmann functional integral~\cite{CWFGI,CWPCA,CWNEW,CWFPPCA}. 
It is based on the observation that the Hilbert space for fermions in the occupation number basis is identical to the Hilbert space for generalized Ising models. 
The map is established if the step evolution operator obtained from the Grassmann functional integral equals the one from the generalized Ising model.

\section{Dirac automaton}
\label{sec:dirac_automaton}

For the Dirac automaton we supplement the Majorana--Weyl automaton for right-movers by a corresponding one for left-movers.
The action of the generalized Ising model reads, with $\beta\to\infty$,
\begin{align}
    \label{eq:D1}
    S = \, -\beta\sum_{t,x}\big\{& s_R(t+\eps,x+\eps) s_R(t,x) \nn \\
        & + s_L(t+\eps,x-\eps) s_L(t,x) - 2 \big\}\,.
\end{align}
It involves two species of Ising spins $s_R$ and $s_L$, with the index
indicating right-movers or left-movers.

At a given $t$ one has $4^{N_x}$ spin configurations $\tau$. 
We consider first a real wave function $q$ which stands for $\tilde{q}(t)$ or $\bar{q}'(t)$.
We can represent $q$ in a direct product basis~\eqref{eq:C10},~\eqref{eq:C11}, with four component basis vectors $e(x)$,
\begin{equation}
    \label{eq:D2}
    e_{11} = \begin{pmatrix} 1 \\ 0 \\ 0 \\ 0 \end{pmatrix}\,,\ \
    e_{10} = \begin{pmatrix} 0 \\ 1 \\ 0 \\ 0 \end{pmatrix}\,,\ \
    e_{01} = \begin{pmatrix} 0 \\ 0 \\ 1 \\ 0 \end{pmatrix}\,,\ \
    e_{00} = \begin{pmatrix} 0 \\ 0 \\ 0 \\ 1 \end{pmatrix}\,.
\end{equation}
For $e_{11}$ the site is occupied by both a right-mover and a left-mover, or
$s_R = s_L = 1$. For $e_{10}$ only a right-mover is present, $s_R = 1$, $s_L =
    -1$, while for $e_{01}$ one has a single left-mover, $s_R = -1$, $s_L = 1$. For
$e_{00}$ the site is empty $s_R = s_L = -1$. We introduce the $4\times
    4$-matrices
\begin{align}
    \label{eq:D3}
    \tilde\tau_1 = & \, \tau_1\otimes\tau_1 =
    \begin{pmatrix}
        0 & 0 & 0 & 1 \\
        0 & 0 & 1 & 0 \\
        0 & 1 & 0 & 0 \\
        1 & 0 & 0 & 0
    \end{pmatrix}\,,\nn                       \\
    \tilde\tau_3 = & \, \tau_3\otimes1 =
    \begin{pmatrix}
        1 & 0 & 0  & 0  \\
        0 & 1 & 0  & 0  \\
        0 & 0 & -1 & 0  \\
        0 & 0 & 0  & -1
    \end{pmatrix}\,,\quad \{\tilde\tau_3,\tilde\tau_1\} = 0\,,
\end{align}
and generalize eqs.~\eqref{eq:YYA},~\eqref{eq:C14} for $T_1$ and $T_3$ by
replacing $\tau_1\to\tilde\tau_1$, $\tau_3\to\tilde\tau_3$. For an odd number of
sites $T_1$ and $T_3$ anticommute.
Similarly, we generalize $D$ to $\tilde{D} = D\otimes D$ and $\tilde{B} = \tilde{D} T_1$.
The complex wave function $\varphi$ is defined by eq.~\eqref{eq:C13}.

The step evolution operator for the Dirac automaton takes the form of a direct
product,
\begin{equation}
    \label{eq:D4}
    \Shat = \Shat^{(R)}\otimes\Shat^{(L)}\,,
\end{equation}
where $\Shat^{(R)}$ moves the right-movers one position to the right, and
$\Shat^{(L)}$ moves the left-movers one position to the left. A general
configuration $\tau$ can be written as a double index $\tau = (\tau_R,\tau_L)$.
In this notation one has
\begin{equation}
    \label{eq:D5}
    \Shat_{(\tau_R,\tau_L),(\rho_R,\rho_L)} q_{(\rho_R,\rho_L)} =
    \Shat^{(R)}_{\tau_R\rho_R} \Shat^{(L)}_{\tau_L\rho_L} q_{(\rho_R,\rho_L)}\,.
\end{equation}
One may interpret the wave function as a matrix $q$ with elements
$q_{\tau_R,\tau_L}$, such that
\begin{equation}
    \label{eq:D6}
    \Shat(q) = \Shat^{(R)} q (\Shat^{(L)})^T\,.
\end{equation}

We have now two sets of annihilation operators $a_R(j)$, $a_L(j)$, and corresponding creation operators $a^\dagger_R(j)$, $a^\dagger_L(j)$. These
fermionic operators for the right-movers and left-movers anticommute if we
define
\begin{equation}
    \label{eq:D7}
    a_R(j) = a(j)\otimes1\,,\quad a_L(j) = T_3\otimes a(j)\,.
\end{equation}
The Fourier transform
acts separately on the fermionic operators for right-movers and left-movers. We
can represent
\begin{equation}
    \label{eq:90A}
    \Shat = \exp\big\{ -i\eps(H^{(R)} + H^{(L)})\big\}\,,
\end{equation}
where $H=H^{(R)} + H^{(L)}$,
\begin{align}
    \label{eq:D8}
    H^{(R)} = & \, \sum_k\frac{2\pi k}{N_x\eps} a_R^\dagger(k) a_R(k)\,,\nn \\
    H^{(L)} = & \, -\sum_k\frac{2\pi k}{N_x\eps} a_L^\dagger(k) a_L(k)\,.
\end{align}
Since $q$ and $\hat q$ obey the same evolution law eqs.~\eqref{eq:D5},~\eqref{eq:D6} generalize to the complex picture, with $q\to\varphi$.
This also holds for eq.~\eqref{eq:D8}.

\section{Parity and time reversal}
\label{sec:parity_and_time_reversal}
We realize the discrete transformations of parity ($P$) and time reversal ($T$) by suitable transformations of the Ising spins in the generalized Ising model for the overall probability distribution.
The corresponding transformations of the step evolution operator, Hamiltonian and wave functions are inferred from the formalism of sect.~\ref{sec:generalized_ising_models}.
If the action \eqref{eq:G4},\eqref{eq:G5} or \eqref{eq:D1} is invariant under the spin transformations corresponding to $P$ or $T$, then the dynamics is $P$- or $T$-invariant.
Non-invariant wave functions are induced if the boundary terms are not invariant.
For PCA this approach derives the symmetries $P$ and $T$ characteristic for quantum field theories from simple discrete transformations in generalized Ising models.
On the other hand, symmetries as $P$ and $T$ can be useful in the wide field of general classical statistical models.

\subsection*{Parity}

The parity transformation reverses the sign of $j$ or $x$. We implement it here
as a discrete transformation of the Ising spins, keeping the lattice of sites
$(t,x)$ unchanged,
\begin{equation}
    \label{eq:PT1}
    P s(t,x) = s(t,-x)\,.
\end{equation}
This transformation is performed for the action and the boundary terms, and
therefore for the overall probability distribution. The action~\eqref{eq:G4}
for the Ising model is parity invariant, the action~\eqref{eq:G5} for the
Majorana--Weyl automaton is not. Parity changes a right-mover to a left-mover.
The action~\eqref{eq:D1} for the Dirac automaton conserves parity provided one
also changes the species of Ising spins
\begin{equation}
    \label{eq:PT2}
    Ps_R(t,x) = s_L(t,-x)\,,\quad Ps_L(t,x) = s_R(t,-x)\,.
\end{equation}
For both eqs.~\eqref{eq:PT1} and~\eqref{eq:PT2} one has
\begin{equation}
    \label{eq:eq:93A}
    P^2 = 1\,.
\end{equation}
\indent Parity invariance of the action does not imply parity invariance of the overall probability distribution since the boundary terms may violate parity. 
Since parity does not affect the time coordinate one can directly infer its action on the wave function at a given $t$.
We may denote by $\tau_P$ the configuration which obtains from $\tau$ by the transformation~\eqref{eq:PT1} or~\eqref{eq:PT2}.
The wave functions $\tilde q$ and $\bar q'$ transform as
\begin{equation}
    \label{eq:PT3}
    P \gl \tilde q_\tau(t) \gr = \tilde q_{\tau_P}(t)\,,\quad P \gl\bar
    q'_\tau(t)\gr = \bar q'_{\tau_P}(t)\,.
\end{equation}
Indeed, the parity transformation of $\tilde f$ and $\bar f$ in
eqs.~\eqref{eq:G9}~\eqref{eq:G10} obtain by inserting eqs.~\eqref{eq:PT1}
or~\eqref{eq:PT2}. The basis functions transform $h_\tau\to h_{\tau_P}$, and
this carries over to the wave functions~\eqref{eq:PT3} if one keeps fixed basis
functions. For the Dirac automaton one has $(\tau_R,\tau_L)_P =
    (\tau_{L,P},\tau_{R,P})$ and the wave function transforms as
\begin{equation}
    \label{eq:PT3A}
    P:\, q_{(\tau_R,\tau_L)}(t) \to q_{(\tau_{L,P},\tau_{R,P})}(t)\,.
\end{equation}

\indent For the particular case of one-particle wave functions one has $\tau
    = x$, with $x$ the position of the particle, and $\tau_P = -x$. In case of
the two species of right- and left-movers the one-particle wave function has two
components $q_R(t,x)$ and $q_L(t,x)$. The parity transformation~\eqref{eq:PT2}
results in
\begin{equation}
    \label{eq:PT4}
    P \begin{pmatrix} q_R(t,x) \\ q_L(t,x) \end{pmatrix} = \begin{pmatrix}
        q_L(t,-x) \\ q_R(t,-x) \end{pmatrix}\,.
\end{equation}
This carries over to the complex wave function.

\indent The fermionic operators transform as $a_R(j)\to a_L(-j)$, with Fourier transforms $a_R(k) \to a_L(-k)$.
In eq.~\eqref{eq:90A} this results in $H^{(R)} \to H^{(L)}$, such that $\Shat$ is invariant.
The invariance of the step evolution operator follows, of course, also directly from the invariance of the action.

\subsection*{Time reversal}
Time reversal $T$ transforms in the overall probability distribution (action and
boundary terms)
\begin{equation}
    \label{eq:PT5}
    T\gl s(t,x)\gr = s(-t,x)\,.
\end{equation}
Here we have chosen for convenience a labeling of the time points for which
$t\subt{in} = -t\subt{f}$. (More generally, time reversal describes a reflection
at the point $(t\subt{f} + t\subt{in})/2$.) The action~\eqref{eq:G4} of the
next-neighbor Ising model is invariant under $T$, while the action~\eqref{eq:G5}
of the Majorana--Weyl automaton is not. The action of the Dirac automaton is time
reversal invariant provided one transforms
\begin{equation}
    \label{eq:PT6}
    T\gl s_R(t,x)\gr = s_L(-t,x)\,,\quad T\gl s_L(t,x)\gr = s_R(-t,x)\,.
\end{equation}
For both transformations~\eqref{eq:PT5} and~\eqref{eq:PT6} one has
\begin{equation}
    \label{eq:PT7}
    T^2 = 1\,.
\end{equation}

The action of time reversal on the wave functions is more complex than for
parity since $T$ exchanges $\tilde q(t)$ with $\bar q(-t)$,
\begin{align}
    \label{eq:PT7A}
    T\gl \tilde q(t)\gr & = \,\bar q^{(T)}(-t)\nn   \\
    T\gl \bar q(t)\gr   & = \,\tilde q^{(T)}(-t)\,.
\end{align}
For a $T$-invariant action one has $\tilde q^{(T)}(t) = \tilde q(t)$, $\bar q^{(T)}(t) = \bar q(t)$.
In order to see this we first note that $T$ interchanges the boundary terms
$\mathcal{B}\subt{f}$ and $\mathcal{B}\subt{in}$ for the overall probability
distribution~\eqref{eq:G1}. The transformation
\begin{equation}
    \label{eq:PT8}
    T\gl \mathcal{B}\subt{f}[s(t\subt{f},x)]\gr =
    \mathcal{B}\subt{f}[s(t\subt{in},x)]
\end{equation}
results in the replacement
\begin{align}
    \label{eq:PT9}
    \mathcal{B}\subt{in}[s(t\subt{in},x)] \to & \,
    \mathcal{B}\subt{f}[s(t\subt{in},x)]\,,\nn     \\
    \mathcal{B}\subt{f}[s(t\subt{f},x)] \to   & \,
    \mathcal{B}\subt{in}[s(t\subt{f},x)]\,.
\end{align}
For a $T$-invariant action a time reversal invariant overall probability distribution is only realized if $\mathcal{B}\subt{f}$ and $\mathcal{B}\subt{in}$ are the same functions of the corresponding spin configurations.

We next observe that $T$ maps
\begin{align}
    \label{eq:PT10}
    S_<(t) \to & \, S_>^{(T)}(-t)\,,\nn \\
    S_>(t) \to & \, S_<^{(T)}(-t)\,.
\end{align}
Here $S_<^{(T)}(t)$ obtains from $S_<(t)$ by exchanging in each factor
$\mathcal{L}(t')$ the arguments $t'$ and $t'+\eps$ of the spins. If the action
is time reversal invariant one has $S_{<(>)}^{(T)}(t) = S_{<(>)}(t)$.
Eqs.~\eqref{eq:PT9},~\eqref{eq:PT10} imply
\begin{equation}
    \label{eq:PT11}
    \tilde f(t) \to \bar f^{(T)}(-t)\,,\quad \bar f(t) \to \tilde f^{(T)}(-t)\,,
\end{equation}
and the definition~\eqref{eq:G12} results in eq.~\eqref{eq:PT7A}. 

For the nearest-neighbor Ising model \eqref{eq:G4} one has $\mathcal{L}[s(t+\eps,x),s(t,y)] = \mathcal{L}[s(t,x),s(t+\eps,y)]$ in accordance with the $T$-invariance of the action. 
In contrast, for our models of PCA the switch of the time arguments in $\mathcal{L}(t)$ leads to the map
\begin{equation}
    \label{eq:PT12}
    T:\, s(t+\eps,x+\eps)s(t,x) \to s(t,x+\eps)s(t+\eps,x)\,.
\end{equation}
Time reversal maps a right-mover to a left-mover.
The Majorana--Weyl automaton is not $T$-invariant.
For the Dirac automaton the time reversal \eqref{eq:PT6} maps
\begin{equation}
    \label{eq:PT13}
    T:\, \tilde{q}_{(\tau_R,\tau_L)}(t) \leftrightarrow \bar{q}_{(\tau_L,\tau_R)}(-t)\,,
\end{equation}
which implies for the one-particle wave function
\begin{equation}
    \label{eq:PT14}
    T\begin{pmatrix} \tilde{q}_R(t,x) \\ \tilde{q}_L(t,x) \end{pmatrix} =
    \begin{pmatrix} \bar{q}_L(-t,x) \\ \bar{q}_R(-t,x)\end{pmatrix}\,.
\end{equation}

\indent For the Ising model, the step evolution operator transforms under time reversal by transposition,
\begin{equation}
    \label{eq:PT15}
    T:\, \Shat(t) \to \Shat^T(-(t+\eps))\,.
\end{equation}
This follows directly from eq.~\eqref{eq:G13} with
\begin{equation}
    \label{eq:169A}
    \bar{q}(-(t+\eps)) = \hat{S}^T(-(t+\eps)) \bar{q}(-t)\,.
\end{equation}
For the Ising model $\Shat$ does not depend on $t$ and $\Shat$ is symmetric, $\Shat^T = \Shat$.
This opens the possibility to identify
\begin{equation}
    \label{eq:169B}
    \bar{q}'(t) = \bar{q}(-t)\,.
\end{equation}
If one chooses this option the complex wave function $\varphi(t)$ employs information from both the initial and final boundary.
In particular, the initial wave function $\varphi(t_{in})$ involves both $\tilde{q}(t_{in})$ and $\bar{q}(t_f)$.
For the computation of expectation values of time-local observables we still need the conjugate wave function $\bar{q}(t)$ or the associated complex wave function $\bar{\varphi}(t)$.
For the latter the choice \eqref{eq:169B} can be supplemented by $\tilde{q}'(t) = \tilde{q}(-t)$.
From the information contained in $\varphi(t)$ and $\bar{\varphi}(t)$ one can compute in this case expectation values of local observables at $t$ and $-t$.
Another option identifies $\bar{q}'(t)$ with the $CPT$-transform of $\tilde{q}(t)$, see next section.

\indent For the Dirac automaton the two factors in eq.~\eqref{eq:D5} transform as
\begin{align}
    \label{eq:PT16}
    T:\, \Shat^{(R)}(t) \to & \, \gl\Shat^{(L)}\gr^T(-(t+\eps))\nn \\
    \Shat^{(L)}(t) \to      & \, \gl\Shat^{(R)}\gr^T(-(t+\eps))\,,
\end{align}
since $T$ involves the additional map between right- and left-movers. 
For this model $\Shat^{(R,L)}$ are independent of $t$ and obey $\gl\Shat^{(R,L)}\gr^T = \gl\Shat^{(R,L)}\gr^{-1} = \Shat^{L,R}$, in accordance with the $T$-invariance of the action.
With $\Shat^{(R,L)} = \exp\gl-i\eps H^{(R,L)}\gr$ this can be achieved by
\begin{equation}
    \label{eq:PT17}
    T:\, H^{(R)} \leftrightarrow -H^{(L)}\,.
\end{equation}
For the total Hamiltonian $H = H^{(R)} + H^{(L)}$ one infers
\begin{equation}
    \label{eq:PT18}
    T:\, H \to -H\,,
\end{equation}
as appropriate for the map $\partial_t \to -\partial_t$ in the Schrödinger
equation.

\indent For the complex picture time reversal transforms the complex wave function of the Dirac automaton as
\begin{equation}
    T:\,\varphi_{(\tau_R,\tau_L)}(t) = \varphi_{(\tau_L,\tau_R)}(-t)\,.
\end{equation}
Eqs.~\eqref{eq:PT15}--\eqref{eq:PT18} remain valid in the complex picture.
While $T$-invariance of the action leads to eq.~\eqref{eq:PT18}, it does not imply direct consequences for the spectrum of $H$ since time runs backwards in the $T$-transformed evolution equation.

\subsection*{$PT$-symmetry}
\indent One can combine parity and time reversal to the $PT$-transformation. The
reflection of both $x$ and $t$ correspond to a rotation by $\pi$ in the $(t,x)$
plane. For the Dirac automaton this transformation does not switch between
right- and left-movers,
\begin{equation}
    \label{eq:PT19}
    PT:\, \tilde{q}_{(\tau_R,\tau_L)}(t) \to \bar{q}_{(\tau_{R,P},\tau_{L,P})}(-t)\,.
\end{equation}
For the single particle wave functions this yields
\begin{equation}
    \label{eq:PT20}
    PT:\, \tilde{q}_{R,L}(t,x) \to \bar{q}_{R,L}(-t,-x)\,.
\end{equation}
In the complex picture the $PT$-transformation results in
\begin{equation}
    PT:\,\varphi_{(\tau_R,\tau_L)}(t) = \varphi_{(\tau_{R,P},\tau_{L,P})}(-t)\,.
\end{equation}
For the Hamiltonian of the Dirac automaton one finds
\begin{equation}
    \label{eq:PT21}
    PT:\, H^{(R)} \to -H^{(R)}\,,\quad H^{(L)} \to -H^{(L)}\,,\quad H\to-H\,.
\end{equation}

The action~\eqref{eq:G5} of the Majorana--Weyl automaton is invariant under the
combined $PT$-transformation. As compared to the $PT$-transformation of the Dirac
automaton one simply omits the parts for the left-movers. The
$PT$-transformation changes the sign for both the energy $H$ and the momentum
$\hat P$. The relation~\eqref{eq:F16} $H=\hat P$ is maintained.

\indent With $\Shat$ independent of $t$ and $\Shat_{\tau,P;\rho,P} = \Shat_{\tau,\rho}$ one has
\begin{equation}
    \label{eq:177A}
    \tilde{q}_\tau (t+\eps) = \Shat_{\tau\rho} \tau_\rho\,,\quad
    \bar{q}'_{\tau,P} (-(t+\eps)) = \Shat_{\tau,P;\rho,P} \bar{q}'_{\rho,P}(-t)\,.
\end{equation}
This allows for the option to identify $\bar{q}'$ with the $PT$-transform of $\tilde{q}$,
\begin{equation}
    \label{eq:177B}
    \bar{q}'_\tau(t) = \bar{q}_{\tau,P}(-t)\,.
\end{equation}
If we choose the association \eqref{eq:177B}, the identification $B\varphi = \varphi$ or $\bar{q}' = \tilde{q}$ amounts to a projection on $PT$-invariant wave functions.
For such a projection the expectation values of observables at $t$ and $-t$ are related.
More generally, the map $\tilde{q} \to \bar{q}'$ or $\varphi \to B\varphi$ describes a $PT$-transformation in this case.
In the next section we discuss the possibility to use instead the $CPT$ transform of $\tilde{q}$ for defining $\bar{q}'$.
This choice will reveal more natural in view of the particle physics vacuum, see sect.~\ref{sec:vacuum}.

\section{Charges and Charge conjugation}
\label{sec:charges_and_charge_conjugation}

\indent There is another simple discrete symmetry of generalized Ising models, namely the switch of sign of all Ising spins.
We associate this symmetry to charge conjugation.
For our choice of a complex structure charge conjugation is closely related to complex conjugation of the wave function.
Charge conjugation reverses the sign of suitable charges.
We discuss different charge-observables for which this is the case.

\subsection*{Charge conjugation}
We identify the charge conjugation $C$ with the particle-hole conjugation.
It is realized by the spin flip
\begin{equation}
    C:\, s_\gamma(t,x) \to -s_\gamma(t,x)\,.
\end{equation}
The action is invariant under $C$ if $\mathcal{L}(t)$ involves an even number of Ising spins. This is the case for our examples.

\indent In the real picture $C$ induces the transformation
\begin{equation}
    \label{eq:TCP4}
    C:\, \tilde q(t) \to B \tilde q(t)\,,\quad
    \hat q(t) \to B \hat q(t) = \bar{q}'(t)\,.
\end{equation}
This results in the complex picture in the map
\begin{equation}
    \label{eq:CC1}
    C:\, \varphi(t) \to B \varphi^*(t)\,.
\end{equation}
Due to the complex conjugation, $C$ acts as an antilinear transformation in the complex picture.
In the real picture it amounts to a linear transformation.
For the annihilation and creation operators in the complex picture one has (for real $a(j)$)
\begin{align}
    \label{eq:TCP10}
    C a(j) C         & = B a(j) B \phantom{{}^\dagger} =  a^\dagger(j)\,, \nn \\
    C a^\dagger(j) C & = B a^\dagger(j) B =  a(j)\,,
\end{align}
in accordance with eq.~\eqref{eq:TCP9}.
Similarly, one has
\begin{equation}
    \label{eq:ZA0}
    C a(k) C = B a^{*}(k) B = a^\dagger(k)\,.
\end{equation}

\indent The step evolution operator transforms as
\begin{equation}
    \label{eq:CC2}
    C:\, V(t) \to B V^*(t) B\,,
\end{equation}
in accordance with the complex conjugation of eq.~\eqref{eq:C2}.
For the particle-hole invariant action of the next-neighbor Ising model, the Majorana--Weyl automaton or the Dirac automaton the particle-hole conjugate wave functions $\tilde q^{(c)}$, $\bar q'^{(c)}$ obey the same evolution law as $\tilde q$, $\bar q'$. 
In the occupation number basis one has $V = \Shat = \Shat^*$, with $C$-invariant real step evolution operator $B \Shat B = \Shat$.

\indent The eigenstates of $C$ are given by
\begin{equation}
    \label{eq:138A}
    \varphi_{+} = \frac{1}{2} (\varphi + B\varphi^{*})\,,\quad \varphi_{-} = \frac{1}{2} (\varphi - B\varphi^{*})\,,
\end{equation}
obeying
\begin{equation}
    \label{eq:CC10}
    C\varphi_{+} = \varphi_{+}\,,\quad C\varphi_{-} = -\varphi_{-}\,\quad \varphi_{\pm}^{*} = B \varphi_{\pm}\,.
\end{equation}
According to eq.~\eqref{eq:YY4} the eigenstates $\varphi_{+}$ and
$\varphi_{-}$ can be expressed in terms of $\tilde q$ and $\bar q'$.
The eigenstates $\varphi_{\pm}$ of $C$ differ from the eigenstates $\varphi^{\pm}$ of $B$ in eq.~\eqref{eq:49A}.
For $B \varphi = \varphi$ or $\bar{q}' = \tilde{q}$ they correspond to the real and imaginary parts of $\varphi$.

\subsection*{Energy spectrum of PCA}
A real step evolution operator in the complex picture has important consequences for the energy spectrum of PCA.
If $\varphi(t)$ is a solution of $\varphi(t+\eps) = \Shat(t)\varphi(t)$, the complex conjugate $\varphi^{*}(t)$ is another solution of the same evolution equation.
In particular, if $\varphi_n(t)$ is an eigenstate of $H$ with energy $E_n$,
\begin{equation}
    \label{eq:CC3}
    \varphi_{n,\tau}(t) = b_\tau^{(n)}\exp(-iE_nt)\,,
\end{equation}
then $\varphi_n^*(t)$ is an eigenstate with energy $-E_n$.
\begin{equation}
    \label{eq:CC4}
    \varphi_{n,\tau}^*(t) = b_\tau^{(n)\ast}\exp(iE_nt)\,.
\end{equation}
In consequence, the spectrum of the Hamiltonian consists of pairs of eigenvalues $(E_n, -E_n)$.
It has necessarily negative eigenvalues.
This is in accordance with the fact that the eigenvalues of the real matrix $\Shat$ come in pairs $(\lambda, \lambda^*)$.

\indent For probabilistic cellular automata a sufficient (but not necessary) condition for $H$ in the occupation number basis reads
\begin{equation}
    \label{eq:CC5}
    B H B = H\,.
\end{equation}
For the Hamiltonian \eqref{eq:H7} or \eqref{eq:D8} this is realized, cf.\ eq.~\eqref{eq:YY11}.
With $H^{*} = -H$ this implies that $H$ changes sign under charge conjugation
\begin{equation}
    C:\, H \to C H C = B H^{*} B = -H\,.
\end{equation}
The relation
\begin{equation}
    \label{eq:YY15}
    C (i H) C = - i B H^{*} B = - i H^{*}\,,
\end{equation}
is compatible with the complex conjugate of the Schrödinger equation~\eqref{eq:C24}.
We recall, however, that the choice of $H$ is not unique and eq.~\eqref{eq:CC5} does not necessarily hold in the occupation number basis.
Furthermore, a change of basis can change the $C$-transformation properties of $\Shat$ and $H$.
Nevertheless, the properties of the eigenvalues of $\Shat$ and $H$ do not change, in particular the occurrence of both $E_n$ and $-E_n$ in the spectrum of $H$.

\subsection*{Charge operators}
In the real picture one may define a local charge operator which reads out the Ising spin $s(j)$ for a given configuration $\tau$,
\begin{equation}
    \label{eq:CC7}
    \tilde{Q}(j) = \hat{n}(j) - \frac{1}{2} = \frac{1}{2} \hat{s}(j)\,,
\end{equation}
with total charge
\begin{equation}
    \label{eq:CC8}
    \tilde{Q} = \sum_{j}\tilde{Q}(j)\,.
\end{equation}
Charge conjugation switches the sign of $\tilde{Q}(j)$ and $\tilde{Q}$,
\begin{equation}
    \label{eq:CC9}
    \{C,\tilde Q(j)\} = 0\,,\quad \{C,\tilde Q\} = 0\,.
\end{equation}
There exist therefore no simultaneous eigenstates to $C$ and $\tilde Q$ for $\tilde{Q}\neq 0$.

\indent The charge observables $\tilde{Q}(j)$ or $\tilde{Q}$ are not compatible with our choice of a complex structure if $\tilde{Q}$ acts in the same way on $\tilde{q}$ and $\bar{q}'$.
In this case it anticommutes with the matrix $I$.
The basis vectors $E_\tau$ in eq.~\eqref{eq:C10} are eigenstates for all local charges $\tilde Q(j)$.
If $\varphi$ has only one non-zero component $\varphi_\tau$, possible eigenstates of $\tilde{Q}(j)$ can be inferred from eq.~\eqref{eq:C18E} by setting either $\tilde{q}_{\tau}$ or $\hat{q}_\tau$ to zero.
For $\hat{q}_\tau = 0$ one has $\tilde Q(j)\varphi_\tau = \frac{1}{2}\varphi_\tau$ if for the configuration $\tau$ the spin $s(j)$ is up, and $\tilde Q(j)\varphi_\tau = -\frac{1}{2}\varphi_\tau$ if it is down.
For eigenstates with $\tilde{q}_{\tau}=0$ the sign of $\tilde Q(j)$ is switched.
The conditions $\tilde{q}_\tau = 0$ or $\hat{q}_\tau=0$ mix $\varphi_\tau$ and $\varphi_\tau^*$.
The operators $\tilde{Q}(j)$ mix $\varphi$ and $\varphi^*$ and have therefore no expression as linear operators acting on $\varphi$.
Momentum eigenstates rotate between eigenstates with opposite values of $\tilde Q$.
The charge $\tilde Q$ does not commute with the momentum operator $\hat P$.

\indent We want to define a local charge operator $Q(j)$ which is compatible with our complex structure, such that wave functions with a single non-zero component $\tau$ are eigenstates of $Q(j)$ for arbitrary phases of $\varphi_\tau$.
For this purpose one may choose
\begin{equation}
    \label{eq:reduced_number_operator}
    Q(j) = \hat{N}(j)-\frac{1}{2} = a^\dagger(j) a(j) -\frac{1}{2}\,,
\end{equation}
with $a(j)$ given by eq.~\eqref{eq:F5} acting on the complex wave function.
The eigenvalues of $Q(j)$ are $\pm \frac{1}{2}$.
The total charge,
\begin{equation}
    Q' = \sum_j Q(j)\,,
\end{equation}
has half-integers eigenvalues if $N_x$ is odd.
For an eigenstate to the eigenvalue $Q(j)=\frac{1}{2}$ only configurations with $s(j)=1$ contribute to $\tilde{q}(t)$, while only configurations with $s(j)=-1$ contribute to $\bar{q}'(t)$.
In the real picture the charge $Q(j)$ acts as
\begin{equation}
    \label{eq:195A}
    Q(j) \begin{pmatrix} \tilde{q} \\ \bar{q}' \end{pmatrix} = \begin{pmatrix}
    \tilde{Q} & 0 \\ 0 & -\tilde{Q} \end{pmatrix} \begin{pmatrix} \tilde{q} \\ \bar{q}' \end{pmatrix}\,.
\end{equation}
The charges \eqref{eq:reduced_number_operator} are not compatible with the identification $\bar{q}' = \tilde{q}$.
We will discuss below an alternative charge which is compatible with this identification.

\indent The local charges and the total charge are flipped by charge conjugation,
\begin{equation}
    \label{eq:TCP8}
    \{Q(j),C\} = 0\,,\quad \{Q',C\} = 0\,.
\end{equation}
Indeed, the relation $CQ(j)C=-Q(j)$ follows from
\begin{equation}
    \label{eq:TCP9}
    \begin{aligned}
        C \hat{N}(j) C \varphi & = C a^\dagger(j) a(j) C \varphi = B \gl a^\dagger(j) a(j) B \varphi^* \gr^{*} \\
                           & = B a^\dagger(j) a(j) B \varphi = a(j) a^\dagger(j) \varphi                 \\
                           & = (1-\hat{N}(j))\varphi\,.
    \end{aligned}
\end{equation}
This implies that the $C$-eigenstates $\varphi_{+}$ and $\varphi_{-}$ in eq.~\eqref{eq:CC10} cannot be charge eigenstates. 

\indent With
\begin{equation}
    \label{eq:ZA1}
    Q' = \sum_j \left(a^\dagger(j) a(j) - \frac{1}{2}\right)
    = \sum_k \left(a^\dagger(k) a(k) - \frac{1}{2}\right)\,,
\end{equation}
one establishes that $Q'$ commutes with the momentum operator and therefore with $H$
\begin{equation}
    \label{eq:ZA2}
    [Q',\hat{P}] = 0\,,\quad
    [Q',H] = 0\,.
\end{equation}
The charge $Q'$ is therefore a conserved quantity.
A free massless Weyl fermion can be constructed from simultaneous eigenstates of $\hat{P}$ and $Q'$.
We observe the relations
\begin{alignat}{3}
    \label{eq:ZA3}
    [Q',a(k)] & = -a(k)\,, &  & \quad [Q',a^\dagger(k)] &  & = a^\dagger(k)\,,\nn \\
    [Q',a(j)] & = -a(j)\,, &  & \quad [Q',a^\dagger(j)] &  & = a^\dagger(j)\,,
\end{alignat}
and
\begin{equation}
    \label{eq:ZA4}
    [\hat{P},a(k)] = -\frac{2\pi k}{L} a(k)\,,\quad
    [\hat{P},a^\dagger(k)] = \frac{2 \pi k}{L} a^\dagger(k)\,.
\end{equation}

\indent Using eqs.~\eqref{eq:H7}\eqref{eq:63C} one also infers for $k>0$
\begin{equation}
    \label{eq:ZA5}
    [\hat{P},b^\dagger(k)] = \frac{2 \pi k}{L} b^\dagger(k)\,,\quad
    [Q',b^\dagger(k)] = -b^\dagger(k)\,.
\end{equation}
Thus $a^\dagger(k)$ creates a particle with $Q'=1$ and momentum $2\pi k/L$, while $b^\dagger(k)$ creates a particle with $Q' = -1$ and the same positive momentum.
This is the antiparticle of the particle with $Q' = 1$.
The precise notion of particles depends on the vacuum, as we will discuss below.

\indent One may subtract for the definition of the total charge the \qq{vacuum contribution} for $k=0$,
\begin{equation}
    \begin{aligned}
        Q & = \sum_{k\neq 0} \left(a^\dagger(k) a(k)-\frac{1}{2}\right)  =  Q' - a^\dagger(0)\,a(0) + \frac{1}{2}    \\
          & = \sum_{k>0} \left(a^\dagger(k) a(k) - b^\dagger(k) b(k)\right)\,.
    \end{aligned}
\end{equation}
Every occupied $k$-mode for $k>0$ contributes a charge one, while the antiparticle, or empty $k$-mode for $k<0$, contributes a charge minus one.
The eigenvalues of $Q$ are integer.
The relation \eqref{eq:ZA2} continues to hold if one replaces $Q'\to Q$.
This is also the case for the first relation \eqref{eq:ZA3} and \eqref{eq:ZA5} for $k\neq 0$.
In the limit $N_x\to \infty$ the difference between $Q'$ and $Q$ can be neglected.

\subsection*{CPT}
Charge conjugation can be combined with $P$ or $T$ in order to define $CP$ or $CT$ transformations.
The combined transformation $CPT$ induces in the complex picture the map
\begin{equation}
    \label{ZA6}
    CPT:\, \varphi_\tau(t)\to B \varphi_{\tau_P}^*(-t)\,.
\end{equation}
The Majorana--Weyl automaton or the Dirac automaton show a $CPT$-invariant evolution.
In particular, the one particle wave functions transform as
\begin{equation}
    \label{eq:ZA7}
    CPT:\, \varphi^{(1)}(t,x) \to B \varphi^{(1){*}}(-t,-x)\,.
\end{equation}
A plane-wave solution, $\varphi^{(1)} = u \exp(-i\omega t + i p x)$, with $Bu^{*}=u$, is invariant under $CPT$.

\indent If we choose the identification \eqref{eq:177B}, $\bar{q}' = PT(\tilde{q})$, the $CPT$-transform $\tilde{q}\to CPT(\tilde{q})$ amounts to the map $\tilde{q} \to \hat{q}$.
Thus $\varphi \to CPT(\varphi)$ is equivalent to $\varphi \to \varphi^*$.
Complex conjugation of the complex wave function is equivalent to a $CPT$-transformation in this case.
An interesting alternative is the identification of $\bar{q}'$ with the $CPT$-transform of $\tilde{q}$.
In this case $\varphi\to B\varphi$ operates the $CPT$-transform.
Now $\hat{q}$ stands for the $PT$-transform of $\tilde{q}$ and $\varphi \to \varphi^*$ encodes the $PT$-transformation.

\indent The transformation $CT$ involves a complex conjugation and is therefore antilinear,
\begin{equation}
    \label{eq:ZA8}
    CT:\, \varphi_{\tau}(t) \to B \varphi^{*}_\tau (-t)\,,
\end{equation}
or for the Dirac automaton
\begin{equation}
    \label{eq:ZA9}
    CT:\, \varphi_{(\tau_R,\tau_L)}(t) \to B \varphi_{(\tau_{L},\tau_R)}^{*}(-t)\,.
\end{equation}
The Hamiltonian \eqref{eq:D8} of the Dirac automaton is invariant under $CT$.
This is the reason why $CT$ is often taken as the definition of time reversal in particle physics.

\indent If we choose $\tilde{T} = CT$ for time reversal, the transformation $CP\tilde{T}$ amounts to $PT$.
The possible identification \eqref{eq:177B} chooses for $\bar{q}'$ the $CP\tilde{T}$-transform of $\tilde{q}$.
This identification is possible for all generalized Ising models for which the action is invariant under $PT = CP\tilde{T}$.
In particle physics the Lorentz symmetry requires invariance under $CP\tilde{T}$.
For models which are candidates for a Lorentz-invariant continuum limit we may require $CP\tilde{T}$ invariance of the action.
If we choose for $\bar{q}'$ the $CPT$-or $P\tilde{T}$-transform of $\tilde{q}$, the complex conjugation $\varphi \to \varphi^*$ stands for $CP\tilde{T} = PT$.
We will argue in the next section that this seems to be the most natural choice.

\subsection*{Alternative charge conjugations}
For the definition of local charges one may use different sets of annihilation and creation operators, as for example $A(j)$ in eq.~\eqref{eq:AA9A}.
The local charges, 
\begin{equation}
    \label{eq:207A}
    Q(j) = A^\dagger(j) A(j) - \frac{1}{2}\,,
\end{equation}
are compatible with the identification $\bar{q}' = \tilde{q}$ since $[B,Q(j)] = 0$.
A charge conjugation operator adapted to this set is given by
\begin{equation}
    \label{eq:ACC1}
    C:\, \varphi\to B_c \varphi^*\,,\quad
    B_c^* B_c = 1\,.
\end{equation}
With $B_c$ obeying a relation analogous to eq.~\eqref{eq:TCP10},
\begin{equation}
    \label{eq:ACC2}
    C A(j) C = B_c A^*(j) B_c^* = A^\dagger(j)\,,
\end{equation}
this charge conjugation switches the sign of $Q(j)$ according to eq.~\eqref{eq:TCP8}.
With $A(j) = W a(j) W^\dagger$ and $W$ given by eq.~\eqref{eq:WoperatorSets} one finds
\begin{equation}
    \label{eq:209A}
    \begin{aligned}
        B_c &= W B W^T \\
            &= \tau_3 \otimes -i\tau_2 \otimes\tau_3 \otimes -i\tau_2\otimes \dots\otimes \tau_3\,.
    \end{aligned}
\end{equation}

\indent More generally, the discrete transformations $P$, $T$ and $C$ may be supplemented by multiplication of matrices which preserve the properties $P^2=1$, $T^2=1$, $C^2=1$.
For extended models this may be necessary in order to guarantee covariance with respect to certain global or local continuous symmetry transformations.

\section{Vacuum}
\label{sec:vacuum}

\indent There exist many wave functions that are invariant under translations in space and time.
One example is the totally empty vacuum $\ket{0}_E$.
Another example is the total equipartition state $\ket{0}_{eq}$ for which all components of the wave function are equal, $\tilde{q}_{\tau} = 2^{-N_x}$, or $\tilde{q}_\tau = \bar{q}'_\tau = 2^{-(N_x+1)}$ for $\bar{q}' \neq \tilde{q}$.
This state is also invariant under $C$ and has $\langle Q \rangle = 0$.
For the Majorana--Weyl automaton or the Dirac automaton these states are possible candidates for a vacuum state.

\indent In the view of adding interactions we will focus for probabilistic cellular automata on the particle physics vacuum $\ket{0}_{p}$.
It corresponds to a minimum of the Hamiltonian with ground state energy $E_0$.
All particle excitations of this vacuum have energies larger than $E_0$.
If the Hamiltonian continues to have a minimum at some $E_0$ in the presence of interactions, one may expect stable particle excitations of the particle physics vacuum for a PCA describing interacting fermions.
The particle excitations have necessarily a higher energy than the vacuum and cannot \qq{decay} into states with energy below $E_0$.
The energy is a conserved quantity.
This would, however, not prevent instabilities if energies smaller than the vacuum energy are associated to accessible states.
The exclusion of this possibility for a vacuum with minimal energy is a powerful agent for stability.

\indent In the next section we will discuss particle excitations of the particle physics vacuum.
The associated Feynman propagator of the discrete fermionic quantum field theory becomes the standard Feynman propagator for Weyl or Dirac fermions in the continuum limit.
This will not be the case for an arbitrary choice of the vacuum.

\subsection*{Majorana--Weyl vacuum}
For the Majorana--Weyl automaton the particle physics vacuum obeys for $k>0$ the relations
\begin{equation}
    \label{eq:V1}
    a(k)\ket{0}_p = 0\,,\quad
    b(k)\ket{0}_p = 0\,.
\end{equation}
According to eq.~\eqref{eq:63D} the spectrum of the Hamiltonian has a minimum at $E_0$.
This minimum is realized for the particle physics vacuum due to eq.~\eqref{eq:V1}.
From eq.~\eqref{eq:V1} one also infers that this vacuum has vanishing charge
\begin{equation}
    \label{eq:V2}
    Q\ket{0}_p = 0\,,
\end{equation}
and
\begin{equation}
    \label{eq:V3}
    \bra{0}_p Q' \ket{0}_p = 0\,.
\end{equation}

\indent Under a time translation by $\eps$ the particle physics vacuum is not invariant but rather picks up a phase
\begin{equation}
    \label{eq:V4}
    \ket{0}_p(t+\eps) = \Shat(t) \ket{0}_p (t) = \exp(-i\eps E_0) \ket{0}_p(t)\,.
\end{equation}
This phase involves the vacuum energy $E_0$.
As familiar from quantum mechanics the vacuum energy will not matter for particle excitations.
The dynamics of particle excitations for a given vacuum only involves the energy difference $H-E_0$.
At a given $t$ the translation of $\ket{0}_p$ in $x$ by $\eps$ involves the same phase factor
\begin{equation}
    \label{eq:V5}
    \ket{0}_p \to \Shat \ket{0}_p = \exp(-i\eps E_0) \ket{0}_p\,.
\end{equation}
Again, such a phase shift of the vacuum will not matter for particle excitations.

\indent In view of the continuum limit we may actually consider shifts by $8\eps$ in time or space, replacing effectively
\begin{equation}
    \label{eq:V6}
    8E_0 \to \frac{2\pi}{L}\,,
\end{equation}
by identifying $8E_0$ with $8E_0+2\pi/\eps$.
Thus $8 E_0$ vanishes in the infinite volume limit $L\to \infty$.
(For an even number of sites one has \cite{CWPW2020} $8E_0\to 0$.)
For $L \to \infty$ the particle physics vacuum is then invariant under time- and space-translations by $8\eps$.

\indent A vacuum without time translation invariance may perhaps seem somewhat unfamiliar.
This is, however, the general case for a quantum theory with a non-zero ground state energy.
For a quantum theory the overall phase of the wave function drops out in the expression for the expectation values of observables.
Despite the factor $\exp(-i E_0 t)$ in the wave function, the expectation values of observables in the ground state are time independent.
In our case this applies to all observables which are compatible with the complex structure, see appendix \ref{app:B}.
Their expectation values in the particle physics vacuum are invariant under discrete translations by $\eps$ in time or space.
For observables which are not compatible with the complex structure this does no longer hold.
An overall phase change rotates between particles and holes according to eq.~\eqref{eq:49A0}.
For $L \to \infty$ the expectation values of \qq{incompatible observables} are invariant only with respect to translations by $8 \eps$.

\subsection*{Half filling}
\indent Let us assume some choice of fermionic operators $a(j)$, $a^\dagger(j)$ in position space such that the local occupation number $\hat{N}(j) = a^\dagger(j) a(j)$ is compatible with the complex structure.
The particle physics vacuum $\ket{0}_p$ is half-filled with respect to $\hat{N}(j)$.

\indent The relations for $k>0$,
\begin{equation}
    \label{eq:V7}
    a^\dagger(k)a(k)\ket{0}_p = 0\,,\quad
    a^\dagger(-k)a(-k)\ket{0}_p = \ket{0}_p\,,
\end{equation}
have the simple interpretation that all $(N_x-1)/2$ modes with negative $k$ are filled, and all modes with positive $k$ are empty.
For $k=0$ one employs the condition
\begin{equation}
    \label{eq:V8}
    \bra{0} a^\dagger(k=0) a(k=0) \ket{0} = \frac{1}{2}\,.
\end{equation}
Defining $\hat{N} = \sum_{j} \hat{N}(j) = Q' + N_x/2$ yields
\begin{equation}
    \label{eq:V9}
    \bra{0} \hat{N} \ket{0} = \frac{N_x}{2}\,.
\end{equation}

\indent For each individual site $j$ one has
\begin{equation}
    \label{eq:V10}
    \langle \hat{N}(j)\rangle = \bra{0} \hat{N}(j) \ket{0} = \frac{1}{2}\,.
\end{equation}
This follows from
\begin{equation}
    \label{eq:V11}
    \begin{aligned}
        \langle \hat{N}(j) \rangle
         & = \bra{0} a^\dagger(j) a(j) \ket{0}                                                                               \\
         & = \frac{1}{N_x} \sum_{k,l}\exp\left\lbrace\frac{2\pi i j(l-k)}{N_x}\right\rbrace \bra{0} a^\dagger(k)a(l) \ket{0} \\
         & = \frac{1}{N_x} \left\lbrace \sum_{k<0} \bra{0} a^\dagger(k)a(k) \ket{0} + \frac{1}{2}\right\rbrace\,.
    \end{aligned}
\end{equation}
All these relations only involve the Fourier expression of $a(k)$ and therefore generalize if one replaces $a(j)$ in eq.~\eqref{eq:F5} by $A(j)$ as in eq.~\eqref{eq:AA9A}, or some other fermionic operators obeying the anticommutation relations \eqref{eq:F4} 

\indent We can construct the particle physics vacuum $\Ket{0}_p$ at $t=0$ from the totally empty state $\Ket{0}_E$ as
\begin{equation}
    \label{186Y}
    \Ket{0}_p = \prod_{k<0} a^\dagger(k) \frac{1+a^\dagger(0)}{\sqrt{2}} \ket{0}_E\,,
\end{equation}
where the operators in the product are ordered with smaller $k$ to the left. It is normalized, ${\braket{0|0}}{}_p=1$, and obeys (with $N_x=1\mod 4$)
\begin{equation}
    \label{eq:186Z}
    \gl a^\dagger(k=0)+a(k=0)\gr \ket{0}_p = \ket{0}_p\,,
\end{equation}
which implies eq.~\eqref{eq:V8}.
This vacuum is a superposition of states with $(N_x-1)/2$ and $(N_x+1)/2$ occupied bits.
Inserting the definition \eqref{eq:H3} the part with $\tilde{N} = (N_x-1)/2$ occupied bits reads
\begin{equation}
    \label{eq:186Y2}
    \begin{aligned}
        \ket{0}_{p-} = & \, \frac{1}{\sqrt{2}}\prod_{k>0}\!\left( \sum_{j_k} \frac{1}{\sqrt{N_x}}\exp\left\lbrace-\frac{2\pi i}{N_x}k j_k\right\rbrace a^\dagger(j_k)\right)\!\ket{0}_E \\
        =              & \,           \frac{1}{\sqrt{2}}N_x^{-\tilde{N}/2} \sum_{j_1}\dots\sum_{j_{\tilde{N}}}\exp\bigg\lbrace-\frac{2\pi i}{N_x}  (j_1 + 2j_2 + \dots                  \\
                       & \dots+\tilde{N}j_{\tilde{N}})\bigg\rbrace\,a^\dagger(j_{\tilde{N}})\dots a^\dagger(j_1)\ket{0}_E\,.
    \end{aligned}
\end{equation}
A given configuration with $\tilde{N}$ occupied sites sums up the terms with creation operators at these sites.
This involves a sum over different phase factors, with positive or negative signs due to the anticommutators necessary to bring the creation operators to the given order.
The second term with $(N_x+1)/2 = \tilde{N}+1$ occupied sites involves a similar product of sums, with an additional factor given by a sum over $j_0$.
While the properties of the particle physics vacuum are simple in momentum space, they get rather complex in the space of Ising spin configurations (position space).

\indent The particle physics vacuum is an eigenstate of the discrete transformation $\varphi \to B\varphi$,
\begin{equation}
    \label{eq:186A}
    B \ket{0}_p = \ket{0}_p\,.
\end{equation}
The condition \eqref{eq:V1} is invariant under $\varphi\to B\varphi$ due to the relation \eqref{eq:YY10a},
\begin{equation}
    \label{eq:186B}
    B a(k) B = b(k)\,,
\end{equation}
and this also holds for eq.~\eqref{eq:186Z}.
The Majorana--Weyl vacuum $\ket{0}_p$ is compatible with the identification $\bar{q}' = \tilde{q}$.
If one uses this identification one may choose $A(j)$ according to eq.~\eqref{eq:AA9A} for the general annihilation operator $a(j)$.
The local occupation number $\hat{N}(j)$ is compatible with the constraint $[B.\hat{N}_j]=0$.

\indent If the vacuum wave function would be real, the vacuum would be an eigenstate of charge conjugation.
This cannot be since for $k>0$ one has $a^\dagger(k) C \ket{0}_p=0$, such that a $C$-invariant vacuum would have to obey simultaneously $a(k)\ket{0}_p=0$ and $a^\dagger(k) \ket{0}_p = 0$ for $k>0$.
We conclude $C\ket{0}_p = \ket{0}_p^* \neq \ket{0}_p$, in accordance with eq.~\eqref{eq:186Y2}.
In the real picture the transformation $\varphi\to B\varphi$ reads
\begin{equation}
    \label{eq:186D}
    \varphi\to B\varphi \;\;\hateq\;\; \tilde{q}\to\bar{q}' = B\hat{q}\,.
\end{equation}
For a $B$-invariant vacuum this implies
\begin{equation}
    \label{eq:186E}
    \langle\tilde{q}\,\hat{n}(j)\,\tilde{q}\rangle = \langle{\bar{q}'}\hat{n}(j)\,\bar{q}'\rangle\,,
\end{equation}
in accordance with $\langle\hat{N}(j)\rangle = 1/2$, cf.\ eq.~\eqref{eq:103A}.

\subsection*{Dirac vacuum}

\indent For the Dirac automaton the particle physics vacuum is a direct product of the corresponding vacua for right-movers and left-movers.
For the left-movers one replaces $k$ by $-k$.
The particle physics vacuum for left-movers obeys for $k>0$ the constraints
\begin{equation}
    \label{eq:186F}
    a_L(-k)\ket{0}_D = a^\dagger_L(k)\ket{0}_D = 0\,.
\end{equation}
The energy of the vacuum amounts to $2E_0$.
On the other hand, the momentum of the vacuum vanishes
\begin{equation}
    \label{eq:186G}
    \hat{P} \ket{0}_D = 0\,.
\end{equation}
The operator $\exp(-i\eps \hat{P})$ translates all configurations in the positive $x$-direction by $\eps$.
Thus eq.~\eqref{eq:186G} implies that $\ket{0}_D$ is invariant by discrete space translations.
This contrasts with the particle physics vacuum for the Majorana-Weyl automaton where $\hat{P}\ket{0}_p = E_0 \ket{0}_p$, such that discrete space translations generate an overall phase.

\indent The Dirac vacuum is invariant under space-translations by $\eps$, as well as under the discrete symmetries $P$, $CT$, and $CPT$.
It is not invariant under $C$, $T$ or $PT$.
Thus the Dirac vacuum is compatible with the identification of $\bar{q}'$ with the $CPT$-transform of $\tilde{q}$, but not for the identifications with the $T$- or $PT$-transformations of $\tilde{q}$.
It seems therefore most natural to identify $\bar{q}'$ with the $CPT$-transform of $\tilde{q}$.

\section{One-particle states}
\label{sec:one-particle-states}

\indent One-particle states obtain by applying creation or annihilation operators to a vacuum states. 
They should be considered as excitations from a given vacuum.
In general, the properties of particles depend on the vacuum state.
We explore here the particle physics vacuum and free massless Weyl fermions as one particle states.
For this purpose we introduce field operators that are defined on continuous time- and space-variables.
This allows us to formulate one-particle wave functions in continuous spacetime.
The evolution obeys the continuous Dirac equation (or its restriction to Weyl fermions) with the associated Lorentz symmetry.
The discreteness of the lattice formulation shows up only in the commutation relations for the field operators.
We also discuss correlations of field operators.
The Feynman propagator is a discretized version of the continuum propagator, allowing for a straightforward continuum limit.
The discrete Feynman propagator for the chiral Weyl fermions shows no sign of fermion doubling.

\indent The definition of the particle physics vacuum in the preceding section only involves the operators $a(k)$ and $a^\dagger(k)$ in Fourier space.
They can be defined for different versions of annihilation and creation operators in position space.
From now on we assume for $a(j)$ and $a^\dagger(j)$ only the anticommutation relations \eqref{eq:F4} and leave it open if one chooses eq.~\eqref{eq:F5} or \eqref{eq:AA9A} or some other representation.
The implications for Ising spin configurations in position space may depend on this choice, however.
For any choice the relation to the operators in momentum space is given by eq.~\eqref{eq:H3}.

\subsection*{Dirac equation}
We focus on the Dirac automaton and the particle physics vacuum.
One-particle states are characterized by an additional particle or hole which may be a right-mover or a left-mover.
In position space the complex one-particle wave function has two components $\tilde{\varphi_R}(t,x)$ and $\tilde{\varphi}_L(t,x)$, where $x$ denotes the position of the particle or antiparticle.
The updating of the automaton moves all right-movers one position to the right and all left-movers one position to the left.
One expects the relation
\begin{equation}
    \label{eq:OPS1}
    \tilde{\varphi}^{(1)}(t+\eps,x) 
    = \begin{pmatrix} \tilde{\varphi}_R(t+\eps,x) \\ \tilde{\varphi}_L(t+\eps,x) \end{pmatrix} 
    = \begin{pmatrix} \tilde{\varphi}_R(t,x-\eps) \\ \tilde{\varphi}_L(t,x+\eps) \end{pmatrix}\,.
\end{equation}
In terms of lattice derivatives
\begin{align}
    \label{eq:OPS2}
    \partial_t \tilde{\varphi}(t,x) &= \frac{1}{2\eps} \gl \tilde{\varphi}(t+\eps,x) - \tilde{\varphi}(t-\eps,x) \gr\,,\nn \\
    \partial_x \tilde{\varphi}(t,x) &= \frac{1}{2\eps} \gl \tilde{\varphi}(t,x+\eps) - \tilde{\varphi}(t,x-\eps) \gr
\end{align}
this reads
\begin{equation}
    \label{eq:OPS3}
    i \partial_t \tilde{\varphi}^{(1)}(t,x) = -i \partial_x \tau_3 \tilde{\varphi}^{(1)}(t,x)\,.
\end{equation}

\indent With Dirac matrices
\begin{equation}
    \label{eq:OPS4}
    \gamma^0 = -i \tau_2\,,\quad \gamma^1 = \tau_1\,,
\end{equation}
this takes the form of the Dirac equation for free massless fermions,
\begin{equation}
    \label{eq:OPS5}
    \gamma^\mu \partial_\mu \tilde{\varphi}^{(1)}(t,x) = 0\,,
\end{equation}
where $\partial_0 = \partial_t$ and $\partial_1 = \partial_x$.
For smooth enough wave functions the lattice derivatives can be replaced by partial continuum-derivatives.
Smooth wave functions remain smooth under the evolution of the Dirac automaton.
The continuum limit is straightforward in this case and results in the continuous Dirac equation for free massless fermions in two dimensions.
The continuum limit realizes Lorentz symmetry for the one-particle states.

\indent The one-particle wave function describes both particles and antiparticles.
They are related by complex conjugation or charge conjugation.
Furthermore, the particle physics vacuum is not invariant under time translations but rather picks up a phase $\exp(-2 i E_0 t)$.
One therefore needs some care for the precise definition of the wave function which obeys eq.~\eqref{eq:OPS1}. Right-movers and left-movers can be discussed separately.
We focus here on right moving Weyl fermions for the Dirac automaton. 
The discussion for the Majorana--Weyl automaton only differs by the value of the vacuum energy.

\subsection*{One-particle states with positive energy}
The one-particle state vector in Hilbert space for right-movers reads
\begin{align}
    \label{eq:OPS6}
    \varphi_R^{(1)}(t) &= \sum_{k} \tilde{\varphi}(t,k) \left(a^\dagger(k) + a(k)\right)  
    \ket{0} \\
    &= \left[ \sum_{k>0} \tilde{\varphi}(t,k) a^\dagger(k) + \sum_{k<0} \tilde{\varphi}(t,k) a(k) \right] \ket{0} \nn\,.
\end{align}
The one-particle wave function $\tilde{\varphi}(t,k)$ in momentum space describes particles as well as holes or antiparticles.
Since $\tilde{\varphi}(t,k=0)$ does not contribute in eq.~\eqref{eq:OPS6} we set it to zero.

\indent With eqs.~\eqref{eq:63C}, \eqref{eq:63D} the Hamiltonian acts on $\varphi_R^{(1)}(t)$ as 
\begin{align}
    \label{eq:OPS7}
    \Big( H^{(R)} -& 2 E_0 \Big) \varphi_R^{(1)}(t) \nn\\
        &= \frac{2\pi}{L} \sum_{k>0} k \left[ \tilde{\varphi}(t,k) a^\dagger(k) + \tilde{\varphi}(t,-k) a(-k) \right] \ket{0} \nn\\
        &= \frac{2\pi}{L} \sum_{k} |k| \tilde{\varphi}(t,k) \left( a^\dagger(k) + a(k) \right) \ket{0}\,,
\end{align}
with $H^{(R)}\ket{0} = 2E_0\ket{0}$ for the Dirac vacuum.
The appearance of the absolute value $|k|$ reflects the different phases in eq.~\eqref{eq:H9}.
The relative energy $H^{(R)} - 2E_0$ of the particle or antiparticle is positive, as expected.
On the other hand, the presence of the absolute value $|k|$ does not allow to perform a simple Fourier transformation.

\indent The one-particle wave function \eqref{eq:OPS6} is a sum of particle and antiparticle parts, which are eigenstates of the charge $Q$ with opposite eigenvalues
\begin{align}
    \label{eq:OPS8}
    \varphi^{(1)}_R(t) &= \varphi^{(1)}_p(t) + \varphi^{(1)}_a(t)\,, \\
    \varphi^{(1)}_p(t) &= \sum_{k>0} \tilde{\varphi}(t,k) a^\dagger(k) \ket{0}\,,& Q\varphi^{(1)}_p(t) &= \varphi^{(1)}_p(t) \,,\nn \\
    \varphi^{(1)}_a(t) &= \sum_{k>0} \tilde{\varphi}(t,-k) b^\dagger(k) \ket{0}\,,& Q\varphi^{(1)}_a(t) &= -\varphi^{(1)}_a(t)\,.\nn
\end{align}
Momentum eigenstates with eigenvalue $p=2\pi \bar{k}/L$, $L=N_x \eps$, $\bar{k}>0$ obtain for $\tilde{\varphi}(t,k) = c_p \delta_{k,\bar{k}}$ or $\tilde{\varphi}(t,k) = c_a \delta_{k,-\bar{k}}$ for particles and antiparticles, respectively.
Those are energy eigenstates with $E - 2E_0 = p$.
In accordance with eq.~\eqref{eq:ZA2} one can realize simultaneous eigenstates of $Q$, $\hat{P}$ and $H$.

\indent For the time evolution one employs eq.~\eqref{eq:H9}
\begin{align}
    \label{eq:OPS9}
    \varphi^{(1)}_p (t+\eps) 
    &= \Shat \varphi^{(1)}_p(t) \\
    =&\; \sum_{k>0} \tilde{\varphi}(t,k) \exp\left\{-2i\eps\left(E_0 + \frac{\pi k}{L}\right)\right\} a^\dagger(k) \ket{0},\nn
\end{align}
or, for $k>0$,
\begin{equation}
    \label{eq:OPS10}
    \tilde{\varphi}(t+\eps,k) = \exp\left\{-2i\eps\left(E_0 + \frac{\pi k}{L}\right)\right\} \tilde{\varphi}(t,k)\,.
\end{equation}
With a similar relation for $\varphi^{(1)}_a(t+\eps)$ one finds for $k<0$
\begin{equation}
    \label{eq:OPS11}
    \tilde{\varphi}(t+\eps,k) = \exp\left\{-2i\eps\left(E_0 - \frac{\pi k}{L}\right)\right\} \tilde{\varphi}(t,k)\,.
\end{equation}
The different phase factors reflect the appearance of the absolute value $|k|$ in eq.~\eqref{eq:OPS7}.

\indent For wave functions in position space we employ a different Fourier transform for particles and antiparticles.
For $k>0$ we take 
\begin{align}
    \label{eq:OPS12}
    \tilde{\varphi}(t,k) &= \sum_{j} D(k,j) \tilde{\varphi}_p(t,j)\,,\nn\\
    \varphi^{(1)}_p(t) &= \sum_{j} \tilde{\varphi}_p(t,j) a^\dagger(j) \ket{0}\,,
\end{align}
while for $k<0$ we use instead
\begin{align}
    \label{eq:OPS13}
    \tilde{\varphi}(t,k) &= \sum_j D^*(k,j) \tilde{\varphi}_a^*(t,j)\,,\nn\\
    \varphi^{(1)}_a(t) &= \sum_j \tilde{\varphi}_a^* (t,j) a(j) \ket{0}\,.
\end{align}
Here $\tilde{\varphi}_p(t,j)$ and $\tilde{\varphi}_a(t,j)$ are defined by
\begin{align}
    \label{eq:OPS14}
    \tilde{\varphi}_p(t,j) &= \sum_{k>0} D^{-1}(j,k) \tilde{\varphi}(t,k)\,,\nn\\
    \tilde{\varphi}_a(t,j) &= \sum_{k<0} D^{-1}(j,k) \tilde{\varphi}^*(t,k)\,.
\end{align}

\subsection*{Wave function for the Dirac equation}
We are now ready to define the one-particle wave function which obeys the Dirac equation.
The difference in the time evolution and the Fourier transform of the modes with $k>0$ and $k<0$ has its root in the different transformation properties of $a^\dagger(k)$ and $a(k)$ in eq.~\eqref{eq:OPS6}.
One can find a simpler description by a new wave function $\tilde{\varphi}_R(t,k)$ which combines parts of $\tilde{\varphi}(t,k)$ and its complex conjugate $\tilde{\varphi}^*(t,k)$.
Since $\tilde{\varphi}^*(t,k)$ has opposite energy eigenvalues as compared to $\tilde{\varphi}(t,k)$ the wave function $\tilde{\varphi}_R(t,k)$ will exhibit positive and negative energies.
The \qq{physical energies} of the particle or antiparticle excitations are positive as visible in eq.~\eqref{eq:OPS7}.

\indent For the definition of $\tilde{\varphi}_R$ we combine the Fourier components of $\tilde{\varphi}$ and $\tilde{\varphi}^*$ for positive and negative $k$ respectively,
\begin{equation}
    \label{eq:OPS15}
    \tilde{\varphi}_R (t,k) = \begin{cases}
        \exp(2 i t E_0) \tilde{\varphi}(t,k) & \text{for} \quad k>0 \\
        \exp(-2 i t E_0) \tilde{\varphi}^*(t,k) & \text{for} \quad k<0
    \end{cases}\,.
\end{equation}
This yields a uniform evolution for all $k$
\begin{equation}
    \label{eq:OPS16}
    \tilde{\varphi}_R(t+\eps,k) = \exp\left\{ -\frac{2\pi i k \eps}{L}\right\} \tilde{\varphi}_R(t,k)\,.
\end{equation}
In position space one has a uniform Fourier transform
\begin{equation}
    \label{eq:OPS16A}
    \tilde{\varphi}_R(t,j) = \sum_k D^{-1}(j,k) \tilde{\varphi}_R(t,k)\,.    
\end{equation}
If one omits the phase factors involving the ground state energy $E_0$ one has $\tilde{\varphi}_R(t,j) = \tilde{\varphi}_p(t,j) + \tilde{\varphi}_a (t,j)$. The evolution in position space obeys
\begin{equation}
    \label{eq:OPS17}
    \tilde{\varphi}_R(t+\eps,j) = \tilde{\varphi}_R(t,j-1)\,.
\end{equation}
This is precisely the evolution of $\tilde{\varphi}_R(t,x)$ in eq.~\eqref{eq:OPS1}.

\indent The evolution for $\tilde{\varphi}_L(t,j)$ in \eqref{eq:OPS1} follows in complete analogy, replacing $k\to-k$.
We may use the Hamilton operator for defining a continuous evolution for time in-between the discrete time points.
We can then use the continuous Dirac equation \eqref{eq:OPS3} for interpolating the wave function to continuous $x$.
The continuous wave function $\tilde{\varphi}^{(1)}(t,x)$ obeys therefore by definition the Dirac equation.
It coincides with the wave function $\tilde{\varphi}^{(1)}(t,j)$ for the PCA for the discrete space and time points on the lattice, $t=m_t\eps$, $x=j\eps$.
As is well known, the Dirac equation has solutions with positive and negative energies.
Our construction, which is in complete analogy to constructions in relativistic particle physics, shows that this does not contradict to the positivity of the relative energy $E-2E_0$ for the one-particle excitations of the Dirac automaton.

\indent The complex conjugate $\tilde{\varphi}_a^*(t,j)$ has the opposite charge as $\tilde{\varphi}_a(t,j)$.
The complex wave function $\tilde{\varphi}_R(t,j)$ corresponds therefore to an eigenstate of charge with $Q=1$.

\section{Field operators}
\label{sec:field-operators}
Relativistic quantum field theory is often formulated in terms of field operators.
They permit a systematic discussion of Lorentz symmetry and the construction of the $S$-matrix \cite{weinberg2014}.
Field operators which depend on continuous time and space variables can be constructed for the discrete setting of PCA as well.
For a given reference time, say $t=0$, they interpolate between the annihilation and creation operators $a(j)$, $a^\dagger(j)$ on the lattice points $x=j\eps$.
Switching to Heisenberg operators the field operators become time-dependent.
Heisenberg operators can the formulated for continuous time using the Hamilton operator.
This permits a continuous interpolation between the discrete time steps.
The Hamiltonian can be expressed in terms of the continuous field operators.
The discreteness of the PCA remains manifest, however, in deviations from the canonical commutation relations for positions away from the discrete lattice points.
In the continuum limit these deviations become irrelevant.

\indent Field operators define continuous families of operators and observables even for a discrete setting without a continuum limit, for which the Hilbert space remains finite dimensional.
This permits a formulation of continuous symmetries, as continuous translations, Lorentz symmetry or chiral symmetry, for discrete generalized Ising models.

\subsection*{Field operators for right movers}
We will discuss here the field operators for the right-movers, with straightforward extension to the left-movers.
We define field operators in terms of the annihilation and creation operators in Fourier space
\begin{align}
    \label{eq:FO1}
    \Psi_+(t,x) & = \sum_k u(t,x;\,k) \,a(k)\,,\nn      \\
    \Psi_-(t,x) & = \sum_k v(t,x;\,k) \,a^\dagger(k)\,.
\end{align}
For the coefficient functions $u$ and $v$ we take
\begin{align}
    \label{eq:FO2}
    u(t,x;\,k) & = \frac{1}{\sqrt{L}} \exp\left\{\frac{2\pi i k}{L}(x-t)\right\}\nn  \\
    v(t,x;\,k) & = \frac{1}{\sqrt{L}} \exp\left\{-\frac{2\pi i k}{L}(x-t)\right\}\,.
\end{align}
One has
\begin{equation}
    \label{eq:FO8}
    \Psi_-(t,x) = \Psi_+^\dagger(t,x)\,.
\end{equation}
For $t=0$, $x=\eps j$, $j$ integer, one finds
\begin{equation}
    \label{eq:FO3}
    \Psi_+ (0,\eps j) = \frac{1}{\sqrt{\eps}} a(j)\,,
    \quad\Psi_-(0,\eps j) = \frac{1}{\sqrt{\eps}} a^\dagger(j)\,.
\end{equation}

\indent We extend field operators to continuous $x$ and $t$.
They will interpolate between the operators at the discrete sites of the lattice.
Denoting $\Psi_{\pm} (x) = \Psi_{\pm}(0,x)$ we define for continuous $x$
\begin{equation}
    \label{eq:225A}
    \Psi_+(x)=\sqrt{\eps} \sum_j \tilde{\delta}(x-\eps j) a(j)\,,
\end{equation}
and similar for $\Psi_-(x)$ with $a(j)$ replaced by $a^\dagger(j)$.
Here the interpolation function
\begin{equation}
    \label{eq:FO10}
    \tilde{\delta}(x) = \frac{1}{L} \sum_k \exp\left(\frac{2\pi i k x}{L}\right)\,,
\end{equation}
is a type of modified $\delta$-function, with 
\begin{align}
    \label{eq:IP2}
    \tilde{\delta}^*(x) &= \tilde{\delta}(-x) = \tilde{\delta}(x)\,,\quad
    \tilde{\delta}(x+L) = \tilde{\delta}(x)\,, \nn\\
    \int_x\tilde{\delta}(x) &= \int_{-L/2}^{L/2}\mathrm{d}x\, \tilde{\delta}(x) = 1\,.
\end{align}
For $\eps\to 0$ and $L\to \infty$ the function $\tilde{\delta}(x)$ approaches the distribution $\delta(x)$.
The interpolation function $\tilde{\delta}(x)$ has the well known explicit form
\begin{equation}
    \label{eq:IP2A}
    \tilde{\delta}(x) = \frac{\sin(\pi x/\eps)}{L\sin(\pi x/L)}\,.
\end{equation}
It obeys the relations ($j$ integer)
\begin{align}
    \label{eq:IP3}
    \tilde{\delta}(0) =& \frac{1}{\eps}\,,\quad
    \sum_j \tilde{\delta}(x-\eps j) = \frac{1}{\eps}\,, \nn\\
    \tilde{\delta}(x=\eps j) &= 0 \quad\text{for}\quad j\ne0\mod N_x\,,
\end{align}
and for $M\ge 1$,
\begin{equation}
    \label{eq:IP4}
    \int_x \tilde{\delta}(x-\eps j) \prod_{i=1}^M \tilde{\delta}(x - \eps j'_i) = \eps^{-M} \prod_{i=1}^M \hat{\delta}(j'_i-j)\,.
\end{equation}
One also observes the identity for $j\neq 0$
\begin{equation}
    \label{eq:IP5}
    \partial_x \tilde{\delta}(x)(x=j\eps) = \frac{\pi}{\eps^2 N_x} \frac{(-1)^j}{\sin\left(\frac{\pi j}{N_x}\right)}\,,
\end{equation}
while $\partial_x \tilde{\delta}(x)|_{x = 0}=0$.
And for $j'\neq j$ one has
\begin{equation}
    \label{eq:IP6}
    \int_x \tilde{\delta}(x-\eps j') \partial_x \tilde{\delta}(x-\eps j) = \frac{\pi}{\eps^2 N_x} \frac{(-1)^{j'-j}}{\sin\left(\frac{\pi (j'-j)}{N_x}\right)}\,,
\end{equation}
while the r.h.s.\ of eq.~\eqref{eq:IP6} vanishes for $j'=j$.
The operators $\Psi_{\pm} (t,x)$ will be the Heisenberg operators associated to $\Psi_{\pm}(x)$.

\indent The anticommutator of field operators reads
\begin{align}
    \label{eq:FO9}
    \big\{\Psi_{\pm}^\dagger(t,x),\Psi_{\pm}(t',x')\big\} &= \tilde{\delta}(x'-x-t'+t)\,,\nn\\
    \big\{ \Psi_{\pm}(t,x), \Psi_{\pm}(t',x')\big\} &= 0\,.
\end{align}
For $(t,x)$-points situated on the lattice the anticommutator \eqref{eq:FO9} differs from zero only on the light cone $x'=x+t'-t$.
For $\eps \to 0$ at fixed $L$ the anticommutator vanishes for all continuous $(t,t',x,x')$ not obeying the light cone condition.
For right-movers the light cone only consists of the part $x'=x+t'-t$, while for left-movers it will be the other part $x'=x-t'+t$.

\subsection*{Hamiltonian in terms of field operators}
One can express the Hamiltonian in terms of the continuous field operators.
This redundant description will be the basis of a discussion of continuous symmetries for generalized Ising models.
For the Majorana--Weyl automaton the Hamiltonian can be expressed as a bilinear of the field operators as
\begin{align}
    \label{eq:FO7}
    H = \hat{P} & = \int\mathrm{d}x\,\Psi_+^\dagger(t,x) (-i\partial_x)\Psi_+(t,x) \nn \\
                & = \int\mathrm{d}x\,\Psi_-^\dagger(t,x) (i\partial_x)\Psi_-(t,x)\,.
\end{align}
This expression of the Hamiltonian in terms of field operators brings the PCA close to the more standard formulation of quantum field theories.
The Hamiltonian is now formulated in continuous space, with a continuous integral $\int\mathrm{d}x$.
The time dependence of the field operators drops out.
For the Dirac automaton one adds to $H$ a similar piece for the left-movers, with $\partial_x$ replaced by $-\partial_x$.

\indent One can use the field operators for a description of continuous one-particle states, see below.
For this purpose we establish the commutator of field operators with the momentum or charge operator.
The commutators with the momentum operator \eqref{eq:F15} read
\begin{equation}
    \label{eq:FO4}
    [\hat{P},\Psi_{\pm}] = i\partial_x \Psi_{\pm}\,.
\end{equation}
The relation
\begin{equation}
    \label{eq:FO5}
    e^{-i\Delta \hat{P}} \Psi_{\pm}(t,x) e^{i\Delta \hat{P}}
    = \Psi_{\pm}(t,x+\Delta) = \Psi_{\pm}(t-\Delta,x)\,,
\end{equation}
expresses a translation of the field operators from $x$ to $x+\Delta$.
(For $\Delta=\eps$ one recovers for the Majorana--Weyl automaton $\exp(-i\eps\hat{P})=\Shat$.)
The field operators obey the continuous wave equation
\begin{equation}
    \label{eq:FO6}
    (\partial_t + \partial_x) \Psi_{\pm} (t,x) = 0\,.
\end{equation}
One may also use the continuous operator expression for $\hat{P}$ in eq.~\eqref{eq:FO7} for showing eq.~\eqref{eq:FO4} by use of the identities for the interpolation function.

\indent With eq.~\eqref{eq:ZA3} one finds for the commutator with the charge operator
\begin{equation}
    \label{eq:FO11}
    \left[Q',\Psi_{\pm}(t,x) \right] = \mp \Psi_{\pm}(t,x)
\end{equation}
or
\begin{equation}
    \label{eq:FO12}
    Q' \Psi_{\pm}(t,x) = \Psi_{\pm}(t,x)(Q'\mp 1)\,.
\end{equation}
For the similar relation,
\begin{equation}
    \label{eq:FO13}
    Q \Psi_+(t,x) = \Psi_+(t,x)(Q-1) + \frac{1}{\sqrt{L}}a(k=0)\,,
\end{equation}
one can neglect the last term for $L \to \infty$.
For a vacuum with $Q\ket{0}=0$ this limit implies
\begin{equation}
    \label{eq:FO14}
    Q\Psi_+(t,x)\ket{0} = -\Psi_+(t,x)\ket{0}\,.
\end{equation}
Thus a non-zero $\Psi_+(t,x)\ket{0}$ is an eigenstate of $Q$ with eigenvalue $-1$.
The operation of $\Psi_+^\dagger(t,x)$ on the vacuum creates a particle with charge $Q=1$,
\begin{equation}
    \label{eq:FO15}
    Q\Psi_+^\dagger(t,x)\ket{0} = \Psi_+^\dagger(t,x)\ket{0}\,.
\end{equation}

\subsection*{Heisenberg operators}
We want to show that the continuous field operators \eqref{eq:FO1} are the Heisenberg operators associated to the operators  \eqref{eq:225A} at a reference time $t=0$.
As familiar in quantum mechanics this permits the computation of expectation values of observables at arbitrary $t$ from the wave function at $t=0$.
The probabilistic information contained in the wave function at $t=0$ is therefore sufficient for finding expectation values at other times.
The Heisenberg operators are typically off-diagonal even for operators which are diagonal at $t=0$.
Heisenberg operators at different times $t_1$ and $t_2$ do typically not commute.
They provide for a simple example of the use of non-commuting operators for observables in a classical statistical setting \cite{CWPW2020}.

\indent For a time-independent operator $\hat{B}$ and Hamiltonian $H$ the time-dependent Heisenberg operator is defined by
\begin{equation}
    \label{eq:HO1}
    \hat{B}_H(t) = \exp(i H t) \hat{B} \exp(-i H t)\,.
\end{equation}
It permits to compute the expectation value of $\hat{B}$ for arbitrary time from the wave function at $t=0$,
\begin{equation}
    \label{eq:HO2}
    \langle B(t) \rangle = \langle \varphi^\dagger(t) \hat{B} \varphi(t)\rangle
    = \langle \varphi^\dagger(0) \hat{B}_H(t) \varphi(0)\rangle\,.
\end{equation}
The field operators \eqref{eq:FO1} are the Heisenberg operators corresponding to the time-independent interpolating annihilation and creation operators $\Psi_{\pm}(x) = \Psi_{\pm} (t=0,x)$ i.e.
\begin{equation}
    \label{eq:HO3}
    \Psi_{\pm}(t,x) = \exp(i H t) \Psi_{\pm}(x) \exp(-i H t)\,.
\end{equation}

\indent In order to show this we may first consider discrete time points $t=m_t \eps$, $m_t$ integer, where $\exp(-i H t) = \Shat^{m_t}$.
Eq.~\eqref{eq:H9} yields for the Heisenberg operator associated to $\Psi_+(x)$
\begin{equation}
    \label{eq:HO4}
        \begin{aligned}
            \Psi_{+}(t,x) &= \Shat^{-m_t} \sum_k \frac{1}{\sqrt{L}} \exp\left(\frac{2\pi i k x}{L}\right) a(k) \Shat^{m_t} \\
          &= \sum_k \frac{1}{\sqrt{L}} \exp\left(\frac{2\pi i k (x-m_t \eps)}{L}\right) a(k)\,,
        \end{aligned}
\end{equation}
which coincides with eq.~\eqref{eq:FO1}.
The analogue argument holds for $\Psi_-$.

\indent For continuous $t$ we extend the relation \eqref{eq:H9} to
\begin{align}
    \label{eq:HO5}
    \exp\left(-i H^{(R)}(t)\right) &a(k) = \nn\\
    &\exp\left\{\frac{2\pi i k t}{L}\right\} a(k) \exp(-i H^{(R)}t)\,,\nn\\
    \exp\left(-i H^{(R)}(t)\right) &a^\dagger(k) = \nn\\
    &\exp\left\{-\frac{2\pi i k t}{L}\right\}a^\dagger(k) \exp(-i H^{(R)}t)\,.
\end{align}
This implies eq.~\eqref{eq:HO3} for arbitrary $t$ and $x$.
The relation \eqref{eq:HO5} follows from eq.~\eqref{eq:H7} using
\begin{equation}
    \label{eq:HO6}
    [\hat{n}(k'),a(k)] = -a(k) \delta_{k',k}\,,\quad
    [\hat{n}(k'),a^\dagger(k)] = a^\dagger(k) \delta_{k',k}\,.
\end{equation}
This concludes the proof that the continuous field operators \eqref{eq:FO1} are the Heisenberg operators for the interpolating field operators at $t=0$.

\subsection*{Continuous one-particle states}
The field operators can be employed for the definition of one-particle states that depend continuously on $t$ and $x$.
We extend eq.~\eqref{eq:OPS12} at $t=0$ to
\begin{equation}
    \label{eq:HO7}
    \varphi_p^{(1)}(0) = \int \mathrm{d}x\, \tilde{\varphi}_p(x) \Psi_+^\dagger(x) \ket{0}\,.
\end{equation}
With
\begin{equation}
    \label{eq:HO8}
    \tilde{\varphi}_p(x) = \frac{1}{\sqrt{L}} \sum_{k>0} \exp \left( \frac{2\pi i k x}{L}\right) \tilde{\varphi}(k)\,,
\end{equation}
one has
\begin{equation}
    \label{eq:HO9}
    \varphi^{(1)}_p(0) = \sum_{k>0} \tilde{\varphi}(k) a^\dagger(k) \ket{0}\,,
\end{equation}
and
\begin{equation}
    \label{eq:HO10}
    \varphi^{(1)\dagger}_p(0) \varphi^{(1)}_p(0) = \sum_{k>0} \tilde{\varphi}^*(k) \tilde{\varphi}(k) = \int \mathrm{d}x\, \tilde{\varphi}_p^*(x) \tilde{\varphi}_p(x)\,.
\end{equation}

\indent For the evolution of the wave function with time in the Schrödinger picture we employ
\begin{equation}
    \label{eq:HO11}
    \begin{aligned}
        \varphi^{(1)}_p(t) &= \exp(-i H t) \varphi^{(1)}_p(0) \\
        &= \exp(-2 i E_0 t) \int \mathrm{d}x\, \tilde{\varphi}_p(x) \Psi_+^\dagger(-t,x)\ket{0}\,.
    \end{aligned}
\end{equation}
We can therefore construct the continuous one-particle wave function in the Schrödinger picture by applying a suitable time-dependent field operator to the vacuum.
The time-argument $-t$ of the field operator in eq.~\eqref{eq:HO11} reflects (in an abstract notation without space and internal \qq{indices}) the relation
\begin{equation}
    \label{eq:272A}
    \begin{aligned}
        \varphi(t) &= e^{-i H t}\varphi(0)
    = e^{-i H t} \Psi^\dagger(0) \ket{0} \\
                   &= \Psi^\dagger(-t) e^{-i H t} \ket{0}
                   = e^{-2i E_0 t} \Psi^\dagger(-t) \ket{0}\,.
\end{aligned}
\end{equation}
(For the Majorana--Weyl automaton one replaces $2E_0 \to E_0$.)

\indent One may factor out the global phase rotation of the vacuum, 
\begin{equation}
    \label{eq:272B}
    \hat{\varphi}_p^{(1)}(t) = \exp(2 i E_0 t) \varphi^{(1)}_p(t)\,.
\end{equation}
With eq.~\eqref{eq:FO4} the momentum operator $\hat{P}$ acts as
\begin{equation}
    \begin{aligned}
        \label{eq:FO17}
        \hat{P}\hat{\varphi}_p^{(1)}(t) = 
        \int_x \tilde{\varphi}_p(x) \Big\{(i\partial_x) &\Psi_+^\dagger(-t,x)\ket{0}\\+ &\Psi_+^\dagger(-t,x)\hat{P}\ket{0}\Big\}\,,
    \end{aligned}
\end{equation}
or, with $\hat{P}\ket{0}=P_0\ket{0}$,
\begin{equation}
    \label{eq:FO18}
    (\hat{P}-P_0)\hat{\varphi}_p^{(1)}(t) = \int_x \gl -i\partial_x \tilde{\varphi}_p(x)\gr \Psi_+^\dagger(-t,x)\ket{0}\,.
\end{equation}
For fixed basis functions $\Psi_+(-t,x)\ket{0}$ its action on the wave function $\tilde{\varphi}_p(x)$ is given by the usual continuum operator $\hat{P} = -i\partial_x$. 

\indent In the Schrödinger picture we define the one-particle wave function $\tilde{\varphi}_p(t,x)$ for continuous $t$ and $x$ by 
\begin{equation}
    \label{eq:HO12}
    \varphi^{(1)}_p(t) = \int \mathrm{d}x\, \tilde{\varphi}_p(t,x) \Psi_+^\dagger(x)\ket{0}\,.
\end{equation}
This employs fixed basis functions $\Psi_+^\dagger(x)\ket{0}$.
For the wave functions $\tilde{\varphi}_p(t,x)$ in this basis the evolution equation is the continuous equation 
\begin{equation}
    \label{eq:HO13}
    i \partial_t \tilde{\varphi}_p(t,x) = -i \partial_x \tilde{\varphi}_p(t,x)\,.
\end{equation}
\indent The discussion of the antiparticle part $\tilde{\varphi}_a(t,x)$ is analogous, now applying $\Psi_-^\dagger$ to the vacuum state.
Similar results obtain for the left-movers.
As a result, the continuous wave function $\tilde{\varphi}^{(1)}(t,x)$ obeys the Dirac equation \eqref{eq:OPS5}. For the continuous wave function no continuum limit has to be taken in order to obtain the continuous Dirac equation.
The discreteness of the formulation appears only in the non-canonical commutation relation \eqref{eq:FO9}.
This becomes canonical in the continuum limit where $\tilde{\delta}(y) \to \delta(y)$. 
We also observe the difference in the normalization of the discrete and continuous one-particle wave function $\tilde{\varphi}^{(1)}(t,j)$ or $\tilde{\varphi}^{(1)}(t,x)$ by a factor $\sqrt{\eps}$,
\begin{align}
    \sum_j \tilde{\varphi}^{(1)\dagger}(t,j) \tilde{\varphi}^{(1)}(t,j) &= 1\,,\nn\\
    \int \mathrm{d}x\, \tilde{\varphi}^{(1)\dagger}(t,x) \tilde{\varphi}^{(1)}(t,x) &= 1\,.
\end{align}
If we associate the dimension of length to $\eps$, the continuum wave function $\tilde{\varphi}^{(1)}(t,x)$ has dimension $(\text{length})^{-1/2}$, in line with $\int \mathrm{d}x \hateq \eps \sum_j$.

\indent Multiparticle states obtain by applying products of field operators to the particle physics vacuum.
For example, a two-particle state is given by
\begin{equation}
    \label{eq:w277A}
    \varphi^{(2)}_p(t) = \int_{x,y} \tilde{\varphi}_p^{(2)}(t,x,y) \Psi_+^\dagger(x) \Psi_+^\dagger(y) \ket{0}\,.
\end{equation}
As appropriate for fermions the two-particle wave function $\tilde{\varphi}_p^{(2)}(t,x,y)$ is antisymmetric in $(x,y)$ by virtue of the commutation relation \eqref{eq:FO9}.

\subsection*{Correlation functions}
\indent Field operators are convenient for the formulation of correlation functions.
We employ for the particle physics vacuum
\begin{equation}
    \label{eq:CF1}
    \Bra{0} \Psi_{\pm}(t,x)\Psi_{\pm}(t',x')\ket{0}_p = 0 \,,
\end{equation}
and
\begin{equation}
    \label{eq:CF2}
    \Bra{0} \Psi_{\pm}(t,x)\Psi_{\pm}^\dagger(t',x')\ket{0}_p = \tilde{\delta}_+(x-x'-t+t') + \frac{1}{2L}\,.
\end{equation}
The function $\tilde{\delta}_+(y)$ is defined by
\begin{equation}
    \label{eq:281A}
    \tilde{\delta}_+ (y) = \frac{1}{L}\sum_{k>0} \exp\left\{\frac{2\pi i k y}{L}\right\}\,,
\end{equation}
and obeys
\begin{equation}
    \label{eq:CF3}
    \tilde{\delta}_+^{*}(y) + \tilde{\delta}_+(y) = \tilde{\delta}(y)-  \frac{1}{L}\,.
\end{equation}

\indent In view of eq.~\eqref{eq:OPS6} we subtract for the Whiteman function the constant \qq{vacuum piece} which vanishes for $L\to\infty$,
\begin{equation}
    \label{eq:CF4}
    W_+(t,x) = \Bra{0} \Psi_+(t,x)\Psi_+^\dagger(0,0)\ket{0}_p - \frac{1}{2L}\
    = \tilde{\delta}_+(x-t)\,.
\end{equation}
According to eq.~\eqref{eq:281A} its Fourier transform has support only for positive momenta $p>0$
\begin{equation}
    \label{eq:CF5}
    W_+(p) = \int_x e^{-i p x} W_+(0,x) = \frac{2\pi}{L} \sum_{k>0} \delta\left(p-\frac{2\pi k}{L}\right)\,.
\end{equation}
The Whiteman function for discrete field operators obtains by evaluating eq.~\eqref{eq:CF4} for $t$ and $x$ on the sites of the lattice.
(The interpolating continuous field operators involve no smoothing by integrating out momenta within the Brillouin zone.
They interpolate between the discrete lattice sites, rather than averaging over lattices sites.)

\indent The continuum limit of the Whiteman function corresponds to the limit $\eps \to 0$, $L \to \infty$. 
One finds, with $\tilde{\eps} = \eps/\pi > 0$,
\begin{equation}
    \begin{aligned}
        \label{eq:CF6}
        \lim_{\eps \to 0} \lim_{L \to \infty} \tilde{\delta}_+(x-t)
        &= \delta_+(x-t) \\
        &= \frac{1}{2\pi} \int_{0}^\infty \mathrm{d}p\, e^{i p (x-t)} \\
        &= - \frac{1}{2\pi i} \lim_{\tilde{\eps} \to 0} \left( \frac{1}{x-t+i\tilde{\eps}} \right)\,.
    \end{aligned}
\end{equation}
This is the standard continuum expression for a massless free Weyl fermion in two dimensions.

\indent The precise meaning of the continuum limit \eqref{eq:CF6} needs some specification.
For non-zero $\eps$ and $L \to \infty$ the upper bound in the $p$-integral is $\pi/\eps$.
This yields \cite{LUE}
\begin{equation}
    \label{eq:CDE1}
    \begin{aligned}
    \lim_{L\to \infty} \tilde{\delta}_+(y) &= \frac{1}{2\pi} \int_0^{\pi/\eps}\mathrm{d}p\, e^{ipy} \\
                                           &= \frac{1}{2\pi i y}\left[\exp\left(\frac{i\pi y}{\eps}\right) - 1\right]\,.
\end{aligned}
\end{equation}
One may also obtain this expression by the limit $L\to \infty$ of the geometric series
\begin{equation}
    \label{eq:CDE2}
    \begin{aligned}
        \tilde{\delta}_+(y) &= \frac{1}{L} \sum_{k=1}^{(N_x-1)/2} \left[\exp\left(\frac{2\pi i y}{L}\right)\right]^k \\ 
                            &= \frac{1}{L} \left[\exp\left(-\frac{2\pi i y}{L}\right) - 1\right]^{-1} \\
                            &\quad\; \times\left[1-\exp\left(i\pi y\left(\frac{1}{\eps} - \frac{1}{L}\right)\right)\right]\,.
    \end{aligned}
\end{equation}
For $y=x-t$ and $x$, $t$ on the lattice sites one has $y=n\eps$.
One finds for integer $n\neq 0$
\begin{equation}
    \label{eq:CDE3}
    \tilde{\delta}_+(n\eps) = \begin{cases}
        0 & \text{for $n$ even} \\
        - \left(i \pi \eps n\right)^{-1} & \text{for $n$ odd}
        \end{cases}\,,
\end{equation}
and $\tilde{\delta}_+(0) = 1/(2\eps)$.
If one averages over $n$ and $n+1$ (for odd positive $n$) one obtains the continuum expression \eqref{eq:CF6}.
(For odd negative $n$ one averages over $n$ and $n-1$.)

\indent For the hermitian conjugate operator one has
\begin{equation}
    \begin{aligned}
        \label{eq:CF7}
        W_-(t,x) 
        &= \Bra{0} \Psi_+(0,0) \Psi_+^\dagger(t,x)\ket{0}_p - \frac{1}{2L} \\
        &= \Bra{0} \Psi_+(t,x) \Psi_+^\dagger(0,0)\ket{0}_p - \frac{1}{2L} \\
        &= W_+^{*}(t,x)
        = \tilde{\delta}_-(x-t)\,.
    \end{aligned}
\end{equation}
This yields the continuum limit
\begin{equation}
    \begin{aligned}
        \label{eq:CF8}
        \lim_{\eps\to 0} \lim_{L \to \infty} W_-(t,x) 
        &= \delta_-(x-t)
        = \delta_+^{*}(x-t)\\
        &= \frac{1}{2\pi i} \lim_{\tilde{\eps} \to 0} \left( \frac{1}{x-t-i\tilde{\eps}} \right)\,.
    \end{aligned}
\end{equation}
We observe the characteristic changes of signs between eqs.~\eqref{eq:CF8} and \eqref{eq:CF6}.

\subsection*{Feynman propagator}
\indent With a minus sign in the time ordering for fermions the Feynman propagator is given by the time-ordered correlation function
\begin{equation}
    \label{eq:CF9}
    G_F(t,x;t',x') = \begin{cases}
        W_+(t-t',x-x') & \text{for} \quad t > t' \\
        -W_-(t-t',x-x') & \text{for} \quad t < t'
    \end{cases}\,.
\end{equation}
By virtue of the continuous definition of the field operators the correlations $W_{\pm}$ and $G_F$ obey (for fixed $t'$, $x'$) the continuous evolution equation for right-moving Weyl fermions
\begin{equation}
    \label{eq:CF10}
    (\partial_t + \partial_x) W_{\pm} = 0\,, \quad
    (\partial_t + \partial_x) G_F = 0\,.
\end{equation}
This holds for the discrete setting without taking the continuum limit $\eps\to 0$.

\indent A similar discussion can be performed for the left-movers, replacing $t-x$ by $t+x$.
The two-component Feynman propagator obeys the continuous Dirac equation
\begin{equation}
    \label{eq:CF11}
    \gamma^\mu \partial_\mu \begin{pmatrix}
        G_F^{(R)} \\
        G_F^{(L)}
    \end{pmatrix} = 0\,.
\end{equation}
This makes the Lorentz symmetry for the formulation with continuous field operators manifest.
The discreteness of the original formulation resides only in the non-standard anticommutation relation \eqref{eq:FO9}, reflected by the difference of $\tilde{\delta}_\pm(y)$ from the continuum limit $\delta_\pm(y)$.

\indent Acting with the Feynman propagator on any smooth function the continuum limit \eqref{eq:CF6} is valid.
In this limit the Feynman propagator takes the standard form for free massless Weyl or Dirac fermions.
One-particle states which are energy eigenstates have a relative energy $E=2\pi |k| / (\eps N_x)$ as compared to the vacuum.
With maximal values $|k| = (N_x-1)/2$ the only states with small $E$ correspond to small $|k|$.
For eigenstates with small $|k|$ the plane wave function is smooth and the continuum limit \eqref{eq:CF6} applies.
One may associate the continuum limit with an effective theory for low energies.
Wave functions with low expectation value of the relative energy are sufficiently smooth for the continuum limit to apply.
In particular, there are no one-particle wave functions with small energies and strong oscillations in space similar to the staggered fermions in lattice theories.
For the chiral Weyl fermions the Feynman propagator shows no direct sign of lattice doublers similar to the one which would occur if we replace in the Hamiltonian \eqref{eq:H7} $k$ by $\frac{N_x}{2\pi}\sin\left(\frac{2\pi k}{N_x}\right)$ or similar.
This may be a way to avoid the constraints of ref.~\cite{NINI1,NINI2,FRI}.

\indent We may transform the Feynman propagator to energy-momentum space by taking a discrete Fourier transform over the space-volume $L$ and a time interval $\Delta t$, which we take to be a multiple of $L$.
For the right-movers this results in
\begin{align}
    \label{eq:GF1}
        G_F(E,p;E'\!,p') &= \!\int_t\! \int_{t'}\!\! \int_x\! \int_{x'}\!\! e^{i \left(E t - E'\!t'\! - px + p'\!x'\!\right)} G(t,x;t'\!,x') \nn\\
                       &= \frac{i L \Delta t}{E - p} \hat{\delta} (p-p') \hat{\delta}(E - E')\,,
\end{align}
with periodic discrete $\delta$-functions.
In the continuum limit one has
\begin{equation}
    \label{eq:GF2}
    \hat{\delta}(p) = \frac{2\pi}{L}\delta(p)\,,\quad
    \hat{\delta}(E) = \frac{2\pi}{\Delta t} \delta(E)\,,
\end{equation}
leading to
\begin{equation}
    \label{eq:GF3}
    G_F(E,p;E',p') = \frac{i}{E-p} 2\pi \delta(p-p') 2\pi \delta(E-E')\,.
\end{equation}
This is the standard Feynman propagator for a free massless Weyl fermion without doublers.
Including the left movers one obtains 
\begin{equation}
    \label{eq:GF4}
    G_F = i \frac{(E+p\tau_3)}{E^2 - p^2} = i\gamma^0 \frac{\gamma^\mu p_\mu}{p^\mu p_\mu} = i \gamma^0 (\gamma^\mu p_\mu)^{-1}\,,
\end{equation}
with $\gamma^0 = -i\tau_2$, $\gamma^1 = \tau_1$, $p_0 = E$, $p_1 = p$, $p^\mu p_\mu = p^2 - E^2$.
The factor $\gamma^0$ may be absorbed by defining the propagator for the two-point function involving $\bar{\psi} = \psi^\dagger \gamma^0$ instead of $\psi^\dagger$.
One finds the standard propagator for a single massless Dirac fermion with the characteristic Lorentz-covariance of the operator $\gamma^\mu p_\mu$.

\subsection*{Density correlation function}
\indent The density correlation function is defined by 
\begin{equation}
    \label{eq:GF5}
    G_n = \left\langle n(t,x) n(0,0)\right\rangle - \left\langle n(t,x) \right\rangle \left\langle n(0,0)\right\rangle\,,
\end{equation}
with
\begin{equation}
    \label{eq:GF6}
    n(t,x) = \Psi^\dagger(t,x) \Psi(t,x)\,,\quad 
    \left\langle n(t,x)\right\rangle = \frac{1}{2\eps}\,.
\end{equation}
One finds for the particle physics vacuum
\begin{equation}
    \label{eq:GF7}
    G_n = \left(\tilde{\delta}_+ (x-t) + \frac{1}{2L}\right)^2\,.
\end{equation}
In the limit $L\to \infty$ this results for $x \neq t$ in 
\begin{equation}
    \label{eq:GF8}
    \begin{aligned}
        G_n = -\frac{1}{4\pi^2 (x-t)^2} \bigg[&1-2\exp\left(\frac{\pi i(x-t)}{\eps}\right) \\
                                              &+ \exp \left(\frac{2\pi i (x-t)}{\eps}\right)\bigg]\,.
    \end{aligned}
\end{equation}
The oscillating terms drop out if one averages the continuous function in $t$ or $x$ over an interval $2\eps$, leading to the standard result.
Evaluating $G_n$ on lattice sites $x-t = n\eps$ leads to 
\begin{equation}
    \label{eq:GF9}
    G_n = -\frac{2}{4\pi^2 n^2 \eps^2} \big(1-(-1)^n\big)\,.
\end{equation}
Taking the discrete average over two neighboring sites with $n$ and $n+1$ yields a factor two as compared to the continuous average.
This reflects that for rapidly oscillating functions the average can depend on the precise way of averaging.
The leading term for $x-t\to 0$ can be found from 
\begin{equation}
    \label{eq:GF10}
    G_n = -\frac{1}{4\pi^2(x-t+i \tilde\eps)^2}\,,\quad
    \tilde\eps = \frac{\eps}{\pi} \to 0\,,
\end{equation}
with $G_n(x=t) = 1/(4\eps^2)$.

\subsection*{Continuous symmetries}
The Hamiltonian for the Dirac automaton extends eq.~\eqref{eq:FO7},
\begin{equation}
    \label{eq:SYM1}
    H = \int\mathrm{d}x\, \Psi^\dagger(-i\partial_x) \tau_3 \Psi\,,
\end{equation}
where
\begin{equation}
    \label{eq:SYM2}
    \Psi = \begin{pmatrix}
        \Psi_{R+}(t,x) \\
        \Psi_{L-}(t,x)
        \end{pmatrix}\,.
\end{equation}
This form is convenient for a discussion of the continuous symmetries of the Dirac automaton.
Those symmetries are not easily visible on the level of the action for the Ising spins.

\indent A first global continuous symmetry are global phase rotations
\begin{equation}
    \label{eq:SYM3}
    \Psi \to e^{i\alpha} \Psi \,,\quad
    \Psi^\dagger \to e^{-i\alpha} \Psi^\dagger\,.
\end{equation}
As usual, this symmetry is related to charge conservation, in our case the charge $Q$.
Phase rotations can be performed independently for $\Psi_R$ and $\Psi_L$, resulting in the chiral symmetry
\begin{equation}
    \Psi_R \to e^{i\alpha_R} \Psi_R\,,\quad
    \Psi_L \to e^{i\alpha_L} \Psi_L\,.
\end{equation}
This corresponds to separate conserved charges $Q_R$ and $Q_L$ for the right-movers and left-movers.
The chiral symmetry forbids a mass term for the fermions unless it is broken explicitly or spontaneously by additional interactions, see sect.~\ref{sec:interactions}.

\indent The Hamiltonian \eqref{eq:SYM1} is invariant under continuous translations in space 
\begin{equation}
    \label{eq:SYM5}
    \Psi(t,x) \to \Psi(t, x+a)
\end{equation}
The presence of this continuous symmetry is linked to the possibility of choosing a continuous family of field operators.
Furthermore, the Hamiltonian \eqref{eq:SYM1} generates an evolution for the continuous one- (and multi-) particle wave functions which is invariant under continuous Lorentz transformations.
Without going into details, this is visible by the observation that eq.~\eqref{eq:SYM1} denotes precisely the Hamiltonian of a Lorentz-invariant continuum theory for the massless Dirac fermions.
All these symmetries are realized also for Weyl fermions by restricting the model to $\Psi_R$ or $\Psi_L$.

\indent The presence of continuous symmetries for generalized Ising models reflects a redundancy in the choice of fermionic operators or field operators.
Different choices of these operators can lead to the same step evolution operator.

\section{Density matrix and thermal equilibrium}
\label{sec:Density_matrix_thermal_equilibrium}

\indent The general probabilistic setting for quantum mechanics is based on the density matrix.
This concept extends to classical statistics.

\subsection*{Classical density matrix}
\indent In the real formulation we can construct from the wave functions of generalized Ising models the real pure state density matrix
\begin{equation}
    \label{eq:DM1}
    \rho'_{\tau \rho}(t) = \tilde{q}_\tau(t)\bar{q}_\rho(t)\,.
\end{equation}
It incorporates information from the \qq{initial boundary} at $t_{\text{in}}$ via $\tilde{q}(t)$, as well as information from the \qq{final boundary} at $t_f$ in terms of $\bar{q}(t)$.
These wave functions obtain from $\tilde{q}(t_{\text{in}})$ and $\bar{q}(t_f)$ by use of the evolution equation \eqref{eq:G13}.
The evolution of the density matrix obeys
\begin{equation}
    \label{eq:DM2}
    \rho'(t+\eps) = \Shat(t) \rho'(t) \Shat^{-1}(t)\,.
\end{equation}
Similar to quantum mechanics we can generalize to a mixed state density matrix as a weighted sum over pure state density matrices $\rho^{\prime (\alpha)}(t)$ given by eq.~\eqref{eq:DM1} for wave functions $\tilde{q}^{(\alpha)}(t)$, $\bar{q}^{(\alpha)}(t)$,
\begin{equation}
    \label{eq:DM3}
    \rho'(t) = \sum_\alpha w_{\alpha} \rho^{\prime (\alpha)}(t)\,,\quad
    w_{\alpha} \geq 0\,,\quad \sum_\alpha w_{\alpha} = 1\,.
\end{equation}
The evolution equation \eqref{eq:DM2} extends to mixed state density matrices.
With $\tr\left\{\rho^{\prime(\alpha)}(t)\right\} = \sum_\tau \rho^{\prime(\alpha)}_{\tau\tau}(t) = \sum_{\tau} p^{(\alpha)}_{\tau}(t) = 1$ one has
\begin{equation}
    \label{eq:DM4}
    \tr\left\{\rho'(t)\right\} = 1\,.
\end{equation}
Expectation values of time-local observables are given by
\begin{equation}
    \label{eq:DM5}
    \langle \hat{A}(t)\rangle = \tr\big\{\hat{A}\,\rho'(t)\big\}\,.
\end{equation}
This \qq{quantum rule} follows directly from eq.~\eqref{eq:OP1}.

\indent Let us consider a time independent step evolution operator.
(This may hold in a \qq{coarse grained} sense of combining several time  steps.)
The density matrix is time-independent if it commutes with the step evolution operator.
This is realized if $\rho'$ is a function of an operator $\hat{L}$ which commutes with $\Shat$,
\begin{equation}
    \label{eq:DM6}
    \rho' = f(\hat{L})\,,\quad [\Shat,\hat{L}] = 0\,.
\end{equation}
For arbitrary density matrices of this type the expectation values of all powers $\langle \hat{L}^n \rangle$ are time-independent.
As a result, if a \qq{conserved observable} represented by $\hat{L}$ exists, one may have obstructions for $\rho'(t)$ to approach a universal equilibrium state due to the infinite number of conserved quantities.

\subsection*{Quantum density matrix for PCA}
\indent Let us focus on probabilistic cellular automata and extend the discussion to the complex picture, where the complex density matrix is defined by
\begin{equation}
    \label{eq:296A}
    \rho_{\tau\rho}(t) = \sum_\alpha w_\alpha \varphi_\tau^{(\alpha)}(t)\varphi_\rho^{(\alpha)*}(t)\,,\quad \rho^\dagger(t) = \rho(t)\,,
\end{equation}
and obeys
\begin{equation}
    \rho(t+\eps) = \exp\left(-i\eps H(t)\right)\rho(t)\exp\left(i\eps H(t)\right)\,.
\end{equation}
The density matrix $\rho$ is a positive matrix (all eigenvalues positive or zero).
In particular for any (normalized) function $\rho = f(H)$ the density matrix is conserved.
There are infinitely many conserved quantities $H^n$ which prevent the approach to a unique equilibrium state for $t-t_{\text{in}}\to\infty$ \cite{ABC1, ABC2}.
This extends to other conserved quantities as the charge $Q$ or the total number of occupied bits $N_E$.

\subsection*{Thermal equilibrium state}
The thermal equilibrium state for a PCA is given by the density matrix
\begin{equation}
    \rho = Z_T^{-1} \exp\left(-\frac{H}{T}\right)\,,\quad
    Z_T = \tr\exp\left(-\frac{H}{T}\right)\,.
\end{equation}
Here $T$ is the temperature in energy units.
The thermal equilibrium state is static.
Let us compute the expectation values for the occupation numbers of momentum modes for the Majorana--Weyl automaton for right-movers, $p=2\pi k/L$,
\begin{equation}
    \begin{aligned}
        \langle \hat{n}_R(p)\rangle & = \tr\left\{\hat{n}(p)\rho\right\}                                                      \\
                                    & = Z_T^{-1}\tr\left\{\hat{n}(p)\exp\bigg(-\sum_{p'}\frac{p'\hat{n}(p')}{T}\bigg)\right\} \\
                                    & = \left[ \exp\left(\frac{p}{T}\right)+1\right]^{-1}\,.
    \end{aligned}
\end{equation}
The thermal equilibrium state for this automaton realizes the Fermi--Dirac distribution.
At first sight one may not have suspected that there exist boundary conditions for a PCA which correspond to a thermal equilibrium state with the momentum distribution characteristic for quantum fermions.
In view to the equivalence with a quantum field theory for fermions this is no longer a surprise.

\indent One observes the limits
\begin{align}
    \label{eq:DM10}
    \lim_{T\to\infty} \langle \hat{n}_R(p)\rangle & = \frac{1}{2}\,,\nn \\
    \lim_{T\to 0} \langle \hat{n}_R(p)\rangle     & = 1 - \theta(p)\,.
\end{align}
In the limit $T\to 0$ the thermal equilibrium state approaches the particle physics vacuum with all modes with $p<0$ occupied and all modes with $p>0$ empty.
The complex density matrix for the particle physics vacuum is time-independent.
Indeed, the overall phase of the wave function $\exp(-i E_0 t)$ drops out for the density matrix.
The complex density matrix for the Majorana--Weyl vacuum is also invariant under space translations by $\eps$.

\indent The thermal equilibrium state is half-filled in the sense
\begin{equation}
    \label{eq:DM11}
    \langle\hat{N}(j)\rangle = \frac12\,,\quad
    \langle Q(j)\rangle = 0\,.
\end{equation}
The proof is straightforward: Since $H$ and therefore $\rho$ are diagonal in momentum space one has
\begin{equation}
    \label{eq:DM12}
    \langle a^\dagger(k) a(k')\rangle = 0\quad \text{for}\quad k\neq k'\,.
\end{equation}
This implies
\begin{equation}
    \label{eq:DM13}
    \begin{aligned}
        \langle \hat{N}(j)\rangle & = \langle a^\dagger(j) a(j)\rangle                                                                          \\
                                  & = \sum_{k,k'} \frac{1}{N_x}\exp\left\{-\frac{2\pi i}{N_x}(k-k') j\right\} \langle a^\dagger(k) a(k')\rangle \\
                                  & = \frac{1}{N_x}\sum_k \langle n_R(k)\rangle = \frac{\langle N \rangle}{N_x}\,.
    \end{aligned}
\end{equation}
We employ
\begin{equation}
    \label{eq:DM14}
    \langle n_R(k)\rangle + \langle n_R(-k)\rangle = 1\,,
\end{equation}
in order to establish $\langle N\rangle = N_x/2$.
For left-movers one has $\langle n_L(p)\rangle = \left[\exp\left(-\frac{p}{T}\right)+1\right]^{-1}$, and for the Dirac automaton $\langle n_R(p)\rangle$ and $\langle n_L(p)\rangle$ are the same as for the corresponding Majorana--Weyl automata.

\section{Interactions}
\label{sec:interactions}

\indent For probabilistic cellular automata we have so far only discussed rather trivial \qq{transport automata} that move spin configurations to the right or left.
The corresponding two-dimensional generalized Ising models describe discrete quantum field theories for free massless fermions in one time- and one space-dimension.
The step evolution operator for these transport automata may be denoted by $\Shat_f$, with a corresponding Hamiltonian $H_f$ for free massless fermions.
It is rather straightforward to add some types of local interactions by products of step evolution operators according to
\begin{equation}
    \label{eq:I1}
    \Shat(t) = \Shat_{\text{int}}(t) \Shat_f\,.
\end{equation}
The \qq{interaction part} $\Shat_{\text{int}}$ has to be a unique jump matrix.
Then $\Shat(t)$ is a unique jump matrix, implying a deterministic updating rule for the automaton and therefore a \qq{unitary evolution} $\Shat(t)\Shat^T(t)=1$.
Equivalently, we may consider eq.~\eqref{eq:I1} as an alternating series of step evolution operators, with $\Shat'(t)=\Shat_f$ for even $t=t_{in}+2n\eps$, and $\Shat'(t)=\Shat_{\text{int}}(t)$ for odd $t=t_{in}+(2n+1)\eps$, with integer $n=0,1,2,\dots$.
The operator $\Shat(t)$ in eq.~\eqref{eq:I1} is then defined
for $t=t_{in}+2n\eps$ and corresponds to the combination $\Shat_{\text{int}}(t+\eps)\Shat_{f}$.
Generalized Ising models describing an alternating series of step evolution operators will be discussed below.

\subsection*{Local interactions}
\indent We will focus on local interactions for which $\Shat_{\text{int}}$ can be written as a direct product of local factors.
\begin{equation}
    \label{eq:I2}
    \Shat_{\text{int}}(t) = \prod_j \Shat_i(j)\,.
\end{equation}
Such automata have a simple interpretation: after a transport step the bit configurations at every site $j$ are permuted according to the action of $\Shat_i(j)$.
One may extend this setting by allowing local regions, say the sites $j-1$, $j$ and $j+1$, with $\Shat_i(j)$ acting on the spin configurations in this region, and $\prod_j$ in eq.~\eqref{eq:I2} replaced by a product over disconnected regions.
The possible forms of $\Shat_i(j)$ may be further restricted by imposing symmetries.
In the presence of a complex structure $\Shat_i(j)$ should be compatible with this complex structure.

\indent The matrices $\Shat_i(j)$ and $\Shat_{\text{int}}$ are orthogonal matrices in a real formulation.
We can therefore write them in a Hamiltonian form
\begin{align}
    \label{eq:I3}
    \Shat_i(t,j)          & = \exp(-i\eps H_i(t,j))\,,\nn          \\
    \Shat_{\text{int}}(t) & = \exp(-i\eps H_{\text{int}}(t))\,,\nn \\
    H_{\text{int}}(t)     & = \sum_j H_i (t,j)\,.
\end{align}
Also $\Shat(t)$ in eq.~\eqref{eq:I1} is orthogonal, such that
\begin{equation}
    \label{eq:I4}
    \Shat(t) = \exp(-i\eps H(t))
    = \exp\gl -i\eps H_{\text{int}}(t)\gr \exp(-i\eps H_f)\,.
\end{equation}
Typically $H_{\text{int}}(t)$ and $H_f$ do not commute. As a consequence, the form of $H(t)$, as defined by the product in eq.~\eqref{eq:I4}, is often not easily accessible.
One only knows that a hermitian $H(t)$ exists.
One may speculate that for smooth enough wave functions both $H_f$ and $H_{\text{int}}$ induce only small changes in the wave functions and that the commutator terms $\sim\eps^2[H_{\text{int}},H_f]$ may become negligible.
This would lead to a \qq{naive continuum limit} $H(t) = H_{\text{int}}(t)+H_f$.
The question arises how long a smooth wave function remains sufficiently smooth.
Finding the true continuum limit for $H(t)$ remains a challenge.

\subsection*{Interaction Hamiltonian}
For $\Shat_i(t,j)$ obeying the property
\begin{equation}
    \label{eq:I5}
    \Shat_i^2 = 1\,,
\end{equation}
it is rather straightforward to find $H_i(t,j)$ by employing the operator identity
\begin{equation}
    \label{eq:I6}
    \hat{A}^2 = 1 \;\;\implies\;\; \hat{A} = \exp\left\lbrace\frac{i\pi}{2}(\hat{A}-1)\right\rbrace
\end{equation}
This yields
\begin{equation}
    \label{eq:I7}
    H_i(t,j) = \frac{\pi}{2\eps}(1-\Shat_i(t,j))\,.
\end{equation}
For real $\Shat_i(t,j)$ one finds that $H_i(t,j)$ is real and symmetric, since eq.~\eqref{eq:I5} implies $\Shat_i^T = \Shat_i$.
The interaction part $H_i$ takes the form of a generalized potential.
It diverges, however, for $\eps\to 0$, such that the neglect of commutator terms $\sim\eps^2[H_{\text{int}},H_f]$ is not generally justified.

\indent As a simple example we may consider two possible species of bits per site $j$.
The particular form,
\begin{equation}
    \label{eq:I8}
    \Shat_i = a_1^\dagger a_2 + a_2^\dagger a_1 \pm a_1^\dagger a_1 a_2^\dagger a_2 \pm a_1 a_1^\dagger a_2 a_2^\dagger\,,
\end{equation}
obeys the property \eqref{eq:I5}.
If we associate $a_1 = a_R(j)$ and $a_2 = a_L(j)$ the Hamiltonian $H_{\text{int}}$ involves a piece changing right-movers to left-movers and vice versa, and an \qq{interaction term} containing four fermionic operators.
If we use the same $\Shat_{\text{int}}$ for all odd $t=t_{\text{in}} +(2n+1)\eps$ the updating rule for the automaton leads to a rather simple dynamics.
All fermions change direction at odd $t$ and propagate at even $t$.
More precisely, at odd $t$ all single right-movers become left-movers due to the term $a_L^\dagger a^{\phantom{\dagger}}_R$ in $\Shat_i$, and vice versa due to $a_R^\dagger a^{\phantom{\dagger}}_L$.
The quartic term $(1-\hat{n}_R - \hat{n}_L + 2\hat{n}_R \hat{n}_L)$ equals one for equal $\hat{n}_L$ and $\hat{n}_R$, such that a pair of right- and left-movers remains unchanged, and similarly for a pair of holes.
As a result, every configuration is mapped to itself after four time steps.
More complex updating rules follow if $\Shat_i$ is replaced by $1$ at certain positions $j$ for certain odd time steps.

\indent Instead of a strictly local updating at each odd time step we can also update cells consisting of several neighboring locations.
For example, we can group the three sites $j=-1,0,1$ into a cell located at $j=0$, the sites $j=2,3,4$ into a neighboring cell at $j=3$, and so on.
This is compatible with the periodic boundary condition for $N_x=9\mod 12$.
Different cells have no common sites.
We can therefore define a quasi-local updating by defining $\Shat_{\text{int}}$ as a product of step evolution operators $\Shat_i(\tilde{j})$ for the different cells.
(The cell location is at $x=\tilde{j}\eps$, $\tilde{j}=0\mod 3$.)
Within a given cell the Majorana-Weyl automaton has three spins, and therefore eight spin configurations.
The local evolution operator is a $8\times 8$ matrix.
For the Dirac automaton the six spins in the cell admit $64$ configurations, with $\Shat_i(\tilde{j})$ a $64\times 64$ matrix.
For an automaton $\Shat_i(\tilde{j})$ has to be a unique jump matrix, describing a permutation of the $64$ spin-configurations.
A large variety of probabilistic automata or corresponding discrete quantum field theories for fermions can be constructed in this way.

\indent We may impose additional restrictions on the form of $\Shat_i(\tilde{j})$, as invariance under the discrete symmetries $C$, $P$, $T$.
If $\Shat_i(j)$ is invariant under $C$ and acts in the same way on $\tilde{q}$ and $\bar{q}'$, it is compatible with the complex structure \eqref{eq:C13}.
More precisely, one chooses signs in $\Shat_i(j)$ such that it commutes with $B$ and $T_3$.
Then $\Shat_i(j)$ acts in the complex picture as the same real unique jump matrix as in the real picture.

\subsection*{Modified propagation and particle production}
\indent As an example, we may demonstrate the rich possibilities of interactions by supplementing the Dirac automaton for the propagation by a quasi-local interaction for cells with three sites. The interaction of our example preserves the symmetries $C$, $P$ and $T$.
We denote the configurations of the six spins in a given cell by the occupation numbers $(n_{R-},n_{R0},n_{R+}|n_{L-},n_{L0},n_{L+})$, where $n_{R0}$ denotes the occupation number of right-movers at the central site of the cell $\tilde{j}$, $n_{R-}$ the one at $\tilde{j}-1$, and $n_{R+}$ the one at $\tilde{j}+1$.
The same holds for the left-movers $n_{L0}$, $n_{L-}$, $n_{L+}$.
An updating is a map $(n_{R-},n_{R0},n_{R+}|n_{L-},n_{L0},n_{L+})\to(n_{R-}',n_{R0}',n_{R+}'|n_{L-}',n_{L0}',n_{L+}')$.
The parity transformation maps
\begin{equation}
    \label{UAA}
    \begin{aligned}
        P:\,
        (n & _{R-},n_{R0},n_{R+}|n_{L-},n_{L0},n_{L+})         \\
           & \to(n_{L+},n_{L0},n_{L-}|n_{R+},n_{R0},n_{R-})\,,
    \end{aligned}
\end{equation}
and time reversal acts as
\begin{equation}
    \label{UAB}
    \begin{aligned}
        T:\,
        (n & _{R-},n_{R0},n_{R+}|n_{L-},n_{L0},n_{L+})               \\
           & \to(n'_{L-},n'_{L0},n'_{L+}|n'_{R-},n'_{R0},n'_{R+})\,.
    \end{aligned}
\end{equation}
Charge conjugation changes all $n_i$ to $1-n_i$.
For a specification of $\Shat_i(\tilde{j})$ we have to fix the transformations of all 64 configurations in a cell.
For $\Shat_i^{\,2}(j)=1$ this concerns pairs of configurations whose members are mapped to each other, or invariant configurations.

\indent Let us start with the updating rule
\begin{equation}
    \label{eq:UA}
    (1\,0\,0\,|\,1\,0\,1)\to(0\,1\,0\,|\,1\,1\,0)\,.
\end{equation}
It displaces a single right-moving particle to the right, and one out of two left-moving particles to the left.
Applying P yields
\begin{equation}
    \label{eq:UB}
    (1\,0\,1\,|\,0\,0\,1)\to(0\,1\,1\,|\,0\,1\,0)\,,
\end{equation}
while $PT$-invariance requires
\begin{equation}
    \label{eq:UC}
    (0\,1\,0\,|\,0\,1\,1)\to(0\,0\,1\,|\,1\,0\,1)\,,
\end{equation}
and the $CPT$-transformation implies
\begin{equation}
    \label{eq:UD}
    (1\,0\,1\,|\,1\,0\,0)\to(1\,1\,0\,|\,0\,1\,0)\,.
\end{equation}
Performing a $T$-transformation of the map \eqref{eq:UA} yields the inverse of the map \eqref{eq:UD}.
This implies for all four updatings that the inverse updating is also part of the rule, and $\Shat^{\,2}_{\text{int}}=1$ is realized in this sector.
For all eight updatings \eqref{eq:UA}--\eqref{eq:UD} the number of right-movers and left-movers is separately conserved.
For all four maps \eqref{eq:UA}--\eqref{eq:UD} one right-mover moves to the right and one left-mover to the left.
On the other hand, one of the three particles involved does not move.
As a result, the overall propagation is modified in the three particle sector.
For the inverse maps the directions are reversed.

\indent Next we consider the updatings
\begin{equation}
    \label{eq:UE}
    (0\,0\,1\,|\,0\,1\,1)\;\leftrightarrow\;(0\,1\,1\,|\,0\,0\,1)\,,
\end{equation}
and
\begin{equation}
    \label{eq:UF}
    (1\,1\,0\,|\,1\,0\,0)\;\leftrightarrow\;(1\,0\,0\,|\,1\,1\,0)\,.
\end{equation}
The four configurations appearing in eqs.~\eqref{eq:UE}, \eqref{eq:UF} are all invariant under the $CP$-transformation.
Indeed, eq.~\eqref{eq:UF} is the parity transform of eq.~\eqref{eq:UE}, and applying further $C$ brings back the configurations in eq.~\eqref{eq:UE}.
If a set of configurations is invariant under one of the discrete symmetries $C$, $P$, $T$ or combinations thereof, the updating has to be a permutation matrix within this set.
Otherwise the unique jump property cannot be realized.
If a $CP$-invariant configuration would be mapped to a configuration which is not invariant under $CP$, the $CP$-transform of this updating rule would yield a different configuration obtained by the updating of the same $CP$-invariant configuration.
This contradicts the unique jump property for automata.
The updatings \eqref{eq:UE}, \eqref{eq:UF} preserve the total number of particles, but not separately the numbers of right-movers and left-movers.
Right-movers turn to left-movers and vice versa.
Similar properties hold for the maps for the $CT$-invariant configurations
\begin{equation}
    \label{eq:UG}
    (1\,0\,0\,|\,0\,1\,1)\;\leftrightarrow\;(1\,1\,0\,|\,0\,0\,1)\,,
\end{equation}
and
\begin{equation}
    \label{eq:UH}
    (0\,1\,1\,|\,1\,0\,0)\;\leftrightarrow\;(0\,0\,1\,|\,1\,1\,0)\,.
\end{equation}

\indent For the configurations which are both invariant under $CP$ and $T$ we take the rule
\begin{equation}
    \label{eq:UI}
    (0\,0\,0\,|\,1\,1\,1)\;\leftrightarrow\;(1\,1\,1\,|\,0\,0\,0)\,,
\end{equation}
while the configurations
\begin{equation}
    \label{eq:UJ}
    (0\,1\,0\,|\,1\,0\,1)\,,\;(1\,0\,1\,|\,0\,1\,0)\,,
\end{equation}
which are invariant under $CP$, $PT$ and $CT$, are left unchanged by the updating.
The twenty configurations appearing in the updatings \eqref{eq:UA}--\eqref{eq:UJ} cover all configurations with three particles and three holes.
They all preserve the total number of particles.
The propagation of the particles becomes rather complex, however.

\indent Interactions that change the number of particles may be implemented in the sectors with two or four particles by updatings
\begin{align}
    \label{eq:UK}
    (0\,1\,0\,|\,0\,0\,1)\;\leftrightarrow\;(1\,0\,1\,|\,0\,1\,1)\,,\nn \\
    (1\,0\,0\,|\,0\,1\,0)\;\leftrightarrow\;(1\,1\,0\,|\,1\,0\,1)\,,\nn \\
    (0\,0\,1\,|\,0\,1\,0)\;\leftrightarrow\;(0\,1\,1\,|\,1\,0\,1)\,,\nn \\
    (0\,1\,0\,|\,1\,0\,0)\;\leftrightarrow\;(1\,0\,1\,|\,1\,1\,0)\,.
\end{align}
These maps correspond to the $CPT$-transformation.
Similar $2\leftrightarrow 4$ scatterings can be implemented in the $P$ invariant sector
\begin{align}
    \label{eq:UL}
    (1\,0\,0\,|\,0\,0\,1)\;\leftrightarrow\;(1\,1\,0\,|\,0\,1\,1)\,,\nn \\
    (0\,0\,1\,|\,1\,0\,0)\;\leftrightarrow\;(0\,1\,1\,|\,1\,1\,0)\,,
\end{align}
the $T$-invariant sector
\begin{align}
    \label{eq:UM}
    (1\,0\,0\,|\,1\,0\,0)\;\leftrightarrow\;(1\,1\,0\,|\,1\,1\,0)\,,\nn \\
    (0\,0\,1\,|\,0\,0\,1)\;\leftrightarrow\;(0\,1\,1\,|\,0\,1\,1)\,,
\end{align}
or the $PT$-invariant sector
\begin{equation}
    \label{eq:UN}
    (0\,1\,0\,|\,0\,1\,0)\;\leftrightarrow\;(1\,0\,1\,|\,1\,0\,1)\,.
\end{equation}
All other 26 configurations are left invariant by the local updating.

\indent The automaton with interactions specified by eqs.~\eqref{eq:UA}--\eqref{eq:UN} preserves the discrete symmetries $C$, $P$ and $T$.
Particle number is only conserved modulo two --- an even (odd) number of particles remains even (odd) under the scattering. The propagation is modified if three particles meet in a cell, and scattering with a change of particle number occurs if two or four particles encounter in a cell, with an equal number of right-movers and left-movers.
Due to translation invariance by steps of three lattice sites a coarse grained momentum is conserved.
Even though the Hamilton operator \eqref{eq:I4} is not known explicitly, it constitutes a conserved energy observable.
Rather rich dynamics can develop on the basis of these interactions.
In order to keep track and to cast interactions into a local form on the level of cells we may associate different colors to particles on the different sites of the cell.
The propagation term changes then colors.
We do not know what is a possible continuum limit of this model.
Only in the continuum limit can one decide if one ends with one or several species of propagating fermions.
We have not attempted to maintain Lorentz symmetry for the naive continuum limit of the interactions, our aim being here only the demonstration of the rich possibilities for PCA.

\subsection*{Generalized Ising model for interacting fermionic quantum field theory}
\indent For any given deterministic and invertible updating rule of a cellular automaton we can write down a corresponding generalized Ising model.
The corresponding PCA describes a discretized fermionic quantum field theory.
We focus on the setting of alternating free propagation at even $t$ and (quasi-)local interactions at odd $t$.
The step evolution operator $\Shat(t)$ is encoded in the local factor $\exp\left\{-\mathcal{L}(t)\right\}$, where for even $t$
\begin{align}
    \label{eq:GI1}
    \mathcal{L}(t)      & = \mathcal{L}_f[s(t+\eps),s(t)]\,,\nn               \\
    \mathcal{L}(t+\eps) & = \mathcal{L}_{\text{int}}[s(t+2\eps),s(t+\eps)]\,.
\end{align}
The action for the generalized Ising model is then given by eq.~\eqref{eq:G2}.
For the propagation part $\mathcal{L}_f$ we may have eq.~\eqref{eq:G5} for the Majorana--Weyl automaton or eq.~\eqref{eq:D1} for the Dirac automaton.

\indent As an example for the interaction part we may first look at the updating rule \eqref{eq:I8}.
It changes the local occupation numbers $\gl n_R(x), n_L(x)\gr$ for the Dirac automaton according to
\begin{align}
    \label{eq:GI2}
    (0,0)\to & \,(0,0)\,,\quad (1,1) \to (1,1)\,,\nn \\
    (0,1)\to & \,(1,0)\,,\quad (1,0) \to (0,1)\,.
\end{align}
This updating rule is realized for $\beta\to\infty$ by
\begin{equation}
    \label{eq:GI3}
    \begin{aligned}
        \mathcal{L}_{\text{int}}(t+\eps & ) =                                                            \\
        -\beta\sum_x \Big\{             & s_R(t+2\eps) s_L(t+\eps) + s_L(t+2\eps) s_R(t+\eps)            \\
        +                               & s_R(t+2\eps) s_L(t+2\eps) s_R(t+\eps) s_L(t+\eps) -3 \Big\}\,,
    \end{aligned}
\end{equation}
where all spin variables are taken at $x$.
One may check that $\mathcal{L}_{\text{int}}$ vanishes precisely for the allowed combinations of spin configurations at $t+2\eps$ and $t+\eps$, and diverges otherwise.

\indent Many features can be inferred by realizing the discrete symmetries.
Invariance under particle-hole conjugation ($C$) requires that $\mathcal{L}_{\text{int}}$ is a sum of terms containing each an even number of Ising spins.
Parity ($P$) maps $s_R(x) \leftrightarrow s_L(-x)$ for given $t+\eps$ and $t+2\eps$.
The combination with time reversal $PT$ corresponds to $s_R(t+\eps,x) \leftrightarrow s_R(t+2\eps,-x)$, $s_L(t+\eps,x) \leftrightarrow s_L(t+2\eps,-x)$.

\indent A systematic way of constructing $\mathcal{L}_{\text{int}}$ starts by drawing a list of the updatings of configurations of occupation numbers of a given cell similar to eq.~\eqref{eq:GI2}.
For each member of this list one has a term in $\mathcal{L}_{\text{int}}$ which multiplies factors $n_{\gamma}(t+\eps)$ or $(1-n_{\gamma}(t+\eps))$ according to the values of the \qq{ingoing} occupation numbers being $1$ or $0$ for species $\gamma$ and factors $n_\gamma(t+2\eps)$ or $(1-n_\gamma(t+2\eps))$ for the \qq{outgoing} bits.
The occupation numbers can be replaced by Ising spins using $n=(s+1)/2$.
For $m$ bits in a cell (which may comprise several sites) one has $2^m$ spin-configurations and therefore $2^m$ terms which all take the values $1$ or $0$ for any given configuration of spins at $t+2\eps$ and $t+\eps$.
Subtracting $2^m$ the result vanishes only if all terms take the value one --- this corresponds to the spin configurations at $t+2\eps$ and $t+\eps$ related by the updating rule.
For all other spin configurations the result is negative.
Multiplying by $-\beta$ with $\beta\to\infty$ and summing over all local cells, yields $\mathcal{L}_{\text{int}}$.

\indent One may combine $\mathcal{L}_f(t)$ and $\mathcal{L}_{\text{int}}(t)$ to
\begin{equation}
    \label{eq:GI4}
    \bar{\mathcal{L}}(t) = \sum_{s_\gamma(t+\eps)} \mathcal{L}_{\text{int}}(t+\eps) \mathcal{L}_f(t)\,.
\end{equation}
Thus $\bar{\mathcal{L}}(t)$ depends on $s_\gamma(t+2\eps)$ and $s_\gamma(t)$ and corresponds to the combined step evolution operator $\Shat(t) = \Shat_{\text{int}}(t+\eps)\Shat_f(t)$.
The sum over $s_{\gamma}(t+\eps)$ is easily performed due to $\beta\to\infty$, such that $\bar{\mathcal{L}}(t)$ obtains from $\mathcal{L}_{\text{int}}(t)$ by replacing $s_R(t+\eps,x) \to s_R(t,x-\eps)$ and $s_L(t+\eps,x) \to s_L(t,x+\eps)$.
The action is then a sum over $\bar{\mathcal{L}}(t)$ for even $t$, and the \qq{functional integral} sums only over spin configurations for Ising spins at even $t$.

\subsection*{Local interactions for several species}
Let us consider $\Shat_i$ involving only a single position $j$.
For a single species of right-movers a local interaction is not possible since a complete local operator basis is $\gl 1,a(j),a^\dagger(j),a^\dagger(j)a(j)\gr$.
For a single species of right- and left-movers the interaction term $\sim a_R^\dagger(j) a_R(j) a_L^\dagger(j) a_L(j)$ is the unique independent operator involving four fermionic annihilation or creation operators.
For two different species of right- and left-movers or two colors additional possibilities open up.
As an example we may consider \cite{CWNEW, CWPW2020}
\begin{equation}
    \label{eq:332A}
    \begin{aligned}
        \hat{H}_i(j) =\;& -\left[ a_{R1}^\dagger(j) a_{R2}(j) - a_{R2}^\dagger(j) a_{R1}(j) \right] \\
                      &\phantom{-}\times \left[a_{L1}^\dagger(j) a_{L2}(j) - a_{L2}^\dagger(j) a_{L1}(j)\right] \\
                      &+ \left[\hat{n}_{R1}(j) - \hat{n}_{R2}(j)\right] \left[ \hat{n}_{L1}(j) - \hat{n}_{L2}(j)\right]\,,
    \end{aligned}
\end{equation}
with
\begin{equation}
    \label{eq:332B}
    \Shat\subt{int} = \exp(-i\eps H\subt{int})\,,\quad
    H\subt{int} = \frac{\pi}{2\eps} \sum_j \hat{H}_i(j)\,,\quad
    \Shat\subt{int}^2 = 1\,.
\end{equation}
The updating rule of the corresponding PCA states that whenever a single right-mover and a single left-mover meet in a cell, their colors are switched, $1\leftrightarrow 2$ or $\text{red}\leftrightarrow\text{green}$.

\indent One can express $H_\text{int}$ in terms of the continuous field operators.
For this purpose we define $\mathcal{H}_i(t,x)$ by replacing for $\hat{H}_i(j)$ in eq.~\eqref{eq:332A} $a^\dagger_{R1}(j) \to \psi_{+R1}^\dagger(t,x)$, $a_{R2}(j) \to \psi_{+R2}(t,x)$ etc.
This yields
\begin{equation}
    \label{eq:LIF1}
    H_\text{int} = \frac{\pi}{2} \int_x \mathcal{H}_i(t,x)\,.
\end{equation}
In terms of the continuous field operators one finds a local interaction with a fixed coupling strength $\sim\pi/2$.
There is no explicit $t$-dependence in $\mathcal{H}_i(t,x)$, $\mathcal{H}_i(t,x) = \mathcal{H}_i(0,x)$.

\indent The Lorentz transformations can be viewed as a part acting on spacetime arguments and an \qq{internal} part acting on spinor indices.
The second part is very simple for two-dimensional Dirac fermions.
The infinitesimal \qq{internal transformation} reads $a_R(j) \to \left(1+\frac{v}{2}\right) a_R(j)$, $a_L(j) \to \left(1-\frac{v}{2}\right) a_L(j)$, with infinitesimal $v$.
Invariance of the interaction part under this internal transformation is achieved if every term in $H\subt{int}$ involves an equal number of $a_R$ or $a_R^\dagger$ and $a_L$ or $a_L^\dagger$.
This is the case for eq.~\eqref{eq:332B}.
Indeed, in the naive continuum limit the interaction \eqref{eq:332B} yields a Lorentz invariant generalized two-color Thirring model \cite{CWNEW, CWPW2020}.

\indent The construction of the particle physics vacuum for a model with interactions is much more involved than for free fermions.
One again looks for a minimum of the combined Hamiltonian with ground state energy $E_0$, such that particle excitations have all positive energies $E-E_0$.
A priori we see no argument why such a minimum of $H$ should not exist.
It is just very hard to find due to the non-vanishing commutator $[H_f,H\subt{int}]$.
It is an interesting question if for the particle physics vacuum spontaneous breaking of the chiral symmetry could occur, which would render the fermions massive.

\subsection*{Lorentz invariance} 
The continuous field operators are well suited for the definition of continuous Lorentz transformations for our discrete lattice model.
The continuous coordinates $(t,x)$ used in the definition \eqref{eq:FO1} of field operators label a continuous family of operators and associated observables.
The time coordinate $t$ defines a time-ordering within this family, with $\Psi(t_2,x)$ \qq{later} than $\Psi(t_1,y)$ if $t_2>t_1$.
Similarly, the $x$-coordinate defines a space-ordering.
We can consider the labels $(t,x)$ as a \qq{reference frame} for a family of observables.
The choice of a reference frame is not unique.
We may order the operators according to some different time variable $t'=t'(t,x)$ and similarly for $x'=x'(t,x)$.
The variables $(t',x')$ constitute a different reference frame.
A different family of field operators $\Psi'$ can be defined by replacing in eq.~\eqref{eq:FO5} the labels $(t,x)$ by $(t',x')$, with an additional multiplication by a factor $\eta$.
Using the map $(t,x)\to(t',x')$ we express $\Psi'_{R,L}$ as functions of $t$ and $x$
\begin{equation}
    \label{eq:LT1}
\Psi'_{R,L}(t,x) = \eta_{R,L} \Psi_{R,L} \left(t'(t,x), x'(t,x)\right)\,.
\end{equation}
(Here and in the following we focus on $\Psi_{+R}$ or $\Psi_{+L}$ and omit the label $+$.)

\indent Particularly interesting maps of reference frames are the Lorentz transformations
\begin{align}
    \label{eq:LT2}
    x' &= \gamma(x-vt)\,,\quad
    t' = \gamma(t-vx)\,, \nn\\
    \gamma &= \frac{1}{\sqrt{1-v^2}}\,.
\end{align}
Here $v$ is the relative velocity between two frames in units of the light velocity (which is set here to one) and we consider the limit $L\to\infty$.
The factors $\eta_R$, $\eta_L$ reflect the fact that $\Psi_{R,L}$ transform as spinors, reflecting the transformation of $a_{R,L}$.
We focus on infinitesimal transformations,
\begin{alignat}{2}
    \label{eq:LT3}
    x' &= x + \delta x\,,\quad &\delta x &= -vt\,,\nn\\
    t' &= t + \delta t\,,\quad &\delta t &= -vx\,.\\
    \Psi_{R,L}' &= \Psi_{R,L} + \delta\Psi_{R,L}\,,\quad
                & \delta\Psi_{R,L} &= \delta_s\Psi_{R,L} + \delta_{\mathcal{L}} \Psi_{R,L}\,,\nn
\end{alignat}
with \qq{spacetime part}
\begin{equation}
    \label{eq:LT4}
    \delta_s \Psi_{R,L} = \left(\delta x\partial_x + \delta t \partial_t\right)\Psi_{R,L}
    = -v\left(t \partial_x + x \partial_t\right)\Psi_{R,L}\,,
\end{equation}
and \qq{spinor part} or \qq{internal part} arising from the transformation of $a_{R,L}(j)$,
\begin{equation}
    \label{eq:LT5}
    \delta_{\mathcal{L}} \Psi_{R,L} = \delta \eta_{R,L} \Psi_{R,L}\,,\quad
    \delta \eta_R = -\delta \eta_L = \frac{1}{2}v\,.
\end{equation}

\indent The effects of Lorentz transformations are best discussed in terms of the operator valued Lagrange density, with $H = \int_x\mathcal{H}$,
\begin{equation}
    \label{eq:LT6}
    \hat{\mathcal{L}} = \mathcal{H} - i \left(\Psi_R^\dagger \partial_t \Psi_R + \Psi_L^\dagger \partial_t \Psi_L\right)\,.
\end{equation}
With the identities
\begin{align}
    \label{eq:LT7}
    \delta_s\big(\Psi_R^\dagger(\partial_t + \partial_x) \Psi_R\big) &=\\
                                                                          -v \big(t \partial_x &+ x \partial_t + 1\big) \big(\Psi_R^\dagger(\partial_t + \partial_x) \Psi_R\big)\,, \nn\\
                                                                          \delta_s\big(\Psi_L^\dagger(\partial_t - \partial_x) \Psi_L\big) &=\nn\\ 
                                                                          -v \big(t \partial_x &+ x \partial_t - 1\big) \big(\Psi_L^\dagger(\partial_t - \partial_x) \Psi_L\big)\,,\nn
\end{align}
one finds for the free Dirac automaton
\begin{equation}
    \label{eq:LT8}
    \delta \hat{\mathcal{L}}_f = -v \left(t \partial_x + x \partial_t\right) \hat{\mathcal{L}}_f\,.
\end{equation}
Here we employ $\mathcal{H}_f = \mathcal{H}_R + \mathcal{H}_L$,
\begin{equation}
    \label{eq:LT9}
    \mathcal{H}_R = - i \Psi_R^\dagger \partial_x \Psi_R\,,\quad
    \mathcal{H}_L = i \Psi_L^\dagger \partial_x \Psi_L\,,
\end{equation}
and note the cancellation of the terms $\pm v$ by $\delta_{\mathcal{L}}$. 
As a result, the integral $\int_x \hat{\mathcal{L}}_f$ transforms as a time-derivative
\begin{equation}
    \label{eq:LT10}
    \delta \int_x \hat{\mathcal{L}}_f = - \partial_t \left(\int_x  v x \hat{\mathcal{L}}_f\right)\,.
\end{equation}
The corresponding \qq{action operator} $\hat{A}_S = \int \mathrm{d} t \,\hat{\mathcal{L}}_f$ is invariant under Lorentz transformations up to boundary terms.

\indent In the naive continuum limit for the Dirac automaton with interactions one has $\mathcal{H} = \mathcal{H}_f + \mathcal{H}_{\text{int}}$.
For a local interaction involving an equal number of $\Psi_R$ and $\Psi_L$, as for $\mathcal{H}_{\text{int}} = (\pi/2) \mathcal{H}_i$ in eq.~\eqref{eq:LIF1}, one has $\delta_{\mathcal{L}} \mathcal{H}_{\text{int}} = 0$, $\delta_s \mathcal{H}_{\text{int}} = -v(t \partial_x  + x\partial_t) \mathcal{H}_{\text{int}}$.
Then $\hat{\mathcal{L}}$ transforms as $\delta \hat{\mathcal{L}} = -v (t\partial_x + x\partial_t) \hat{\mathcal{L}}$ similar to eq.~\eqref{eq:LT8}, and the action operator $\hat{A}_S$ is Lorentz invariant.
The true Hamiltonian involves commutator terms $\sim[\mathcal{H}_f,\mathcal{H}_{\text{int}}]$.
Including those the action operator $\hat{A}_S$ is typically no longer Lorentz invariant.
We conclude that continuous Lorentz transformations can be formulated for discrete generalized Ising models on a lattice.
Lorentz invariance holds for suitable PCA with interactions provided that possible Lorentz symmetry violations from the non-vanishing commutators $[\mathcal{H}_f,\mathcal{H}_{\text{int}}]$ can be neglected in the continuum limit.

\indent For the formulation of the continuous Lorentz transformations we have used the continuous field operators $\Psi_{R,L}(t,x)$.
They are the Heisenberg operators for the Hamiltonian of the Dirac automaton for free massless fermions.
One could use alternatively the Heisenberg operators for free massive fermions or, for formal purposes, the Heisenberg operators for the full Hamiltonian.

\section{Complex functional integral}
\label{sec:complex_functional_integral}
Quantum field theories are usually formulated in terms of a complex functional integral, with weight factor $\exp(iS_M)$ instead of the euclidean weight factor $\exp(-S)$ characterizing generalized Ising models.
Fermionic quantum field theories are formulated in terms of a Grassmann functional integral.
In this short section we sketch how the complex Grassmann functional integral emerges in our formulation.

\indent Our starting point is eq.~\eqref{eq:OP2} in the basis \eqref{eq:YY5}.
Expressing for a PCA the step evolution operator by the Hamiltonian,
\begin{equation}
    \label{eq:CFI1}
    \Shat(t) = \exp(-i\eps H(t))\,,
\end{equation}
yields a trace over a product chain of complex matrices.
This generalizes to the complex picture where $\Shat(t)$ is replaced by the unitary matrix $U(t)$.
If we employ the sequence $\bar{S}(t) = \Shat_{\text{int}}(t+\eps)\Shat_f(t)$, the matrix chain extends over even $t$ with $H(t)$ given by eq.~\eqref{eq:I4}.
The product of unitary matrices $\exp(-i\eps H)$ is the basis of Feynman's construction of the complex path integral.

\indent In the fermionic formulation there is a standard way how to translate a chain of matrices expressed in terms of annihilation and creation operators to a Grassmann functional integral.
The sequence of propagation and interaction steps is familiar from the Feynman path integral.
In our case it is realized directly by our formulation of automata for interacting fermions.
In contrast to the Feynman path integral the interaction part of the Hamiltonian does not remain constant in the limit $\eps \to 0$, but rather diverges $\sim \eps^{-1}$ such that $\eps H_{\text{int}}$ is constant.
Due to non-trivial commutators of $H_{\text{int}}$ and $H_f$ the true continuum limit can differ substantially from a naive continuum limit that would neglect these commutators.

\indent Finally, there exists also a direct way to establish the equivalence between certain Grassmann functionals and generalized Ising models \cite{CWPCA, CWNEW, CWFPPCA}.
The Hilbert space for the generalized Ising model is the same as for the fermionic model.
For a given Grassmann functional integral, one can compute the step evolution operator.
If it is identical to the one of the generalized Ising model, the two formulations are equivalent.
Operators in the generalized Ising model translate to Grassmann operators \cite{CWFGI, CWPCA, CWNEW}.
A rather large and diverse set of cellular automata has been constructed from suitable Grassmann functionals.
This includes models with global or local gauge symmetries or lattice diffeomorphism invariance \cite{CWFPPCA, CWCASG}.

\section{Conclusions}
\label{sec:conclusions}

\indent The introduction of a complex wave function further underlines analogies between classical statistical models and quantum theories.
The quantum formalism for classical statistics is relevant for the understanding of boundary problems.
It describes the transport of information between neighboring \qq{time} layers.
This information transport can be cast into the action of the transfer matrix in a particular normalization --- the step evolution operator --- on real wave functions.
The wave functions are probability amplitudes, with time-local probabilities being bilinear in the wave functions.
The evolution law for the wave functions is linear.
This entails the superposition law for solutions --- a corner stone of quantum mechanics.
For the computation of expectation values of time-local observables one introduces operators acting on wave functions.
The expression for expectation values in terms of operators and wave functions is a generalization of the quantum law.
It follows from the classical statistical definition of expectation values without the need of introducing any additional \qq{quantum axioms}.

\indent A complex structure maps real wave functions to complex wave functions.
In the present paper we associate for generalized Ising models the complex conjugation to the change of sign of all Ising spins or the particle-hole conjugation.
Complex wave functions bring the quantum formalism for classical statistics even closer to quantum mechanics.
They allow for complex basis transformations of the wave functions, as the Fourier transform, and the definition of statistical observables as momentum or energy.
Similar to quantum mechanics, the operators for momentum and position do not commute.
In our proposal complex conjugation is closely related to charge conjugation $C$.
We also discuss the discrete symmetries of parity $P$ and time reversal $T$.
The action of the discrete symmetries $C$, $P$ and $T$ on the complex wave functions establishes further analogies between generalized Ising models and quantum field theories.
We discuss the interesting possibility that complex conjugation involves an additional $CPT$-transformation such that it can be associated to the $PT$-transformation.

\indent General Ising-type models can differ from quantum theories in one important aspect:
The time evolution is often not unitary.
Boundary information can be lost as one proceeds to positions inside the bulk, with a possible approach to an equilibrium state for time differences much larger than the correlation time.
There exists, however, a particular family of generalized Ising models, namely probabilistic cellular automata, for which the time evolution is unitary.
This results from the deterministic updating rule --- the probabilistic aspects only enter by probabilistic initial conditions.
These classical statistical systems are discrete quantum systems.

\indent The discussion of complex wave functions for probabilistic cellular automata (PCA) gives further support to the claim that quantum mechanics can be embedded in classical statistics \cite{CWQMCS,CWEM}.
Bell's inequalities for classical correlation functions do not apply to pairs of observables as momentum and position.
A statistical observable as momentum does not have a fixed value for a given configuration of Ising spins.
A classical correlation for momentum and position therefore simply does not exist.
Fermions in two dimensions do not carry spin.
If our PCA can be generalized to fermions in three or more dimensions one expects in addition to the momentum as generator of translations also angular momentum as generator of rotations.
Angular momentum or spin are expected to be associated to statistical observables which do not take fixed values for the configurations of Ising spins.
For correlations of such observables Bell's inequalities do not hold.

\indent The embedding of quantum mechanics into classical statistics does not imply that every quantum system must have a classical probabilistic counterpart.
The formulation as a PCA imposes restrictions on the quantum systems for which this is possible:
There needs to exist a discretization which realizes the unique jump property of the step evolution operator for automata.
Nevertheless, classical statistical PCA have been constructed for simple quantum systems as qubits or a quantum particle in a harmonic potential \cite{CWPW2024}, or for rather rich classes of fermionic quantum field theories \cite{CWFPPCA}.
It remains to be seen if the requirement of an underlying overall classical probability distribution could be a selection criterion for a quantum field theory describing our universe.

\indent More in detail, generalized Ising models for PCA define discrete quantum field theories for fermions.
We focus first on free massless fermions in two dimensions.
One can express the step evolution operator $\Shat$ in terms of fermionic annihilation and creation operators in momentum space.
This expression only requires the usual anticommutation relations and a specific form of the commutators of the fermionic operators with $\Shat$.
It does not fix the choice of fermionic operators uniquely.
Thus the action of these fermionic operators on the complex wave function needs to be specified.
We discuss different choices of annihilation and creation operators which may or may not be related by basis transformations.
Different choices do not affect the discrete time evolution or the continuum limit.
They can matter, however, for the definition of useful observables as fermionic particle numbers.

\indent The translation of the discrete transformations $C$, $P$ and $T$ from the generalized Ising model to the fermionic quantum field theory elucidates further important aspects of the equivalence of both formulations.
The discrete symmetries of the fermionic quantum field theory find a simple root in transformations for the Ising spins.
The invariance of a generalized Ising model under the transformations $T$, $PT$, $CT$ or $CPT$ offers the interesting possibility to define a complex wave function $\varphi(t)$ which combines information both from $t$ and $-t$.
For a particular choice of the complex structure the action of $PT$ or $CP\tilde{T}$ is equivalent to complex conjugation of the wave function.

\indent A real orthogonal step evolution operator $\Shat$ can be expressed in terms of a hermitian Hamiltonian $H$ by $\Shat = \exp(-i \eps H)$, where $\Shat$ and $H$ act on a complex wave function.
The eigenvalues of $\Shat$ occur in pairs whose members are complex conjugate to each other.
As a consequence, also the real eigenvalues $E_n$ of $H$ come in pairs with opposite sign of its members.
Negative energy eigenvalues are a danger for stability once interactions are included.
This can be overcome by the realization of a \qq{particle physics vacuum} for which all excitations have positive energy.
Typically, the energy $E_0$ of this vacuum differs from zero.
For the particle physics vacuum all states with negative energy are occupied, and the states with positive energy are empty.
For free massless particles in two dimensions we construct the particle physics vacuum explicitly.
It is simple in momentum space, but rather complex in position space where the basis states are configurations of Ising spins.

\indent One-particle excitations of the vacuum are described by applying field operators to it.
These field operators depend on continuous time and space coordinates.
They allow to define one-particle wave functions in continuous spacetime.
The one particle wave functions for free massless fermions obey the continuous Dirac equation.
Field operators can also be used to construct multi-particle states.
In summary, for free massless fermions in two dimensions we have presented a complete construction of a PCA which realizes all aspects of the quantum field theory.
The continuum limit for smooth wave functions is straightforward.
Smooth initial wave functions remain smooth under the evolution.
The evolution of one-particle states according to the Dirac equation realizes Lorentz symmetry.

\indent One may view the generalized Ising model for the Majorana--Weyl or Dirac automaton as a regularization for the quantum field theory of free massless fermions in two dimensions.
A quantum field theory in Minkowski space finds a regularization by a euclidean \qq{functional integral} on a lattice.
A Hamiltonian involving field operators in continuous space is regularized by a finite dimensional Hilbert space for finite $L$ and $\eps>0$.
The discrete properties of the lattice for $\eps > 0$ show up only in the modification of the anticommutation relations for the field operators, which are reflected by corresponding modifications of the Feynman propagator for high energies $\sim\eps^{-1}$ or short time intervals $\sim \eps$.
The proposed PCA constitutes a regularization for chiral fermions in two dimensions.
We have found no sign of \qq{lattice doublers}.
In the continuum limit the Feynman propagator takes the standard form for free massless fermions.

\indent The question arises if a fermionic quantum field theory with interactions can be constructed from a PCA in a similar way.
A large family of PCA that lead to quantum field theories for interacting fermions can be formulated in a simple way by sequences of updatings for propagation and interaction.
Typically, the interaction step involves permutations of spin configurations in a local cell.
We have presented an example with particle production by scattering and a modified propagation.
An important open question asks if it is still possible to find a particle physics vacuum for which all excitations have positive energy.

\indent A second crucial question concerns a possible continuum limit for the corresponding discrete quantum field theories.
Are there PCA for which the interacting quantum field theory shows Lorentz symmetry in the continuum limit?
The naive continuum limit neglects non-vanishing commutators between the propagation and interaction parts of the Hamiltonian.
Simple models which show Lorentz symmetry in the naive continuum limit have been constructed \cite{CWPCAQP, CWFPPCA}.
They correspond to discrete versions of generalized Thirring models \cite{THI,KLA,FAIV,FUR,NAO,COL,DNS} or equivalent Gross--Neveu models \cite{GN,WWE,RSHA,RWP,SZKSR} for several flavors.
In the naive continuum limit the Hamiltonian is given by a simple local expression of the continuous field operators even for finite $\eps$.
It is not established, however, what is the true continuum limit of these discrete models.
In particular, one is interested if the chiral symmetries forbidding mass terms can be spontaneously or explicitly broken, and if a mass for the fermions can be generated.
One also would like to find a generalization to massless or massive fermions in four dimensions.

\indent The conceptual equivalence of probabilistic cellular automata and discrete quantum field theories for fermions appears to be well established.
The construction of interesting \qq{realistic} fermionic models is only at its beginning.

\vspace{10pt}
\noindent \textbf{Acknowledgment:}
\vspace{5pt}

The author would like to thank M. L\"uscher for stimulating discussions.


\appendix
\section{General complex structures}
\label{app:A} 

\indent Assume that complex conjugation corresponds in the real picture to $q\to Kq$, $K^2 = 1$.
The real part of $\varphi$ is even under $K$, and the imaginary part odd, and we may write
\begin{equation}
    \label{eq:GC1}
    \varphi = F_R (q+Kq) + i F_I (q-Kq) = \tilde{G} q\,,
\end{equation}
with real matrices $F_R$ and $F_I$.
This description of $\varphi$ is redundant, since $\varphi$ has here the same number of components as $q$.
Since a $\bar{N}$-component real vector should be equivalent to a $\bar{N}/2$-component complex vector, one expects a constraint on $\varphi$.

\indent The map $\varphi \to \varphi'=i\varphi$ should be realized by $q\to q' = I q$, $I^2 = -1$, $\{K,I\}=0$,
\begin{equation}
    \label{eq:GC2}
    \begin{aligned}
        \varphi' & = -F_I(q-Kq) + i F_R (q+Kq)       \\
             & = F_R (Iq+KIq) + i F_I (Iq-KIq)   \\
             & = F_R I (q-Kq) + iF_I I (q+Kq)\,.
    \end{aligned}
\end{equation}
This requires
\begin{equation}
    \label{eq:GC3}
    (F_R - F_I I) (1+K) = 0\,.
\end{equation}
If $F_R$ is invertible the relation \eqref{eq:GC3} reads
\begin{equation}
    \label{eq:GC7}
    (1 - \tilde{F}_I I) (1+K) = 0\,,\quad
    \tilde{F}_I = F_R^{-1} F_I\,,
\end{equation}
or, with constant $c$,
\begin{equation}
    \label{eq:GC8}
    \tilde{F}_I I = 1- c(1-K)\,.
\end{equation}
The normalization
\begin{equation}
    \label{eq:GC4}
    \varphi^\dagger \varphi = q^T q = 1
\end{equation}
is obeyed for
\begin{equation}
    \label{eq:GC5}
    \mathrm{Re}(\tilde{G}^\dagger \tilde{G}) = 1\,,
\end{equation}
with
\begin{equation}
    \label{eq:GC6}
    \tilde{G} = F_R + F_R K + i F_I - i F_I K\,.
\end{equation}

\indent We can now derive the constraint for $\varphi$ by which $\bar{N}-1$ real numbers are sufficient to describe a normalized $\varphi$.
The relation \eqref{eq:GC8} implies
\begin{equation}
    \label{eq:GC12}
    [\tilde{F}_I I, K] = 0\,,\quad
    \{\tilde{F}_I, K\} = 0\,.
\end{equation}
If $K$ commutes with $F_R$ eq.~\eqref{eq:GC1} implies
\begin{equation}
    \label{eq:GC13}
    K \varphi = \varphi\,.
\end{equation}
For a more general product of $K$ and $F_R$ involving a matrix $A$ one has
\begin{equation}
    \label{eq:GC14}
    K F_R = A F_R K \;\;\implies\;\; \varphi = K A \varphi\,.
\end{equation}
Due to these relations not all components of $\varphi$ are independent. 
This realizes the constraint on $\varphi$. 
We emphasize that a map with $F_I = F_R$, $\tilde{F}_I = 1$ is not compatible with the relation \eqref{eq:GC12}.
This relation also implies that the constraint on $\varphi$ is of the type $\varphi = \tilde{A} \varphi$ and not $\varphi^* = \tilde{A} \varphi$.

\indent For our setting of a complex structure the various matrices are given explicitly.
For the case $\bar q' = \tilde q = q$ the structure \eqref{eq:C15} corresponds to
\begin{equation}
    \label{eq:GC9}
    F_R = \frac{1}{2}\,,\quad
    \tilde{F}_I = T_3\,,\quad
    K=B\,,\quad
    I = T_3 B\,,\quad
    c=1\,.
\end{equation}
For $\bar{q}'$ different from $\tilde{q}$ the real vector $q$ reads
\begin{equation}
    \label{eq:GC10}
    q = \begin{pmatrix} \tilde q \\ \bar{q}' \end{pmatrix}.
\end{equation}
In this case one has
\begin{align}
    \label{eq:GC11}
     & F_R = \frac{1}{2}\,,\quad
    \tilde{F}_I = 1\otimes T_3\,,\quad
    K = \tau_1 \otimes B\,,\quad \nn   \\
     & I = \tau_1\otimes T_3 B\,,\quad
    c = 1\,.
\end{align}
For $\bar{q}' = \tilde{q} = q$ the relation \eqref{eq:GC13} yields the constraint \eqref{eq:C18}.
For the case of independent $\bar{q}'$ and $\tilde{q}$ the constraint $K \varphi = \varphi$ relates the lower component of $\varphi$ to the upper component.
The upper component is then an arbitrary (normalized) $2^{N_x}$-component complex vector.

\section{Operators in the complex picture}
\label{app:B}

\indent Not every observable in the real picture can be expressed by an operator in the complex picture such that
\begin{equation}
    \label{eq:OCP1}
    \langle A \rangle = \varphi^\dagger(q) \hat{A}_C \varphi(q)\,,
\end{equation}
with $\hat{A}_C$ a complex operator.
Let us choose in the real picture a basis for the $\bar{N}$-component real wave functions such that
\begin{equation}
    \label{eq:CO1}
    q = \begin{pmatrix} q_1 \\ q_2 \end{pmatrix}\,,\quad
    K = \tilde{K} \otimes 1\,,\quad
    I = \tilde{I} \otimes 1\,,\quad
\end{equation}
with $q_1$, $q_2$ real vectors with $\bar{N}/2$ components and 1 standing for the $\bar{N}/2\otimes\bar{N}/2$ unit matrix.
If a real $\bar{N}\times \bar{N}$-matrix $\hat{A}$ can be written in the form
\begin{equation}
    \label{eq:CO2}
    \hat{A} = 1 \otimes \hat{A}_R + \tilde{I} \otimes \hat{A}_I\,,
\end{equation}
it can be mapped to a complex $\bar{N}/2\times \bar{N}/2$-matrix
\begin{equation}
    \label{eq:CO3}
    \hat{A}_C = \hat{A}_R + i \hat{A}_I\,.
\end{equation}
For two such matrices $\hat{A}$, $\hat{B}$ the real matrix multiplication $\hat{A}\hat{B}$ is mapped to the complex matrix multiplication $\hat{A}_C \hat{B}_C$.
The map $\hat{A} \to K \hat{A} K $ induces in the complex picture the operation of complex conjugation $\hat{A}_C \to \hat{A}_C^*$,
and the map $\hat{A} \to I \hat{A}$ results in $\hat{A}_C \to i \hat{A}_C$.

\indent Using the map to a complex wave function 
\begin{equation}
    \label{eq:CO4}
    q \to \varphi\,,\quad
    K q \to \varphi^*\,,\quad
    I q \to i \varphi\,,\quad
    (1 \otimes A_R) q \to A_R \varphi\,,    
\end{equation}
one finds
\begin{equation}
    \label{eq:CO5}
    \hat{A} q \to \hat{A}_C \varphi\,,\quad
    K \hat{A} q \to \hat{A}_C^* \varphi^*\,.
\end{equation} 
For the expectation value the map implies
\begin{equation}
    \label{eq:CO6}
    \langle A \rangle = q^T \hat{A} q 
    = \frac{1}{2} \varphi^\dagger(\hat{A}_C + \hat{A}_C^\dagger) \varphi\,.
\end{equation}
Symmetric matrices in the real picture are mapped to hermitian matrices in the complex picture
\begin{equation}
    \label{eq:CO7}
    \hat{A}^T = \hat{A} \;\;\implies\;\; \hat{A}_C^\dagger = \hat{A}_C\,.
\end{equation}
Only the symmetric part of $\hat{A}$ or the hermitian part of $\hat{A}_C$ contribute to the expectation values.

\indent Operators which are compatible with the complex structure have to commute with $I$.
The relations \eqref{eq:CO4}--\eqref{eq:CO7} are easily verified for the case
\begin{equation}
    \label{eq:COP1}
    \tilde{K} = \tau_3\,, \quad
    \tilde{I} = - i \tau_2\,.
\end{equation}
We may order $q$ in terms of eigenvalues of $\tilde{K}$, which correspond to the real and imaginary part of $\varphi$,
\begin{equation}
    \label{eq:COP2}
    q = \begin{pmatrix} q_+ \\ q_- \end{pmatrix}\,,\quad
    \varphi = q_+ + i q_-\,,
\end{equation}
verifying easily eqs.~\eqref{eq:CO4}--\eqref{eq:CO7}.
For $\hat{A} = 1$ one obtains for the normalization
\begin{equation}
    \label{eq:COP3}
    q^T q = \varphi^\dagger \varphi\,.
\end{equation}

\indent Orthogonal transformations of $q$ which are compatible with the complex structure result in unitary transformations of $\varphi$.
They form the subgroup $U\left(\frac{\bar{N}}{2}\right)$ of $O(\bar{N})$.
We can use such orthogonal transformations to map the complex structure \eqref{eq:COP1} to other complex structures.
For
\begin{align}
    \label{eq:COP4}
    q' &= O \begin{pmatrix} q_+ \\ q_- \end{pmatrix}\,,&
    K' &= O K O^T\,, \nn\\
    I' &= O I O^T\,, &
    \hat{A}' &= O \hat{A} O^T\,,
\end{align}
one has
\begin{equation}
    \label{eq:COP5}
    \varphi' = U \varphi \,,\quad
    \hat{A}_C' = U \hat{A}_C U^\dagger\,,\quad
\end{equation}
The maps \eqref{eq:CO4}--\eqref{eq:CO7} remain valid after the transformation.
An example is
\begin{equation}
    \label{eq:COP6}
    O = \tilde{D}\otimes 1\,,\quad \tilde{D} = \frac{1}{\sqrt{2}} (1-i\tau_2)\,,
\end{equation}
which transforms eq.~\eqref{eq:COP1} to
\begin{equation}
    \label{eq:COP7}
    \tilde{K}' = \tau_1\,,\quad
    \tilde{I}' = -i\tau_2\,.
\end{equation}
Orthogonal transformations compatible with the complex structure do not change $I$, \eg $I'=I$.
Further complex structures can be obtained by a map $\varphi \to \varphi^*$, as realized by $q \to (\tau_3 \otimes 1) q$.
This maps $I\to -I$, while leaving $K$ invariant.
With $\hat{A}_C \to \hat{A}_C^*$ the relations \eqref{eq:CO5}--\eqref{eq:CO7} continue to hold.
Those relations remain valid for a more extended setting, as for example the redundant description \eqref{eq:GC9} or \eqref{eq:GC11}.

\indent The requirement that an operator $\hat{A}$ compatible with the complex structure has to commute with $I$ is easily understood.
In the complex picture $i$ commutes with every complex operator $\hat{A}_c$.
This can be obtained by a map from the real picture $\hat{A} \to \hat{A}_c$ only if $[\hat{A},I] = 0$.
(For an arbitrary operator $\hat{A}$ we can construct an associated compatible operator
\begin{equation}
    \label{eq:COP8}
    \hat{A}' = \frac{1}{2} \left(\hat{A} - I \hat{A} I\right)\,,\quad
    [\hat{A}',I] = 0\,.
\end{equation}
For $[\hat{A},I] = 0$ one has $\hat{A}' = \hat{A}$.)
For general $\hat{A}$ we define
\begin{align}
    \label{eq:COP9}
    \hat{A}_R &= \frac{1+K}{2} \hat{A} \frac{1+K}{2} + \frac{1-K}{2} \hat{A} \frac{1-K}{2} = \hat{A}_{++} + \hat{A}_{--}\,, \nn\\
    \hat{A}_I &= \frac{1+K}{2} \hat{A} \frac{1-K}{2} + \frac{1-K}{2} \hat{A} \frac{1+K}{2} = \hat{A}_{+-} + \hat{A}_{-+}\,, \nn\\
    \hat{A} &= \hat{A}_R + \hat{A}_I\,.
\end{align}
The compatibility with the complex structure requires
\begin{align}
    \label{eq:COP10}
    [\hat{A},I] &= 0\,,\quad
    [\hat{A}_R,I] = 0\,,\quad
    [\hat{A}_I,I] = 0\,, \nn\\
    \hat{A}_{--} &= \hat{A}_{++}\,,\quad
    \hat{A}_{-+} = \hat{A}_{+-}\,.
\end{align}
If this condition is obeyed, we can identify
\begin{equation}
    \label{eq:COP11}
    A_R = \hat{A}_{++}\,,\quad
    A_I = -\hat{A}_{+-}\,,\quad
    A_C = A_R + i A_I\,.
\end{equation}

\indent We can also implement complex $\bar{N}\times\bar{N}$-matrices $\hat{A}_C$ in the redundant setting where $\varphi$ has the same number of components $\bar{N}$ as $q$,
\begin{equation}
    \label{eq:COP12}
    \varphi = \varphi(q) = \frac{1+K}{2} q - i I \frac{1-K}{2} q\,,\quad
    K\varphi = \varphi\,.
\end{equation}
Here we take $F_R = \frac{1}{2}$, $\tilde{F}_I = I K$, $c=1$ in eqs.~\eqref{eq:GC7},\eqref{eq:GC8}.
We define the complex $\bar{N}\times\bar{N}$-matrix by the map $\hat{A}\to\hat{A}_C(\hat{A})$,
\begin{equation}
    \label{eq:COP13}
    \hat{A}_C = \hat{A}_R - i I \hat{A}_I\,,\quad
    [K, \hat{A}_C] = 0\,.
\end{equation}
It obeys
\begin{equation}
    \label{eq:COP14}
    \varphi(\hat{A} q) = \hat{A}_C \varphi\,,
\end{equation}
and is consistent with the complex matrix multiplication
\begin{equation}
    \label{eq:COP15}
    \hat{A} \hat{B} \to (\hat{A} \hat{B})_C = \hat{A}_C \hat{B}_C\,,
\end{equation}
provided $[\hat{A},I] = [\hat{B},I] = 0$.
One observes
\begin{alignat}{2}
    \label{eq:COP16}
    (K \hat{A} K)_C &= \hat{A}_C^*\,,\quad
                    & K \hat{A}_C K &= \hat{A}_C\,, \nn \\
    (I \hat{A})_C &= i \hat{A}_C\,,\quad
                  & I \hat{A}_C I &= -\hat{A}_C\,.
\end{alignat}

\section{Fermionic operators for composite particles}
\label{app:C}
In this appendix we explore further fermionic annihilation and creation operators that are compatible with the identification $\tilde{q} = \bar{q}' = q$.
By virtue of
eq.~\eqref{eq:C18} they have to obey
\begin{equation}
    \label{eq:F8}
    [A(j),B] = 0\,,\quad [A^\dagger(j),B] = 0\,,
\end{equation}
and we further require the anti-commutation relations
\begin{align}
    \label{eq:F9}
    \{A(j'),A(j)\} =         & \, \{A^\dagger(j'),A^\dagger(j)\} = 0\,,\nn \\
    \{A^\dagger(j'),A(j)\} = & \, \delta_{jj'}\,.
\end{align}
The operators $a(j)$, $a^\dagger(j)$ in eq.~\eqref{eq:F5} do not obey the
constraint~\eqref{eq:F8}.
This reflects the property that the creation of an additional fermion from a configuration $\tau$ corresponds to the annihilation of a fermion in the particle-hole conjugate state.

\indent A possible choice for $j\neq(N_x-1)/2$ is given by
\begin{align}
    \label{eq:F11}
    A(j) = & \tau_3\otimes \tau_3 \otimes \dots \otimes a^{\phantom{T}}\otimes
    \frac{1+\tau_3}{2}\otimes 1\otimes \dots 1\nn                              \\
           & + \tau_3\otimes \tau_3 \otimes \dots \otimes a^T\otimes
    \frac{1-\tau_3}{2}\otimes 1\otimes \dots 1\,.
\end{align}
For $A^\dagger(j)$ the signs in the factors $(1\pm \tau_3)/2$ are switched.
If at the site $j+1$ a fermion is present, $A(j)$ acts as $a(j)$, switching an occupied site at $j$ to an empty site, and annihilating the wave function for an empty site $j$.
If the site $j+1$ is empty, $A(j)$ acts as the creation operator $a^\dagger(j)$.
The relations $A^2(j)=0$ $(A^\dagger(j))^2=0$ follow from the opposite projectors $(1\pm \tau_3)/2$ for the mixed terms $\sim aa^T$.
One finds
\begin{align}
    \label{eq:F12}
    N(j) = \, & A^\dagger(j)A(j)\nn                                                            \\
    =      \, & 1\otimes1\otimes \dots \otimes a^Ta\otimes \frac{1+\tau_3}{2}\otimes1\otimes
    \dots 1\nn                                                                                 \\
              & +1\otimes 1\otimes \dots \otimes aa^T\otimes \frac{1-\tau_3}{2}\otimes1\otimes
    \dots 1\,,
\end{align}
and
\begin{equation}
    \label{eq:F13}
    \{A^\dagger(j),A(j)\} = 1\,.
\end{equation}

\indent The remaining anticommutations~\eqref{eq:F9} for $j'\neq j$, $j+1$ follow from the anticommutation of $\tau_3$ with $a$ and $a^T$. 
In contrast, one finds for neighboring sites
\begin{equation}
    \label{eq:A5A}
    \begin{aligned}
        \{A( & j),A(j+1)\} =                                                                                                      \\
             & \phantom{+} 1\otimes 1\otimes\dots\otimes \tau_3(a^T-a)\otimes \tau_3 a \otimes \frac{1+\tau_3}{2} \otimes 1 \dots \\
             & + 1\otimes 1\otimes\dots\otimes \tau_3(a^T-a)\otimes \tau_3 a^T \otimes \frac{1-\tau_3}{2} \otimes 1 \dots
    \end{aligned}
\end{equation}
We finally define $A\left(\frac{N_x-1}{2}\right)$ similar to eq.~\eqref{eq:F11}, with $a$ and $a^T$ factors at the position $(N_x-1)/2$, and $(1\pm\tau_3)/2$ factors at position $-(N_x-1)/2$. This completes the anticommutation relations~\eqref{eq:F9} for arbitrary $j'$ and $j$ that are not next neighbors.

The operator $N(j)$ can be expressed in terms of the occupation operators at two neighboring sites
\begin{align}
    \label{eq:75A}
    N(j) = & \, \hat n(j)\hat n(j+1) + \gl 1 - \hat n(j)\gr \gl 1 - \hat n(j+1)\gr
    \nn                                                                            \\
    =      & \, \frac12\gl 1 + \hat s(j) \hat s(j+1)\gr\,,
\end{align}
with $\hat s(j) = 2\hat n(j) - 1$ the operators for the Ising spins. Thus $N(j)$
equals one if the two neighboring Ising spins $s(j)$ and $s(j+1)$ are equal,
while $N(j) = 0$ if they are opposite.
We may interpret $N(j)$ for even $j$ as the number of a type of \qq{composite fermion} which involves the two sites $j$ and $j+1$.
Accordingly, we only consider $A(j)$, $A^\dagger(j)$ for even $j$.
They obey the anticommutation relations \eqref{eq:F9}.

Charge conjugation commutes with the operator $N(j)$ in
eqs.~\eqref{eq:F8},~\eqref{eq:75A},
\begin{equation}
    \label{eq:CC12}
    [C,N(j)] = 0\,.
\end{equation}
With eq.~\eqref{eq:H3} and $H^{(R)} = \hat P$ this operator obeys
\begin{align}
    \label{eq:CC13}
     & [\hat P, N(j)] = \sum_{m>0}\frac{i\pi(-1)^m}{L\sin\left( \frac{\pi
    m\eps}{L}\right)} \gl a^\dagger(j-m)a(j)\nn                            \\
     & -a^\dagger(j)a(j+m) - a^\dagger(j+m)a(j) + a^\dagger(j)a(j-m)\gr\,.
\end{align}
This implies for $N\subt{tot} = \sum_jN(j)$,
\begin{equation}
    \label{eq:CC14}
    [\hat P, N\subt{tot}] = 0\,.
\end{equation}
One can therefore construct simultaneous eigenstates to the operators $\hat P$,
$N\subt{tot}$ and $C$.


\nocite{*}
\bibliography{refs.bib}

\begin{thebibliography}{89}%
\makeatletter
\providecommand \@ifxundefined [1]{%
 \@ifx{#1\undefined}
}%
\providecommand \@ifnum [1]{%
 \ifnum #1\expandafter \@firstoftwo
 \else \expandafter \@secondoftwo
 \fi
}%
\providecommand \@ifx [1]{%
 \ifx #1\expandafter \@firstoftwo
 \else \expandafter \@secondoftwo
 \fi
}%
\providecommand \natexlab [1]{#1}%
\providecommand \enquote  [1]{``#1''}%
\providecommand \bibnamefont  [1]{#1}%
\providecommand \bibfnamefont [1]{#1}%
\providecommand \citenamefont [1]{#1}%
\providecommand \href@noop [0]{\@secondoftwo}%
\providecommand \href [0]{\begingroup \@sanitize@url \@href}%
\providecommand \@href[1]{\@@startlink{#1}\@@href}%
\providecommand \@@href[1]{\endgroup#1\@@endlink}%
\providecommand \@sanitize@url [0]{\catcode `\\12\catcode `\$12\catcode `\&12\catcode `\#12\catcode `\^12\catcode `\_12\catcode `\%12\relax}%
\providecommand \@@startlink[1]{}%
\providecommand \@@endlink[0]{}%
\providecommand \url  [0]{\begingroup\@sanitize@url \@url }%
\providecommand \@url [1]{\endgroup\@href {#1}{\urlprefix }}%
\providecommand \urlprefix  [0]{URL }%
\providecommand \Eprint [0]{\href }%
\providecommand \doibase [0]{https://doi.org/}%
\providecommand \selectlanguage [0]{\@gobble}%
\providecommand \bibinfo  [0]{\@secondoftwo}%
\providecommand \bibfield  [0]{\@secondoftwo}%
\providecommand \translation [1]{[#1]}%
\providecommand \BibitemOpen [0]{}%
\providecommand \bibitemStop [0]{}%
\providecommand \bibitemNoStop [0]{.\EOS\space}%
\providecommand \EOS [0]{\spacefactor3000\relax}%
\providecommand \BibitemShut  [1]{\csname bibitem#1\endcsname}%
\let\auto@bib@innerbib\@empty
\bibitem [{\citenamefont {Bell}(1964)}]{BELL}%
  \BibitemOpen
  \bibfield  {author} {\bibinfo {author} {\bibfnamefont {J.~S.}\ \bibnamefont {Bell}},\ }\bibfield  {title} {\bibinfo {title} {{On the Einstein-Podolsky-Rosen paradox}},\ }\href {https://doi.org/10.1103/PhysicsPhysiqueFizika.1.195} {\bibfield  {journal} {\bibinfo  {journal} {Physics Physique Fizika}\ }\textbf {\bibinfo {volume} {1}},\ \bibinfo {pages} {195} (\bibinfo {year} {1964})}\BibitemShut {NoStop}%
\bibitem [{\citenamefont {Clauser}\ \emph {et~al.}(1969)\citenamefont {Clauser}, \citenamefont {Horne}, \citenamefont {Shimony},\ and\ \citenamefont {Holt}}]{CHSH}%
  \BibitemOpen
  \bibfield  {author} {\bibinfo {author} {\bibfnamefont {J.~F.}\ \bibnamefont {Clauser}}, \bibinfo {author} {\bibfnamefont {M.~A.}\ \bibnamefont {Horne}}, \bibinfo {author} {\bibfnamefont {A.}~\bibnamefont {Shimony}},\ and\ \bibinfo {author} {\bibfnamefont {R.~A.}\ \bibnamefont {Holt}},\ }\bibfield  {title} {\bibinfo {title} {{Proposed experiment to test local hidden variable theories}},\ }\href {https://doi.org/10.1103/PhysRevLett.23.880} {\bibfield  {journal} {\bibinfo  {journal} {Phys. Rev. Lett.}\ }\textbf {\bibinfo {volume} {23}},\ \bibinfo {pages} {880} (\bibinfo {year} {1969})}\BibitemShut {NoStop}%
\bibitem [{\citenamefont {Aspect}\ \emph {et~al.}(1982)\citenamefont {Aspect}, \citenamefont {Dalibard},\ and\ \citenamefont {Roger}}]{ASDA}%
  \BibitemOpen
  \bibfield  {author} {\bibinfo {author} {\bibfnamefont {A.}~\bibnamefont {Aspect}}, \bibinfo {author} {\bibfnamefont {J.}~\bibnamefont {Dalibard}},\ and\ \bibinfo {author} {\bibfnamefont {G.}~\bibnamefont {Roger}},\ }\bibfield  {title} {\bibinfo {title} {{Experimental test of Bell's inequalities using time varying analyzers}},\ }\href {https://doi.org/10.1103/PhysRevLett.49.1804} {\bibfield  {journal} {\bibinfo  {journal} {Phys. Rev. Lett.}\ }\textbf {\bibinfo {volume} {49}},\ \bibinfo {pages} {1804} (\bibinfo {year} {1982})}\BibitemShut {NoStop}%
\bibitem [{\citenamefont {Shannon}(1948)}]{SHA}%
  \BibitemOpen
  \bibfield  {author} {\bibinfo {author} {\bibfnamefont {C.~E.}\ \bibnamefont {Shannon}},\ }\bibfield  {title} {\bibinfo {title} {{A mathematical theory of communication}},\ }\href {https://doi.org/10.1002/j.1538-7305.1948.tb01338.x} {\bibfield  {journal} {\bibinfo  {journal} {Bell Syst. Tech. J.}\ }\textbf {\bibinfo {volume} {27}},\ \bibinfo {pages} {379} (\bibinfo {year} {1948})}\BibitemShut {NoStop}%
\bibitem [{\citenamefont {Lenz}(1920)}]{LENZ}%
  \BibitemOpen
  \bibfield  {author} {\bibinfo {author} {\bibfnamefont {W.}~\bibnamefont {Lenz}},\ }\bibfield  {title} {\bibinfo {title} {Beitrag zum verst{\"a}ndnis der magnetischen erscheinungen in festen k{\"o}rpern},\ }\href@noop {} {\bibfield  {journal} {\bibinfo  {journal} {Z. Phys.}\ }\textbf {\bibinfo {volume} {21}},\ \bibinfo {pages} {613} (\bibinfo {year} {1920})}\BibitemShut {NoStop}%
\bibitem [{\citenamefont {Ising}(1925)}]{ISING}%
  \BibitemOpen
  \bibfield  {author} {\bibinfo {author} {\bibfnamefont {E.}~\bibnamefont {Ising}},\ }\bibfield  {title} {\bibinfo {title} {Beitrag zur theorie des ferromagnetismus},\ }\href@noop {} {\bibfield  {journal} {\bibinfo  {journal} {Zeitschrift f{\"u}r Physik}\ }\textbf {\bibinfo {volume} {31}},\ \bibinfo {pages} {253} (\bibinfo {year} {1925})}\BibitemShut {NoStop}%
\bibitem [{\citenamefont {Binder}(2001)}]{BINDER}%
  \BibitemOpen
  \bibfield  {author} {\bibinfo {author} {\bibfnamefont {K.}~\bibnamefont {Binder}},\ }\bibfield  {title} {\bibinfo {title} {Ising model},\ }\href@noop {} {\bibfield  {journal} {\bibinfo  {journal} {Hazewinkel, Michiel, Encyclopedia of Mathematics. Springer}\ } (\bibinfo {year} {2001})}\BibitemShut {NoStop}%
\bibitem [{\citenamefont {Wetterich}(2012)}]{CWPT}%
  \BibitemOpen
  \bibfield  {author} {\bibinfo {author} {\bibfnamefont {C.}~\bibnamefont {Wetterich}},\ }\bibfield  {title} {\bibinfo {title} {{Probabilistic Time}},\ }\href {https://doi.org/10.1007/s10701-012-9675-3} {\bibfield  {journal} {\bibinfo  {journal} {Found. Phys.}\ }\textbf {\bibinfo {volume} {42}},\ \bibinfo {pages} {1384} (\bibinfo {year} {2012})},\ \Eprint {https://arxiv.org/abs/1002.2593} {arXiv:1002.2593} \BibitemShut {NoStop}%
\bibitem [{\citenamefont {Wetterich}(2018{\natexlab{a}})}]{CWIT}%
  \BibitemOpen
  \bibfield  {author} {\bibinfo {author} {\bibfnamefont {C.}~\bibnamefont {Wetterich}},\ }\bibfield  {title} {\bibinfo {title} {{Information transport in classical statistical systems}},\ }\href {https://doi.org/10.1016/j.nuclphysb.2017.12.008} {\bibfield  {journal} {\bibinfo  {journal} {Nucl. Phys. B}\ }\textbf {\bibinfo {volume} {927}},\ \bibinfo {pages} {35} (\bibinfo {year} {2018}{\natexlab{a}})},\ \Eprint {https://arxiv.org/abs/1611.04820} {arXiv:1611.04820} \BibitemShut {NoStop}%
\bibitem [{\citenamefont {Wetterich}(2018{\natexlab{b}})}]{CWQF}%
  \BibitemOpen
  \bibfield  {author} {\bibinfo {author} {\bibfnamefont {C.}~\bibnamefont {Wetterich}},\ }\bibfield  {title} {\bibinfo {title} {{Quantum formalism for classical statistics}},\ }\href {https://doi.org/10.1016/j.aop.2018.03.022} {\bibfield  {journal} {\bibinfo  {journal} {Annals Phys.}\ }\textbf {\bibinfo {volume} {393}},\ \bibinfo {pages} {1} (\bibinfo {year} {2018}{\natexlab{b}})},\ \Eprint {https://arxiv.org/abs/1706.01772} {arXiv:1706.01772} \BibitemShut {NoStop}%
\bibitem [{\citenamefont {Baxter}(1982)}]{BAX}%
  \BibitemOpen
  \bibfield  {author} {\bibinfo {author} {\bibfnamefont {R.~J.}\ \bibnamefont {Baxter}},\ }\href {https://doi.org/10.1142/9789814415255_0002} {\emph {\bibinfo {title} {{Exactly solved models in statistical mechanics}}}}\ (\bibinfo  {publisher} {Academic Press, London, San Diego},\ \bibinfo {year} {1982})\BibitemShut {NoStop}%
\bibitem [{\citenamefont {Suzuki}(1985)}]{SUZ}%
  \BibitemOpen
  \bibfield  {author} {\bibinfo {author} {\bibfnamefont {M.}~\bibnamefont {Suzuki}},\ }\bibfield  {title} {\bibinfo {title} {Transfer-matrix method and monte carlo simulation in quantum spin systems},\ }\href {https://doi.org/10.1103/PhysRevB.31.2957} {\bibfield  {journal} {\bibinfo  {journal} {Phys. Rev. B}\ }\textbf {\bibinfo {volume} {31}},\ \bibinfo {pages} {2957} (\bibinfo {year} {1985})}\BibitemShut {NoStop}%
\bibitem [{\citenamefont {Fuchs}(1990)}]{FUCH}%
  \BibitemOpen
  \bibfield  {author} {\bibinfo {author} {\bibfnamefont {N.~H.}\ \bibnamefont {Fuchs}},\ }\bibfield  {title} {\bibinfo {title} {Transfer-matrix analysis for ising models},\ }\href {https://doi.org/10.1103/PhysRevB.41.2173} {\bibfield  {journal} {\bibinfo  {journal} {Phys. Rev. B}\ }\textbf {\bibinfo {volume} {41}},\ \bibinfo {pages} {2173} (\bibinfo {year} {1990})}\BibitemShut {NoStop}%
\bibitem [{\citenamefont {Wetterich}(2010{\natexlab{a}})}]{CWQPCS}%
  \BibitemOpen
  \bibfield  {author} {\bibinfo {author} {\bibfnamefont {C.}~\bibnamefont {Wetterich}},\ }\bibfield  {title} {\bibinfo {title} {{Quantum particles from classical statistics}},\ }\href {https://doi.org/10.1002/andp.201000088} {\bibfield  {journal} {\bibinfo  {journal} {Annalen Phys.}\ }\textbf {\bibinfo {volume} {522}},\ \bibinfo {pages} {807} (\bibinfo {year} {2010}{\natexlab{a}})},\ \Eprint {https://arxiv.org/abs/0904.3048} {arXiv:0904.3048} \BibitemShut {NoStop}%
\bibitem [{\citenamefont {Wetterich}(2025)}]{CWPW2020}%
  \BibitemOpen
  \bibfield  {author} {\bibinfo {author} {\bibfnamefont {C.}~\bibnamefont {Wetterich}},\ }\href {https://arxiv.org/abs/2011.02867} {\emph {\bibinfo {title} {{The probabilistic world}}}}\ (\bibinfo  {publisher} {Springer Nature},\ \bibinfo {address} {Heidelberg},\ \bibinfo {year} {2025})\ \Eprint {https://arxiv.org/abs/2011.02867} {arXiv:2011.02867} \BibitemShut {NoStop}%
\bibitem [{\citenamefont {von Neumann}(1951)}]{JVN}%
  \BibitemOpen
  \bibfield  {author} {\bibinfo {author} {\bibfnamefont {J.}~\bibnamefont {von Neumann}},\ }\bibinfo {title} {The general and logical theory of automata.}\ (\bibinfo  {publisher} {Wiley},\ \bibinfo {address} {Oxford, England},\ \bibinfo {year} {1951})\ pp.\ \bibinfo {pages} {1--41}\BibitemShut {NoStop}%
\bibitem [{\citenamefont {Ulam}(1950)}]{ULA}%
  \BibitemOpen
  \bibfield  {author} {\bibinfo {author} {\bibfnamefont {S.}~\bibnamefont {Ulam}},\ }\bibfield  {title} {\bibinfo {title} {Random processes and transformations},\ }in\ \href@noop {} {\emph {\bibinfo {booktitle} {Proceedings of the International Congress on Mathematics}}},\ Vol.~\bibinfo {volume} {2}\ (\bibinfo {year} {1950})\ pp.\ \bibinfo {pages} {264--275}\BibitemShut {NoStop}%
\bibitem [{\citenamefont {Zuse}(1969)}]{ZUS}%
  \BibitemOpen
  \bibfield  {author} {\bibinfo {author} {\bibfnamefont {K.}~\bibnamefont {Zuse}},\ }\href@noop {} {\emph {\bibinfo {title} {Rechnender Raum}}}\ (\bibinfo  {publisher} {Vieweg, Teubner Verlag},\ \bibinfo {year} {1969})\ pp.\ \bibinfo {pages} {1--3}\BibitemShut {NoStop}%
\bibitem [{\citenamefont {Hedlund}(1969)}]{HED}%
  \BibitemOpen
  \bibfield  {author} {\bibinfo {author} {\bibfnamefont {G.~A.}\ \bibnamefont {Hedlund}},\ }\bibfield  {title} {\bibinfo {title} {Endomorphisms and automorphisms of the shift dynamical system},\ }\href@noop {} {\bibfield  {journal} {\bibinfo  {journal} {Mathematical systems theory}\ }\textbf {\bibinfo {volume} {3}},\ \bibinfo {pages} {320} (\bibinfo {year} {1969})}\BibitemShut {NoStop}%
\bibitem [{\citenamefont {Gardner}(1970)}]{GAR}%
  \BibitemOpen
  \bibfield  {author} {\bibinfo {author} {\bibfnamefont {M.}~\bibnamefont {Gardner}},\ }\bibfield  {title} {\bibinfo {title} {Mathematical games},\ }\href@noop {} {\bibfield  {journal} {\bibinfo  {journal} {Scientific American}\ }\textbf {\bibinfo {volume} {223}},\ \bibinfo {pages} {120} (\bibinfo {year} {1970})}\BibitemShut {NoStop}%
\bibitem [{\citenamefont {Richardson}(1972)}]{RICH}%
  \BibitemOpen
  \bibfield  {author} {\bibinfo {author} {\bibfnamefont {D.}~\bibnamefont {Richardson}},\ }\bibfield  {title} {\bibinfo {title} {Tessellations with local transformations},\ }\href@noop {} {\bibfield  {journal} {\bibinfo  {journal} {Journal of Computer and System Sciences}\ }\textbf {\bibinfo {volume} {6}},\ \bibinfo {pages} {373} (\bibinfo {year} {1972})}\BibitemShut {NoStop}%
\bibitem [{\citenamefont {Amoroso}\ and\ \citenamefont {Patt}(1972)}]{AMPA}%
  \BibitemOpen
  \bibfield  {author} {\bibinfo {author} {\bibfnamefont {S.}~\bibnamefont {Amoroso}}\ and\ \bibinfo {author} {\bibfnamefont {Y.}~\bibnamefont {Patt}},\ }\bibfield  {title} {\bibinfo {title} {Decision procedures for surjectivity and injectivity of parallel maps for tessellation structures},\ }\href@noop {} {\bibfield  {journal} {\bibinfo  {journal} {Journal of Computer and System Sciences}\ }\textbf {\bibinfo {volume} {6}},\ \bibinfo {pages} {448} (\bibinfo {year} {1972})}\BibitemShut {NoStop}%
\bibitem [{\citenamefont {Lindenmayer}\ and\ \citenamefont {Rozenberg}(1976)}]{LIRO}%
  \BibitemOpen
  \bibfield  {author} {\bibinfo {author} {\bibfnamefont {A.}~\bibnamefont {Lindenmayer}}\ and\ \bibinfo {author} {\bibfnamefont {G.}~\bibnamefont {Rozenberg}},\ }\bibfield  {title} {\bibinfo {title} {Automata, languages, development}\ }(\bibinfo  {publisher} {North Holland},\ \bibinfo {year} {1976})\BibitemShut {NoStop}%
\bibitem [{\citenamefont {Hardy}\ \emph {et~al.}(1976)\citenamefont {Hardy}, \citenamefont {de~Pazzis},\ and\ \citenamefont {Pomeau}}]{HPP}%
  \BibitemOpen
  \bibfield  {author} {\bibinfo {author} {\bibfnamefont {J.}~\bibnamefont {Hardy}}, \bibinfo {author} {\bibfnamefont {O.}~\bibnamefont {de~Pazzis}},\ and\ \bibinfo {author} {\bibfnamefont {Y.}~\bibnamefont {Pomeau}},\ }\bibfield  {title} {\bibinfo {title} {Molecular dynamics of a classical lattice gas: Transport properties and time correlation functions},\ }\href@noop {} {\bibfield  {journal} {\bibinfo  {journal} {Phys. Rev. A}\ }\textbf {\bibinfo {volume} {13}},\ \bibinfo {pages} {1949} (\bibinfo {year} {1976})}\BibitemShut {NoStop}%
\bibitem [{\citenamefont {Toom}(1978)}]{TOOM}%
  \BibitemOpen
  \bibfield  {author} {\bibinfo {author} {\bibfnamefont {A.~L.}\ \bibnamefont {Toom}},\ }\href@noop {} {\emph {\bibinfo {title} {Locally Interacting Systems and their application in Biology}}}\ (\bibinfo  {publisher} {Springer Berlin Heidelberg},\ \bibinfo {year} {1978})\BibitemShut {NoStop}%
\bibitem [{\citenamefont {{R. L. Dobrushin}}(1978)}]{DKT}%
  \BibitemOpen
  \bibfield  {author} {\bibinfo {author} {\bibfnamefont {A.~L.~T.}\ \bibnamefont {{R. L. Dobrushin}}, \bibfnamefont {V.I.~Kryukov}},\ }\href@noop {} {\emph {\bibinfo {title} {Stochastic cellular systems: Ergodicity, Memory, Morphogenesis}}}\ (\bibinfo  {publisher} {Manchester University Press},\ \bibinfo {year} {1978})\BibitemShut {NoStop}%
\bibitem [{\citenamefont {Wolfram}(1983)}]{WOLF}%
  \BibitemOpen
  \bibfield  {author} {\bibinfo {author} {\bibfnamefont {S.}~\bibnamefont {Wolfram}},\ }\bibfield  {title} {\bibinfo {title} {Statistical mechanics of cellular automata},\ }\href {https://doi.org/10.1103/RevModPhys.55.601} {\bibfield  {journal} {\bibinfo  {journal} {Rev. Mod. Phys.}\ }\textbf {\bibinfo {volume} {55}},\ \bibinfo {pages} {601} (\bibinfo {year} {1983})}\BibitemShut {NoStop}%
\bibitem [{\citenamefont {Vichniac}(1984)}]{VICH}%
  \BibitemOpen
  \bibfield  {author} {\bibinfo {author} {\bibfnamefont {G.~Y.}\ \bibnamefont {Vichniac}},\ }\bibfield  {title} {\bibinfo {title} {Simulating physics with cellular automata},\ }\href@noop {} {\bibfield  {journal} {\bibinfo  {journal} {Physica D: Nonlinear Phenomena}\ }\textbf {\bibinfo {volume} {10}},\ \bibinfo {pages} {96} (\bibinfo {year} {1984})}\BibitemShut {NoStop}%
\bibitem [{\citenamefont {Preston}\ and\ \citenamefont {Duff}(1984)}]{PREDU}%
  \BibitemOpen
  \bibfield  {author} {\bibinfo {author} {\bibfnamefont {K.}~\bibnamefont {Preston}}\ and\ \bibinfo {author} {\bibfnamefont {M.~J.~B.}\ \bibnamefont {Duff}},\ }\href@noop {} {\emph {\bibinfo {title} {Modern Cellular Automata}}}\ (\bibinfo  {publisher} {Springer {US}},\ \bibinfo {year} {1984})\ pp.\ \bibinfo {pages} {1--15}\BibitemShut {NoStop}%
\bibitem [{\citenamefont {Creutz}(1986)}]{CREU}%
  \BibitemOpen
  \bibfield  {author} {\bibinfo {author} {\bibfnamefont {M.}~\bibnamefont {Creutz}},\ }\bibfield  {title} {\bibinfo {title} {Deterministic ising dynamics},\ }\href@noop {} {\bibfield  {journal} {\bibinfo  {journal} {Annals of Physics}\ }\textbf {\bibinfo {volume} {167}},\ \bibinfo {pages} {62} (\bibinfo {year} {1986})}\BibitemShut {NoStop}%
\bibitem [{\citenamefont {Toffoli}\ and\ \citenamefont {Margolus}(1990)}]{TOMA}%
  \BibitemOpen
  \bibfield  {author} {\bibinfo {author} {\bibfnamefont {T.}~\bibnamefont {Toffoli}}\ and\ \bibinfo {author} {\bibfnamefont {N.~H.}\ \bibnamefont {Margolus}},\ }\bibfield  {title} {\bibinfo {title} {Invertible cellular automata: A review},\ }\href@noop {} {\bibfield  {journal} {\bibinfo  {journal} {Physica D: Nonlinear Phenomena}\ }\textbf {\bibinfo {volume} {45}},\ \bibinfo {pages} {229} (\bibinfo {year} {1990})}\BibitemShut {NoStop}%
\bibitem [{\citenamefont {Louis}\ and\ \citenamefont {Nardi}(2018)}]{FLN}%
  \BibitemOpen
  \bibfield  {author} {\bibinfo {author} {\bibfnamefont {P.-Y.}\ \bibnamefont {Louis}}\ and\ \bibinfo {author} {\bibfnamefont {F.~R.}\ \bibnamefont {Nardi}},\ }\href {https://doi.org/10.1007/978-3-319-65558-1} {\emph {\bibinfo {title} {Probabilistic Cellular Automata: Theory, Applications and Future Perspectives}}}\ (\bibinfo  {publisher} {Springer International Publishing},\ \bibinfo {year} {2018})\BibitemShut {NoStop}%
\bibitem [{\citenamefont {{'t Hooft}}(2014)}]{GTH}%
  \BibitemOpen
  \bibfield  {author} {\bibinfo {author} {\bibfnamefont {G.}~\bibnamefont {{'t Hooft}}},\ }\href@noop {} {\bibinfo {title} {{The Cellular Automaton Interpretation of Quantum Mechanics. A View on the Quantum Nature of our Universe, Compulsory or Impossible?}}} (\bibinfo {year} {2014}),\ \Eprint {https://arxiv.org/abs/1405.1548} {arXiv:1405.1548} \BibitemShut {NoStop}%
\bibitem [{\citenamefont {Elze}(2014)}]{ELZE}%
  \BibitemOpen
  \bibfield  {author} {\bibinfo {author} {\bibfnamefont {H.-T.}\ \bibnamefont {Elze}},\ }\bibfield  {title} {\bibinfo {title} {{Quantumness of discrete Hamiltonian cellular automata}},\ }\href {https://doi.org/10.1051/epjconf/20147802005} {\bibfield  {journal} {\bibinfo  {journal} {EPJ Web Conf.}\ }\textbf {\bibinfo {volume} {78}},\ \bibinfo {pages} {02005} (\bibinfo {year} {2014})},\ \Eprint {https://arxiv.org/abs/1407.2160} {arXiv:1407.2160} \BibitemShut {NoStop}%
\bibitem [{\citenamefont {{'t Hooft}}(2010)}]{HOOFT2}%
  \BibitemOpen
  \bibfield  {author} {\bibinfo {author} {\bibfnamefont {G.}~\bibnamefont {{'t Hooft}}},\ }\bibfield  {title} {\bibinfo {title} {{Classical cellular automata and quantum field theory}},\ }\href {https://doi.org/10.1142/S0217751X10050469} {\bibfield  {journal} {\bibinfo  {journal} {Int. J. Mod. Phys. A}\ }\textbf {\bibinfo {volume} {25}},\ \bibinfo {pages} {4385} (\bibinfo {year} {2010})}\BibitemShut {NoStop}%
\bibitem [{\citenamefont {{'t Hooft}}(2021)}]{HOOFT3}%
  \BibitemOpen
  \bibfield  {author} {\bibinfo {author} {\bibfnamefont {G.}~\bibnamefont {{'t Hooft}}},\ }\bibfield  {title} {\bibinfo {title} {{Fast Vacuum Fluctuations and the Emergence of Quantum Mechanics}},\ }\href {https://doi.org/10.1007/s10701-021-00464-7} {\bibfield  {journal} {\bibinfo  {journal} {Found. Phys.}\ }\textbf {\bibinfo {volume} {51}},\ \bibinfo {pages} {63} (\bibinfo {year} {2021})},\ \Eprint {https://arxiv.org/abs/2010.02019} {arXiv:2010.02019} \BibitemShut {NoStop}%
\bibitem [{\citenamefont {Hooft}(2021)}]{HOOFT4}%
  \BibitemOpen
  \bibfield  {author} {\bibinfo {author} {\bibfnamefont {G.~t.}\ \bibnamefont {Hooft}},\ }\href@noop {} {\bibinfo {title} {{Explicit construction of Local Hidden Variables for any quantum theory up to any desired accuracy}}} (\bibinfo {year} {2021}),\ \Eprint {https://arxiv.org/abs/2103.04335} {arXiv:2103.04335} \BibitemShut {NoStop}%
\bibitem [{\citenamefont {Elze}(2025)}]{TEL2}%
  \BibitemOpen
  \bibfield  {author} {\bibinfo {author} {\bibfnamefont {H.-T.}\ \bibnamefont {Elze}},\ }\bibfield  {title} {\bibinfo {title} {The dirac equation, mass and arithmetic by permutations of automaton states},\ }\href {https://doi.org/10.3390/e27040395} {\bibfield  {journal} {\bibinfo  {journal} {Entropy}\ }\textbf {\bibinfo {volume} {27}},\ \bibinfo {pages} {395} (\bibinfo {year} {2025})}\BibitemShut {NoStop}%
\bibitem [{\citenamefont {Verhagen}(1976)}]{VER}%
  \BibitemOpen
  \bibfield  {author} {\bibinfo {author} {\bibfnamefont {A.~M.~W.}\ \bibnamefont {Verhagen}},\ }\bibfield  {title} {\bibinfo {title} {An exactly soluble case of the triangular ising model in a magnetic field},\ }\href {https://doi.org/10.1007/BF01012878} {\bibfield  {journal} {\bibinfo  {journal} {Journal of Statistical Physics}\ }\textbf {\bibinfo {volume} {15}},\ \bibinfo {pages} {219} (\bibinfo {year} {1976})}\BibitemShut {NoStop}%
\bibitem [{\citenamefont {Domany}(1984)}]{DOM}%
  \BibitemOpen
  \bibfield  {author} {\bibinfo {author} {\bibfnamefont {E.}~\bibnamefont {Domany}},\ }\bibfield  {title} {\bibinfo {title} {Exact {Results} for {Two}- and {Three}-{Dimensional} {Ising} and {Potts} {Models}},\ }\href {https://doi.org/10.1103/PhysRevLett.52.871} {\bibfield  {journal} {\bibinfo  {journal} {Physical Review Letters}\ }\textbf {\bibinfo {volume} {52}},\ \bibinfo {pages} {871} (\bibinfo {year} {1984})},\ \bibinfo {note} {publisher: American Physical Society}\BibitemShut {NoStop}%
\bibitem [{\citenamefont {Domany}\ and\ \citenamefont {Kinzel}(1984)}]{DK}%
  \BibitemOpen
  \bibfield  {author} {\bibinfo {author} {\bibfnamefont {E.}~\bibnamefont {Domany}}\ and\ \bibinfo {author} {\bibfnamefont {W.}~\bibnamefont {Kinzel}},\ }\bibfield  {title} {\bibinfo {title} {Equivalence of {Cellular} {Automata} to {Ising} {Models} and {Directed} {Percolation}},\ }\href {https://doi.org/10.1103/PhysRevLett.53.311} {\bibfield  {journal} {\bibinfo  {journal} {Physical Review Letters}\ }\textbf {\bibinfo {volume} {53}},\ \bibinfo {pages} {311} (\bibinfo {year} {1984})},\ \bibinfo {note} {publisher: American Physical Society}\BibitemShut {NoStop}%
\bibitem [{\citenamefont {Grinstein}\ \emph {et~al.}(1985)\citenamefont {Grinstein}, \citenamefont {Jayaprakash},\ and\ \citenamefont {He}}]{GJH}%
  \BibitemOpen
  \bibfield  {author} {\bibinfo {author} {\bibfnamefont {G.}~\bibnamefont {Grinstein}}, \bibinfo {author} {\bibfnamefont {C.}~\bibnamefont {Jayaprakash}},\ and\ \bibinfo {author} {\bibfnamefont {Y.}~\bibnamefont {He}},\ }\bibfield  {title} {\bibinfo {title} {Statistical {Mechanics} of {Probabilistic} {Cellular} {Automata}},\ }\href {https://doi.org/10.1103/PhysRevLett.55.2527} {\bibfield  {journal} {\bibinfo  {journal} {Physical Review Letters}\ }\textbf {\bibinfo {volume} {55}},\ \bibinfo {pages} {2527} (\bibinfo {year} {1985})},\ \bibinfo {note} {publisher: American Physical Society}\BibitemShut {NoStop}%
\bibitem [{\citenamefont {Rujàn}(1987)}]{RU}%
  \BibitemOpen
  \bibfield  {author} {\bibinfo {author} {\bibfnamefont {P.}~\bibnamefont {Rujàn}},\ }\bibfield  {title} {\bibinfo {title} {Cellular automata and statistical mechanical models},\ }\href {https://doi.org/10.1007/BF01009958} {\bibfield  {journal} {\bibinfo  {journal} {Journal of Statistical Physics}\ }\textbf {\bibinfo {volume} {49}},\ \bibinfo {pages} {139} (\bibinfo {year} {1987})}\BibitemShut {NoStop}%
\bibitem [{\citenamefont {Georges}\ and\ \citenamefont {{Le Doussal}}(1989)}]{GD}%
  \BibitemOpen
  \bibfield  {author} {\bibinfo {author} {\bibfnamefont {A.}~\bibnamefont {Georges}}\ and\ \bibinfo {author} {\bibfnamefont {P.}~\bibnamefont {{Le Doussal}}},\ }\bibfield  {title} {\bibinfo {title} {From equilibrium spin models to probabilistic cellular automata},\ }\href {https://doi.org/10.1007/BF01019786} {\bibfield  {journal} {\bibinfo  {journal} {Journal of Statistical Physics}\ }\textbf {\bibinfo {volume} {54}},\ \bibinfo {pages} {1011} (\bibinfo {year} {1989})}\BibitemShut {NoStop}%
\bibitem [{\citenamefont {Lebowitz}\ \emph {et~al.}(1990)\citenamefont {Lebowitz}, \citenamefont {Maes},\ and\ \citenamefont {Speer}}]{LMS}%
  \BibitemOpen
  \bibfield  {author} {\bibinfo {author} {\bibfnamefont {J.~L.}\ \bibnamefont {Lebowitz}}, \bibinfo {author} {\bibfnamefont {C.}~\bibnamefont {Maes}},\ and\ \bibinfo {author} {\bibfnamefont {E.~R.}\ \bibnamefont {Speer}},\ }\bibfield  {title} {\bibinfo {title} {Statistical mechanics of probabilistic cellular automata},\ }\href {https://doi.org/10.1007/BF01015566} {\bibfield  {journal} {\bibinfo  {journal} {Journal of Statistical Physics}\ }\textbf {\bibinfo {volume} {59}},\ \bibinfo {pages} {117} (\bibinfo {year} {1990})}\BibitemShut {NoStop}%
\bibitem [{\citenamefont {Petersen}\ and\ \citenamefont {Alstrøm}(1997)}]{PA}%
  \BibitemOpen
  \bibfield  {author} {\bibinfo {author} {\bibfnamefont {N.~K.}\ \bibnamefont {Petersen}}\ and\ \bibinfo {author} {\bibfnamefont {P.}~\bibnamefont {Alstrøm}},\ }\bibfield  {title} {\bibinfo {title} {Phase transition in an elementary probabilistic cellular automaton},\ }\href {https://doi.org/10.1016/S0378-4371(96)00410-4} {\bibfield  {journal} {\bibinfo  {journal} {Physica A: Statistical Mechanics and its Applications}\ }\textbf {\bibinfo {volume} {235}},\ \bibinfo {pages} {473} (\bibinfo {year} {1997})}\BibitemShut {NoStop}%
\bibitem [{\citenamefont {Fukś}(2003)}]{FU}%
  \BibitemOpen
  \bibfield  {author} {\bibinfo {author} {\bibfnamefont {H.}~\bibnamefont {Fukś}},\ }\bibfield  {title} {\bibinfo {title} {Probabilistic cellular automata with conserved quantities},\ }\href {https://doi.org/10.1088/0951-7715/17/1/010} {\bibfield  {journal} {\bibinfo  {journal} {Nonlinearity}\ }\textbf {\bibinfo {volume} {17}},\ \bibinfo {pages} {159} (\bibinfo {year} {2003})}\BibitemShut {NoStop}%
\bibitem [{\citenamefont {Mairesse}\ and\ \citenamefont {Marcovici}(2014)}]{MM}%
  \BibitemOpen
  \bibfield  {author} {\bibinfo {author} {\bibfnamefont {J.}~\bibnamefont {Mairesse}}\ and\ \bibinfo {author} {\bibfnamefont {I.}~\bibnamefont {Marcovici}},\ }\bibfield  {title} {\bibinfo {title} {Around probabilistic cellular automata},\ }\href {https://doi.org/10.1016/j.tcs.2014.09.009} {\bibfield  {journal} {\bibinfo  {journal} {Theoretical Computer Science}\ }\bibinfo {series} {Non-uniform {Cellular} {Automata}},\ \textbf {\bibinfo {volume} {559}},\ \bibinfo {pages} {42} (\bibinfo {year} {2014})}\BibitemShut {NoStop}%
\bibitem [{\citenamefont {Ray}\ \emph {et~al.}(2024)\citenamefont {Ray}, \citenamefont {Laflamme},\ and\ \citenamefont {Kubica}}]{RLK}%
  \BibitemOpen
  \bibfield  {author} {\bibinfo {author} {\bibfnamefont {A.}~\bibnamefont {Ray}}, \bibinfo {author} {\bibfnamefont {R.}~\bibnamefont {Laflamme}},\ and\ \bibinfo {author} {\bibfnamefont {A.}~\bibnamefont {Kubica}},\ }\bibfield  {title} {\bibinfo {title} {Protecting information via probabilistic cellular automata},\ }\href {https://doi.org/10.1103/PhysRevE.109.044141} {\bibfield  {journal} {\bibinfo  {journal} {Physical Review E}\ }\textbf {\bibinfo {volume} {109}},\ \bibinfo {pages} {044141} (\bibinfo {year} {2024})},\ \bibinfo {note} {publisher: American Physical Society}\BibitemShut {NoStop}%
\bibitem [{\citenamefont {Peschel}\ and\ \citenamefont {Emery}(1981)}]{PE}%
  \BibitemOpen
  \bibfield  {author} {\bibinfo {author} {\bibfnamefont {I.}~\bibnamefont {Peschel}}\ and\ \bibinfo {author} {\bibfnamefont {V.~J.}\ \bibnamefont {Emery}},\ }\bibfield  {title} {\bibinfo {title} {Calculation of spin correlations in two-dimensional {Ising} systems from one-dimensional kinetic models},\ }\href {https://doi.org/10.1007/BF01297524} {\bibfield  {journal} {\bibinfo  {journal} {Zeitschrift für Physik B Condensed Matter}\ }\textbf {\bibinfo {volume} {43}},\ \bibinfo {pages} {241} (\bibinfo {year} {1981})}\BibitemShut {NoStop}%
\bibitem [{\citenamefont {Arrighi}(2019)}]{Arrighi2019}%
  \BibitemOpen
  \bibfield  {author} {\bibinfo {author} {\bibfnamefont {P.}~\bibnamefont {Arrighi}},\ }\bibfield  {title} {\bibinfo {title} {An overview of quantum cellular automata},\ }\href {https://doi.org/10.1007/s11047-019-09762-6} {\bibfield  {journal} {\bibinfo  {journal} {Natural Computing}\ }\textbf {\bibinfo {volume} {18}},\ \bibinfo {pages} {885} (\bibinfo {year} {2019})},\ \Eprint {https://arxiv.org/abs/quant-ph/1904.12956} {arXiv:quant-ph/1904.12956} \BibitemShut {NoStop}%
\bibitem [{\citenamefont {Wetterich}(2010{\natexlab{b}})}]{CWFCS}%
  \BibitemOpen
  \bibfield  {author} {\bibinfo {author} {\bibfnamefont {C.}~\bibnamefont {Wetterich}},\ }\bibfield  {title} {\bibinfo {title} {{Fermions from classical statistics}},\ }\href {https://doi.org/10.1016/j.aop.2010.07.003} {\bibfield  {journal} {\bibinfo  {journal} {Annals Phys.}\ }\textbf {\bibinfo {volume} {325}},\ \bibinfo {pages} {2750} (\bibinfo {year} {2010}{\natexlab{b}})},\ \Eprint {https://arxiv.org/abs/1006.4254} {arXiv:1006.4254} \BibitemShut {NoStop}%
\bibitem [{\citenamefont {Wetterich}(2017)}]{CWFGI}%
  \BibitemOpen
  \bibfield  {author} {\bibinfo {author} {\bibfnamefont {C.}~\bibnamefont {Wetterich}},\ }\bibfield  {title} {\bibinfo {title} {{Fermions as generalized Ising models}},\ }\href {https://doi.org/10.1016/j.nuclphysb.2017.02.012} {\bibfield  {journal} {\bibinfo  {journal} {Nucl. Phys. B}\ }\textbf {\bibinfo {volume} {917}},\ \bibinfo {pages} {241} (\bibinfo {year} {2017})},\ \Eprint {https://arxiv.org/abs/1612.06695} {arXiv:1612.06695} \BibitemShut {NoStop}%
\bibitem [{\citenamefont {Wetterich}(2021{\natexlab{a}})}]{CWPCA}%
  \BibitemOpen
  \bibfield  {author} {\bibinfo {author} {\bibfnamefont {C.}~\bibnamefont {Wetterich}},\ }\bibfield  {title} {\bibinfo {title} {Probabilistic cellular automata for interacting fermionic quantum field theories},\ }\href@noop {} {\bibfield  {journal} {\bibinfo  {journal} {Nuclear Physics B}\ }\textbf {\bibinfo {volume} {963}},\ \bibinfo {pages} {115296} (\bibinfo {year} {2021}{\natexlab{a}})},\ \Eprint {https://arxiv.org/abs/2007.06366} {arXiv:2007.06366} \BibitemShut {NoStop}%
\bibitem [{\citenamefont {Wetterich}(2021{\natexlab{b}})}]{CWFCB}%
  \BibitemOpen
  \bibfield  {author} {\bibinfo {author} {\bibfnamefont {C.}~\bibnamefont {Wetterich}},\ }\href@noop {} {\bibinfo {title} {{Quantum fermions from classical bits}}} (\bibinfo {year} {2021}{\natexlab{b}}),\ \Eprint {https://arxiv.org/abs/2106.15517} {arXiv:2106.15517} \BibitemShut {NoStop}%
\bibitem [{\citenamefont {Wetterich}(2021{\natexlab{c}})}]{CWNEW}%
  \BibitemOpen
  \bibfield  {author} {\bibinfo {author} {\bibfnamefont {C.}~\bibnamefont {Wetterich}},\ }\href@noop {} {\bibinfo {title} {Fermionic quantum field theories as probabilistic cellular automata}} (\bibinfo {year} {2021}{\natexlab{c}}),\ \Eprint {https://arxiv.org/abs/hep-lat/2111.06728} {arXiv:hep-lat/2111.06728} \BibitemShut {NoStop}%
\bibitem [{\citenamefont {Wetterich}(2022{\natexlab{a}})}]{CWPCAQP}%
  \BibitemOpen
  \bibfield  {author} {\bibinfo {author} {\bibfnamefont {C.}~\bibnamefont {Wetterich}},\ }\href@noop {} {\bibinfo {title} {Probabilistic cellular automaton for quantum particle in a potential}} (\bibinfo {year} {2022}{\natexlab{a}}),\ \Eprint {https://arxiv.org/abs/quant-ph/2211.17034} {arXiv:quant-ph/2211.17034} \BibitemShut {NoStop}%
\bibitem [{\citenamefont {Wetterich}(2022{\natexlab{b}})}]{CWFPPCA}%
  \BibitemOpen
  \bibfield  {author} {\bibinfo {author} {\bibfnamefont {C.}~\bibnamefont {Wetterich}},\ }\href {http://arxiv.org/abs/2203.14081} {\bibinfo {title} {Fermion picture for cellular automata}} (\bibinfo {year} {2022}{\natexlab{b}}),\ \Eprint {https://arxiv.org/abs/2203.14081} {arXiv:2203.14081} \BibitemShut {NoStop}%
\bibitem [{\citenamefont {Plechko}(2005)}]{PLECH}%
  \BibitemOpen
  \bibfield  {author} {\bibinfo {author} {\bibfnamefont {V.~N.}\ \bibnamefont {Plechko}},\ }\bibfield  {title} {\bibinfo {title} {{Fermions and correlations in the two-dimensional Ising model}},\ }\href@noop {} {\bibfield  {journal} {\bibinfo  {journal} {Phys. Part. Nucl.}\ }\textbf {\bibinfo {volume} {36}},\ \bibinfo {pages} {S203} (\bibinfo {year} {2005})},\ \Eprint {https://arxiv.org/abs/hep-th/0512263} {arXiv:hep-th/0512263} \BibitemShut {NoStop}%
\bibitem [{\citenamefont {Berezin}(1966)}]{BER1}%
  \BibitemOpen
  \bibfield  {author} {\bibinfo {author} {\bibfnamefont {F.}~\bibnamefont {Berezin}},\ }\bibfield  {title} {\bibinfo {title} {The method of second quantization. nauka{\i}, moscow (1965)},\ }\href@noop {} {\bibfield  {journal} {\bibinfo  {journal} {English translation Academic Press, New York}\ } (\bibinfo {year} {1966})}\BibitemShut {NoStop}%
\bibitem [{\citenamefont {Berezin}(1969)}]{BER2}%
  \BibitemOpen
  \bibfield  {author} {\bibinfo {author} {\bibfnamefont {F.~A.}\ \bibnamefont {Berezin}},\ }\bibfield  {title} {\bibinfo {title} {The plane ising model},\ }\href {https://doi.org/10.1070/rm1969v024n03abeh001346} {\bibfield  {journal} {\bibinfo  {journal} {Russian Mathematical Surveys}\ }\textbf {\bibinfo {volume} {24}},\ \bibinfo {pages} {1} (\bibinfo {year} {1969})}\BibitemShut {NoStop}%
\bibitem [{\citenamefont {Samuel}(1980)}]{SAM}%
  \BibitemOpen
  \bibfield  {author} {\bibinfo {author} {\bibfnamefont {S.}~\bibnamefont {Samuel}},\ }\bibfield  {title} {\bibinfo {title} {{The Use of Anticommuting Integrals in Statistical Mechanics. 1.}},\ }\href {https://doi.org/10.1063/1.524404} {\bibfield  {journal} {\bibinfo  {journal} {J. Math. Phys.}\ }\textbf {\bibinfo {volume} {21}},\ \bibinfo {pages} {2806} (\bibinfo {year} {1980})}\BibitemShut {NoStop}%
\bibitem [{\citenamefont {Itzykson}(1982)}]{ITS}%
  \BibitemOpen
  \bibfield  {author} {\bibinfo {author} {\bibfnamefont {C.}~\bibnamefont {Itzykson}},\ }\bibfield  {title} {\bibinfo {title} {Ising fermions (i). two dimensions},\ }\href {https://doi.org/https://doi.org/10.1016/0550-3213(82)90173-0} {\bibfield  {journal} {\bibinfo  {journal} {Nuclear Physics B}\ }\textbf {\bibinfo {volume} {210}},\ \bibinfo {pages} {448} (\bibinfo {year} {1982})}\BibitemShut {NoStop}%
\bibitem [{\citenamefont {Plechko}(1985)}]{PLE1}%
  \BibitemOpen
  \bibfield  {author} {\bibinfo {author} {\bibfnamefont {V.~N.}\ \bibnamefont {Plechko}},\ }\bibfield  {title} {\bibinfo {title} {Simple solution of two-dimensional ising model on a torus in terms of grassmann integrals},\ }\href@noop {} {\bibfield  {journal} {\bibinfo  {journal} {Theoretical and Mathematical Physics}\ }\textbf {\bibinfo {volume} {64}},\ \bibinfo {pages} {748} (\bibinfo {year} {1985})}\BibitemShut {NoStop}%
\bibitem [{\citenamefont {Kreuzkamp}\ and\ \citenamefont {Wetterich}(2024)}]{Kreuzkamp2024}%
  \BibitemOpen
  \bibfield  {author} {\bibinfo {author} {\bibfnamefont {A.}~\bibnamefont {Kreuzkamp}}\ and\ \bibinfo {author} {\bibfnamefont {C.}~\bibnamefont {Wetterich}},\ }\href@noop {} {\bibinfo {title} {{Quantum Systems from Random Probabilistic Automata}}} (\bibinfo {year} {2024}),\ \Eprint {https://arxiv.org/abs/quant-ph/2405.09829} {arXiv:quant-ph/2405.09829} \BibitemShut {NoStop}%
\bibitem [{\citenamefont {Nielsen}\ and\ \citenamefont {Ninomiya}(1981{\natexlab{a}})}]{NINI1}%
  \BibitemOpen
  \bibfield  {author} {\bibinfo {author} {\bibfnamefont {H.~B.}\ \bibnamefont {Nielsen}}\ and\ \bibinfo {author} {\bibfnamefont {M.}~\bibnamefont {Ninomiya}},\ }\bibfield  {title} {\bibinfo {title} {{Absence of Neutrinos on a Lattice. 1. Proof by Homotopy Theory}},\ }\href {https://doi.org/10.1016/0550-3213(82)90011-6} {\bibfield  {journal} {\bibinfo  {journal} {Nucl. Phys. B}\ }\textbf {\bibinfo {volume} {185}},\ \bibinfo {pages} {20} (\bibinfo {year} {1981}{\natexlab{a}})},\ \bibinfo {note} {[Erratum: Nucl.Phys.B 195, 541 (1982)]}\BibitemShut {NoStop}%
\bibitem [{\citenamefont {Nielsen}\ and\ \citenamefont {Ninomiya}(1981{\natexlab{b}})}]{NINI2}%
  \BibitemOpen
  \bibfield  {author} {\bibinfo {author} {\bibfnamefont {H.~B.}\ \bibnamefont {Nielsen}}\ and\ \bibinfo {author} {\bibfnamefont {M.}~\bibnamefont {Ninomiya}},\ }\bibfield  {title} {\bibinfo {title} {{Absence of Neutrinos on a Lattice. 2. Intuitive Topological Proof}},\ }\href {https://doi.org/10.1016/0550-3213(81)90524-1} {\bibfield  {journal} {\bibinfo  {journal} {Nucl. Phys. B}\ }\textbf {\bibinfo {volume} {193}},\ \bibinfo {pages} {173} (\bibinfo {year} {1981}{\natexlab{b}})}\BibitemShut {NoStop}%
\bibitem [{\citenamefont {Friedan}(1982)}]{FRI}%
  \BibitemOpen
  \bibfield  {author} {\bibinfo {author} {\bibfnamefont {D.}~\bibnamefont {Friedan}},\ }\bibfield  {title} {\bibinfo {title} {{A Proof of the Nielsen-Ninomiya Theorem}},\ }\href {https://doi.org/10.1007/BF01403500} {\bibfield  {journal} {\bibinfo  {journal} {Commun. Math. Phys.}\ }\textbf {\bibinfo {volume} {85}},\ \bibinfo {pages} {481} (\bibinfo {year} {1982})}\BibitemShut {NoStop}%
\bibitem [{\citenamefont {Drell}\ \emph {et~al.}(1976)\citenamefont {Drell}, \citenamefont {Weinstein},\ and\ \citenamefont {Yankielowicz}}]{Drell1976}%
  \BibitemOpen
  \bibfield  {author} {\bibinfo {author} {\bibfnamefont {S.~D.}\ \bibnamefont {Drell}}, \bibinfo {author} {\bibfnamefont {M.}~\bibnamefont {Weinstein}},\ and\ \bibinfo {author} {\bibfnamefont {S.}~\bibnamefont {Yankielowicz}},\ }\bibfield  {title} {\bibinfo {title} {{Strong Coupling Field Theories. 2. Fermions and Gauge Fields on a Lattice}},\ }\href {https://doi.org/10.1103/PhysRevD.14.1627} {\bibfield  {journal} {\bibinfo  {journal} {Phys. Rev. D}\ }\textbf {\bibinfo {volume} {14}},\ \bibinfo {pages} {1627} (\bibinfo {year} {1976})}\BibitemShut {NoStop}%
\bibitem [{\citenamefont {Weinberg}(2014)}]{weinberg2014}%
  \BibitemOpen
  \bibfield  {author} {\bibinfo {author} {\bibfnamefont {S.}~\bibnamefont {Weinberg}},\ }\bibfield  {title} {\bibinfo {title} {The quantum theory of fields. 1: Foundations}\ }(\bibinfo  {publisher} {Cambridge Univ. Press},\ \bibinfo {year} {2014})\BibitemShut {NoStop}%
\bibitem [{\citenamefont {Lüscher}()}]{LUE}%
  \BibitemOpen
  \bibfield  {author} {\bibinfo {author} {\bibfnamefont {M.}~\bibnamefont {Lüscher}},\ }\href@noop {} {\bibinfo {title} {private communication}}\BibitemShut {NoStop}%
\bibitem [{\citenamefont {Aarts}\ \emph {et~al.}(2000)\citenamefont {Aarts}, \citenamefont {Bonini},\ and\ \citenamefont {Wetterich}}]{ABC1}%
  \BibitemOpen
  \bibfield  {author} {\bibinfo {author} {\bibfnamefont {G.}~\bibnamefont {Aarts}}, \bibinfo {author} {\bibfnamefont {G.~F.}\ \bibnamefont {Bonini}},\ and\ \bibinfo {author} {\bibfnamefont {C.}~\bibnamefont {Wetterich}},\ }\bibfield  {title} {\bibinfo {title} {{On Thermalization in classical scalar field theory}},\ }\href {https://doi.org/10.1016/S0550-3213(00)00447-8} {\bibfield  {journal} {\bibinfo  {journal} {Nucl. Phys. B}\ }\textbf {\bibinfo {volume} {587}},\ \bibinfo {pages} {403} (\bibinfo {year} {2000})},\ \Eprint {https://arxiv.org/abs/hep-ph/0003262} {arXiv:hep-ph/0003262} \BibitemShut {NoStop}%
\bibitem [{\citenamefont {Aarts}\ \emph {et~al.}(2001)\citenamefont {Aarts}, \citenamefont {Bonini},\ and\ \citenamefont {Wetterich}}]{ABC2}%
  \BibitemOpen
  \bibfield  {author} {\bibinfo {author} {\bibfnamefont {G.}~\bibnamefont {Aarts}}, \bibinfo {author} {\bibfnamefont {G.~F.}\ \bibnamefont {Bonini}},\ and\ \bibinfo {author} {\bibfnamefont {C.}~\bibnamefont {Wetterich}},\ }\bibfield  {title} {\bibinfo {title} {{Exact and truncated dynamics in nonequilibrium field theory}},\ }\href {https://doi.org/10.1103/PhysRevD.63.025012} {\bibfield  {journal} {\bibinfo  {journal} {Phys. Rev. D}\ }\textbf {\bibinfo {volume} {63}},\ \bibinfo {pages} {025012} (\bibinfo {year} {2001})},\ \Eprint {https://arxiv.org/abs/hep-ph/0007357} {arXiv:hep-ph/0007357} \BibitemShut {NoStop}%
\bibitem [{\citenamefont {Wetterich}(2022{\natexlab{c}})}]{CWCASG}%
  \BibitemOpen
  \bibfield  {author} {\bibinfo {author} {\bibfnamefont {C.}~\bibnamefont {Wetterich}},\ }\href@noop {} {\bibinfo {title} {Cellular automaton for spinor gravity in four dimensions}} (\bibinfo {year} {2022}{\natexlab{c}}),\ \Eprint {https://arxiv.org/abs/hep-lat/2211.09002} {arXiv:hep-lat/2211.09002} \BibitemShut {NoStop}%
\bibitem [{\citenamefont {Wetterich}(2010{\natexlab{c}})}]{CWQMCS}%
  \BibitemOpen
  \bibfield  {author} {\bibinfo {author} {\bibfnamefont {C.}~\bibnamefont {Wetterich}},\ }\bibfield  {title} {\bibinfo {title} {{Quantum mechanics from classical statistics}},\ }\href {https://doi.org/10.1016/j.aop.2009.12.006} {\bibfield  {journal} {\bibinfo  {journal} {Annals Phys.}\ }\textbf {\bibinfo {volume} {325}},\ \bibinfo {pages} {852} (\bibinfo {year} {2010}{\natexlab{c}})},\ \Eprint {https://arxiv.org/abs/0906.4919} {arXiv:0906.4919} \BibitemShut {NoStop}%
\bibitem [{\citenamefont {Wetterich}(2009)}]{CWEM}%
  \BibitemOpen
  \bibfield  {author} {\bibinfo {author} {\bibfnamefont {C.}~\bibnamefont {Wetterich}},\ }\bibfield  {title} {\bibinfo {title} {{Emergence of quantum mechanics from classical statistics}},\ }\href {https://doi.org/10.1088/1742-6596/174/1/012008} {\bibfield  {journal} {\bibinfo  {journal} {J. Phys. Conf. Ser.}\ }\textbf {\bibinfo {volume} {174}},\ \bibinfo {pages} {012008} (\bibinfo {year} {2009})},\ \Eprint {https://arxiv.org/abs/0811.0927} {arXiv:0811.0927} \BibitemShut {NoStop}%
\bibitem [{\citenamefont {Wetterich}(2024)}]{CWPW2024}%
  \BibitemOpen
  \bibfield  {author} {\bibinfo {author} {\bibfnamefont {C.}~\bibnamefont {Wetterich}},\ }\bibfield  {title} {\bibinfo {title} {The probabilistic world {II} : Quantum mechanics from classical statistics},\ }\href@noop {} {\  (\bibinfo {year} {2024})},\ \Eprint {https://arxiv.org/abs/2408.06379} {arXiv:2408.06379 [quant-ph]} \BibitemShut {NoStop}%
\bibitem [{\citenamefont {Thirring}(1958)}]{THI}%
  \BibitemOpen
  \bibfield  {author} {\bibinfo {author} {\bibfnamefont {W.~E.}\ \bibnamefont {Thirring}},\ }\bibfield  {title} {\bibinfo {title} {A soluble relativistic field theory},\ }\href {https://doi.org/https://doi.org/10.1016/0003-4916(58)90015-0} {\bibfield  {journal} {\bibinfo  {journal} {Annals of Physics}\ }\textbf {\bibinfo {volume} {3}},\ \bibinfo {pages} {91 } (\bibinfo {year} {1958})}\BibitemShut {NoStop}%
\bibitem [{\citenamefont {Klaiber}(1968)}]{KLA}%
  \BibitemOpen
  \bibfield  {author} {\bibinfo {author} {\bibfnamefont {B.}~\bibnamefont {Klaiber}},\ }\bibfield  {title} {\bibinfo {title} {{FU, The thirring model}},\ }\href@noop {} {\bibfield  {journal} {\bibinfo  {journal} {Lect. Theor. Phys. A}\ }\textbf {\bibinfo {volume} {10}},\ \bibinfo {pages} {141} (\bibinfo {year} {1968})}\BibitemShut {NoStop}%
\bibitem [{\citenamefont {Faber}\ and\ \citenamefont {Ivanov}(2001)}]{FAIV}%
  \BibitemOpen
  \bibfield  {author} {\bibinfo {author} {\bibfnamefont {M.}~\bibnamefont {Faber}}\ and\ \bibinfo {author} {\bibfnamefont {A.}~\bibnamefont {Ivanov}},\ }\href@noop {} {\bibinfo {title} {{On the solution of the massless Thirring model with fermion fields quantized in the chiral symmetric phase}}} (\bibinfo {year} {2001}),\ \Eprint {https://arxiv.org/abs/hep-th/0112183} {arXiv:hep-th/0112183} \BibitemShut {NoStop}%
\bibitem [{\citenamefont {Furuya}\ \emph {et~al.}(1982)\citenamefont {Furuya}, \citenamefont {Saraví},\ and\ \citenamefont {Schaposnik}}]{FUR}%
  \BibitemOpen
  \bibfield  {author} {\bibinfo {author} {\bibfnamefont {K.}~\bibnamefont {Furuya}}, \bibinfo {author} {\bibfnamefont {R.}~\bibnamefont {Saraví}},\ and\ \bibinfo {author} {\bibfnamefont {F.}~\bibnamefont {Schaposnik}},\ }\bibfield  {title} {\bibinfo {title} {Path-integral formulation of chiral invariant fermion models in two dimensions},\ }\href {https://doi.org/https://doi.org/10.1016/0550-3213(82)90191-2} {\bibfield  {journal} {\bibinfo  {journal} {Nuclear Physics B}\ }\textbf {\bibinfo {volume} {208}},\ \bibinfo {pages} {159 } (\bibinfo {year} {1982})}\BibitemShut {NoStop}%
\bibitem [{\citenamefont {Na\'on}(1985)}]{NAO}%
  \BibitemOpen
  \bibfield  {author} {\bibinfo {author} {\bibfnamefont {C.~M.}\ \bibnamefont {Na\'on}},\ }\bibfield  {title} {\bibinfo {title} {Abelian and non-abelian bosonization in the path-integral framework},\ }\href {https://doi.org/10.1103/PhysRevD.31.2035} {\bibfield  {journal} {\bibinfo  {journal} {Phys. Rev. D}\ }\textbf {\bibinfo {volume} {31}},\ \bibinfo {pages} {2035} (\bibinfo {year} {1985})}\BibitemShut {NoStop}%
\bibitem [{\citenamefont {Coleman}(1975)}]{COL}%
  \BibitemOpen
  \bibfield  {author} {\bibinfo {author} {\bibfnamefont {S.}~\bibnamefont {Coleman}},\ }\bibfield  {title} {\bibinfo {title} {Quantum sine-gordon equation as the massive thirring model},\ }\href {https://doi.org/10.1103/PhysRevD.11.2088} {\bibfield  {journal} {\bibinfo  {journal} {Phys. Rev. D}\ }\textbf {\bibinfo {volume} {11}},\ \bibinfo {pages} {2088} (\bibinfo {year} {1975})}\BibitemShut {NoStop}%
\bibitem [{\citenamefont {Damgaard}\ \emph {et~al.}(1992)\citenamefont {Damgaard}, \citenamefont {Nielsen},\ and\ \citenamefont {Sollacher}}]{DNS}%
  \BibitemOpen
  \bibfield  {author} {\bibinfo {author} {\bibfnamefont {P.}~\bibnamefont {Damgaard}}, \bibinfo {author} {\bibfnamefont {H.}~\bibnamefont {Nielsen}},\ and\ \bibinfo {author} {\bibfnamefont {R.}~\bibnamefont {Sollacher}},\ }\bibfield  {title} {\bibinfo {title} {Smooth bosonization: The cheshire cat revisited},\ }\href {https://doi.org/https://doi.org/10.1016/0550-3213(92)90100-P} {\bibfield  {journal} {\bibinfo  {journal} {Nuclear Physics B}\ }\textbf {\bibinfo {volume} {385}},\ \bibinfo {pages} {227 } (\bibinfo {year} {1992})}\BibitemShut {NoStop}%
\bibitem [{\citenamefont {Gross}\ and\ \citenamefont {Neveu}(1974)}]{GN}%
  \BibitemOpen
  \bibfield  {author} {\bibinfo {author} {\bibfnamefont {D.~J.}\ \bibnamefont {Gross}}\ and\ \bibinfo {author} {\bibfnamefont {A.}~\bibnamefont {Neveu}},\ }\bibfield  {title} {\bibinfo {title} {Dynamical symmetry breaking in asymptotically free field theories},\ }\href@noop {} {\bibfield  {journal} {\bibinfo  {journal} {Phys. Rev. D}\ }\textbf {\bibinfo {volume} {10}},\ \bibinfo {pages} {3235} (\bibinfo {year} {1974})}\BibitemShut {NoStop}%
\bibitem [{\citenamefont {Wetzel}(1984)}]{WWE}%
  \BibitemOpen
  \bibfield  {author} {\bibinfo {author} {\bibfnamefont {W.}~\bibnamefont {Wetzel}},\ }\bibfield  {title} {\bibinfo {title} {{Two-loop beta-function for the Gross-Neveu model}},\ }\href {https://cds.cern.ch/record/156504} {\bibfield  {journal} {\bibinfo  {journal} {Phys. Lett. B}\ }\textbf {\bibinfo {volume} {153}},\ \bibinfo {pages} {297} (\bibinfo {year} {1984})}\BibitemShut {NoStop}%
\bibitem [{\citenamefont {Shankar}(1985)}]{RSHA}%
  \BibitemOpen
  \bibfield  {author} {\bibinfo {author} {\bibfnamefont {R.}~\bibnamefont {Shankar}},\ }\bibfield  {title} {\bibinfo {title} {Ashkin-teller and gross-neveu models: New relations and results},\ }\href {https://link.aps.org/doi/10.1103/PhysRevLett.55.453} {\bibfield  {journal} {\bibinfo  {journal} {Phys. Rev. Lett.}\ }\textbf {\bibinfo {volume} {55}},\ \bibinfo {pages} {453} (\bibinfo {year} {1985})}\BibitemShut {NoStop}%
\bibitem [{\citenamefont {Rosenstein}\ \emph {et~al.}(1991)\citenamefont {Rosenstein}, \citenamefont {Warr},\ and\ \citenamefont {Park}}]{RWP}%
  \BibitemOpen
  \bibfield  {author} {\bibinfo {author} {\bibfnamefont {B.}~\bibnamefont {Rosenstein}}, \bibinfo {author} {\bibfnamefont {B.~J.}\ \bibnamefont {Warr}},\ and\ \bibinfo {author} {\bibfnamefont {S.~H.}\ \bibnamefont {Park}},\ }\bibfield  {title} {\bibinfo {title} {Dynamical symmetry breaking in four-fermion interaction models},\ }\href {https://doi.org/10.1016/0370-1573(91)90129-a} {\bibfield  {journal} {\bibinfo  {journal} {Physics Reports}\ }\textbf {\bibinfo {volume} {205}},\ \bibinfo {pages} {59} (\bibinfo {year} {1991})}\BibitemShut {NoStop}%
\bibitem [{\citenamefont {Stoll}\ \emph {et~al.}(2021)\citenamefont {Stoll}, \citenamefont {Zorbach}, \citenamefont {Koenigstein}, \citenamefont {Steil},\ and\ \citenamefont {Rechenberger}}]{SZKSR}%
  \BibitemOpen
  \bibfield  {author} {\bibinfo {author} {\bibfnamefont {J.}~\bibnamefont {Stoll}}, \bibinfo {author} {\bibfnamefont {N.}~\bibnamefont {Zorbach}}, \bibinfo {author} {\bibfnamefont {A.}~\bibnamefont {Koenigstein}}, \bibinfo {author} {\bibfnamefont {M.~J.}\ \bibnamefont {Steil}},\ and\ \bibinfo {author} {\bibfnamefont {S.}~\bibnamefont {Rechenberger}},\ }\href@noop {} {\bibinfo {title} {Bosonic fluctuations in the $( 1 + 1 )$-dimensional gross-neveu(-yukawa) model at varying $\mu$ and $t$ and finite $n$}} (\bibinfo {year} {2021}),\ \Eprint {https://arxiv.org/abs/hep-ph/2108.10616} {arXiv:hep-ph/2108.10616} \BibitemShut {NoStop}%
\end{thebibliography}%

\end{document}